\documentclass{ati}
\usepackage[utf8]{inputenc}
\usepackage[T1]{fontenc}
\usepackage[estonian,english]{babel}
\usepackage{enumerate,amsmath,amssymb,amsthm,graphicx,color,cite,url}

\usepackage[nottoc]{tocbibind}

\usepackage[bookmarks,colorlinks,
  linkcolor=black,citecolor=black,filecolor=black,urlcolor=black,
  pdfhighlight=/O,pdfstartview=FitH,unicode]{hyperref}
\usepackage{epigraph}

\usepackage{adjustbox}
\usepackage{makecell}
\usepackage{colortbl}

\usepackage{color,soul}

\usepackage{tikz, blindtext}

\usepackage{pdfpages}
    
\begin{document}

% == title pages ==
\def\seriesname{Dissertationes informaticae Universitatis Tartuensis}
\def\seriesnumber{28}
\pagestyle{empty}\begingroup\centering\sffamily
\vbox{\small\MakeUppercase{\seriesname}}
\vbox{\bfseries\small\seriesnumber}
\vfill\newpage\null\newpage
\vbox{\small\MakeUppercase{\seriesname}}
\vbox{\bfseries\small\seriesnumber}
\vskip52mm
\vbox{\bfseries\LARGE\MakeUppercase{Dmytro Fishman}}
\vskip12mm
\vbox{\LARGE{Developing a data analysis pipeline for automated protein profiling in immunology}}

\vfill
\vbox{\MakeUppercase{Tartu \the\year}}
\endgroup\newpage

% == reverse of title page ==
\begingroup\setlength\parindent{0pt}\setlength\parskip\baselineskip
Institute of Computer Science, Faculty of Science and Technology, 
University of Tartu, Estonia.

Dissertation has been accepted for the commencement of the degree of
Doctor of Philosophy (PhD) in informatics on May 20, 2021 by the Council of
the Institute of Computer Science, University of Tartu.

\emph{Supervisors}

\begin{tabular}{p{20mm}l}
Dr. & Hedi Peterson\\
& University of Tartu\\
\\\\
Prof. & Jaak Vilo\\
& University of Tartu\\
\\\\
Prof. & P{\"a}rt Peterson\\
& University of Tartu\\

\end{tabular}

\emph{Opponents}

\begin{tabular}{p{20mm}l}
Dr.& Jessica Da Gama Duarte
\\
& Olivia Newton-John Cancer Research Institute, Australia\\
\\
Dr. & Fridtjof Lund-Johansen\\
& Oslo University Hospital, Norway\\
\end{tabular}

The public defense will take place on 28 June, 2021 at 09:15 in Zoom.

The publication of this dissertation was financed by the Institute of
Computer Science, University of Tartu.

Copyright \textcopyright\ \the\year\ by Dmytro Fishman

ISSN 2613-5906\\
ISBN 978-9949-03-624-0 (print)\\
ISBN 978-9949-03-625-7 (PDF)

University of Tartu Press\\
\url{http://www.tyk.ee/}
\endgroup

% == dedication ==

\newpage\null\vskip35mm
\begin{flushright}
\epigraph{\emph{We’re all in it together against Lord Voldemort and the House Slytherin.}} {Robert M Sapolsky} 
\end{flushright}
\newpage\pagestyle{plain}

% == abstract ==
\chapter*{Abstract}
Accurate information about protein content in the organism is instrumental for a better understanding of human biology and disease mechanisms. While the presence of certain types of proteins can be life-threatening, the abundance of others is an essential condition for an individual's overall well-being. Protein microarray is a technology that enables the quantification of thousands of proteins in hundreds of human samples in a parallel manner. In a series of studies involving protein microarrays, we have explored and implemented various data science methods for all-around analysing of these data. This analysis has enabled the identification and characterisation of proteins targeted by the autoimmune reaction in patients with the APS1 condition. We have also assessed the utility of applying machine learning methods alongside statistical tests in a study based on protein expression data to evaluate potential biomarkers for endometriosis. The keystone of this work is a web-tool PAWER. PAWER implements relevant computational methods, and provides a semi-automatic way to run the analysis of protein microarray data online in a drag-and-drop and click-and-play style. The source code of the tool is publicly available. The work that laid the foundation of this thesis has been instrumental for a number of subsequent studies of human disease and also inspired a contribution to refining standards for validation of machine learning methods in biology.

% == table of contents ==

\tableofcontents
\listoffigures
\listoftables
\chapter*{List of abbreviations}
 
\begin{tabbing}
    AIRE \indent\indent \= autoimmune regulator gene \\
    APECED \> autoimmune polyendocrinopathy-candidiasis-ectodermal dystrophy \\
    APS1 \> autoimmune polyendocrine syndrome type 1 \\
    CV\>cross-validation \\
    DBSCAN\>density-based spatial clustering of applications with noise \\
    DNA \> deoxyribonucleic acid \\
    ELISA \> enzyme-linked immunosorbent assays \\
    ENSG \> Ensembl gene ID \\
    FDR\>false discovery rate \\
    GAL\>GenePix array list \\
    GPR\>GenePix result file format \\
    IQR \> interquartile range \\
    PAA\>protein array analyser \\
    PAWER\>protein microarray web-explorer \\
    PMA\>protein microarray analyser \\
    PMD\>protein microarray database \\
    RLM\>robust linear model \\ 
    RNA\> ribonucleic acid \\
    RPPA\> reverse-phase protein arrays \\
    SNP\>single nucleotide polymorphism \\
    TIFF \> tag image file format \\
    T1D \> type 1 diabetes
\end{tabbing}
% == list of publications ==

\chapter*{List of original publications}
\addcontentsline{toc}{chapter}{List of original publications}

\section*{Publications included in the thesis}
\begin{enumerate}[I]
    
    \item Steffen Meyer, Martin Woodward, Christina Hertel, Philip Vlaicu, Yasmin Haque, Jaanika Kärner, Annalisa Macagno, Shimobi C. Onuoha, \textbf{Dmytro Fishman}, Hedi Peterson, Kaja Metsküla, Raivo Uibo, Kirsi Jäntti, Kati Hokynar, Anette S.B. Wolff, Kai Krohn, Annamari Ranki, Pärt Peterson, Kai Kisand, Adrian Hayday, Antonella Meloni, Nicolas Kluger, Eystein S. Husebye, Katarina Trebusak Podkrajsek, Tadej Battelino, Nina Bratanic, and Aleksandr Peet.  AIRE-deficient patients harbor unique high-affinity disease-\linebreak ameliorating autoantibodies. \textbf{Cell}, July 2016.
    \item \textbf{Dmytro Fishman}$^{*}$, Kai Kisand$^{*}$, Christina Hertel, Mike Rothe, Anu Remm, Maire Pihlap, Priit Adler, Jaak Vilo, Aleksandr Peet, Antonella Meloni, Kata-rina Trebusak Podkrajsek, Tadej Battelino, Øyvind Bruserud, Anette S.B. Wolff, Eystein S. Husebye, Nicolas Kluger, Kai Krohn, Annamari Ranki, Hedi Peterson, Adrian Hayday, and Pärt Peterson. Autoantibody repertoire in APECED patients targets two distinct subgroups of proteins. \textbf{Frontiers in Immunology}, August 2017.
    \item \textbf{Dmytro Fishman}, Ivan Kuzmin, Priit Adler, Jaak Vilo, and Hedi Peterson. PAWER: Protein Array Web ExploreR. \textbf{BMC Bioinformatics}, September 2020
    \item Tamara Knific$^{*}$, \textbf{Dmytro Fishman}$^{*}$, Andrej Vogler, Manuela Gstöttner, Rene Wenzl, Hedi Peterson, and Tea Lanisnik Rizner. Multiplex analysis of 40 cytokines do not allow separation between endometriosis patients and controls. \textbf{Scientific Reports}, November 2019. \\
\end{enumerate}

\section*{Publications not included in the thesis}
\begin{enumerate}[I]
    \item Aigar Ottas, \textbf{Dmytro Fishman}, Tiia-Linda Okas, Külli Kingo, and Ursel Soomets. The metabolic analysis of psoriasis identifies the associated metabo-lites while providing computational models for the monitoring of the disease. \textbf{Archives of Dermatological Research}, July 2017.
    \item Aigar Ottas, \textbf{Dmytro Fishman}, Tiia-Linda Okas, Tõnu Püssa, Peeter Toomik, Aare Märtson, Külli Kingo, and Ursel Soomets. Blood serum metabolome ofatopic dermatitis: Altered energy cycle and the markers of systemic inflam-mation. \textbf{PLOS ONE}, November 2017.
    \item William Jones$^{*}$, Kaur Alasoo$^{*}$, \textbf{Dmytro Fishman}$^{*}$, and Leopold Parts. Computational biology: deep learning. \textbf{Emerging Topics in Life Sciences}, November 2017. \\
    \item Ardi Tampuu, Maksym Semikin, Naveed Muhammad, \textbf{Dmytro Fishman}, Tambet Matiisen. A Survey of End-to-End Driving: Architectures and Training Methods. \textbf{IEEE Transactions on Neural Networks and Learning Systems}, March 2020. 
    \item Ian Walsh$^{*}$, \textbf{Dmytro Fishman}$^{*}$, Dario Garcia-Gasulla, Tiina Titma, The ELIXIR Machine Learning focus group, Jennifer Harrow, Fotis E. Psomopoulos and Silvio C.E. Tosatto. DOME: Recommendations for supervised machine learning validation in biology. \textbf{Nature Methods}, May 2021. \\
\end{enumerate}  
\section*{Preprints}
\begin{enumerate}[I]
    \item \textbf{Dmytro Fishman}, Sten-Oliver Salumaa, Daniel Majoral, Samantha Peel, Jan Wildenhain, Alexander Schreiner, Kaupo Palo, Leopold Parts. Segmenting nuclei in brightfield images with neural networks. \textbf{bioRxiv}, August 2019.
    
\end{enumerate}

$^{*}$ -- shared first author.

%%%%%%%%%%%%%%%%%%%%%%%%%%%%%%%%%%%%%%%%%%%%%%%%%%%%%%%%%%%%%%%%%%%%%%%%%%%%%%%%%%%%%%%%%%%%%%%%%%%%%
% INTRODUCTION CHAPTER
%%%%%%%%%%%%%%%%%%%%%%%%%%%%%%%%%%%%%%%%%%%%%%%%%%%%%%%%%%%%%%%%%%%%%%%%%%%%%%%%%%%%%%%%%%%%%%%%%%%%%

\chapter{Introduction}

% Summarising every chapter with a paragraph

Proteins are essential elements of all living organisms. A great number of life-critical functions depend on these complex molecules. The amount of proteins in an organism's cells is strictly regulated, as an excessive amount or sudden shortage can cause unwanted consequences. Abnormal protein levels can be a sign of a serious malfunction. For example, in the presence of certain types of immune proteins, the immunoglobulins may attack the body's own cells and tissues causing various autoimmune conditions, such as diabetes or multiple sclerosis. Therefore, an ability to accurately assess protein concentrations in the body can be the key to the understanding of various important biological processes including disease mechanisms.

Protein microarray is a popular approach for quantifying protein concentrations in a sample. Hundreds or even thousands of protein concentrations can be measured in parallel. Depending on what is captured on the slide, it is possible to measure either the full proteome of the cell or more specifically autoantibodies present in the sample. Hence, term ``protein profiling'' in the title of the thesis refers to the computational analysis of such protein targets, mostly autoantibodies, derived from protein microarray experiments. Although protein microarrays do not always provide precise information about the number of proteins participating in the process in a particular sample, their readings help to steer the analysis towards the most promising targets.

Despite protein microarrays having a lot in common with DNA microarrays, due to different biological assumptions, not all computational methods developed for the latter translate well to the former. Therefore, methods tailored specifically to protein microarrays are absolutely necessary to efficiently use the full capabilities of the platform.

The classical protein data analysis pipeline is complex and consists of a series of computational methods applied sequentially. Methods for reducing technical noise, detecting and removing outlier observations, and normalising resulting signal values are all necessary to ensure the validity of the analysis. Statistical tests, as well as machine learning methods, are used to identify individual proteins as well as their combinations with sufficiently contrasting concentration levels between experimental conditions. Finally, enrichment analysis tools help to put such proteins into the perspective of the most prevalent biological functions. In this thesis, we explored computational methods and optimised the data analysis pipeline applicable to data acquired from protein microarray experiments. As a result, we developed and released a web-tool that helps to perform the entire analysis in a semi-automatic fashion. Methods described in this work were put into practice and validated in several protein microarray-related publications.

The work towards this thesis started with an analysis of protein microarray data from a study that explored the autoimmune content of the blood from patients with autoimmune polyendocrine syndrome type 1 (APS1) \cite{Meyer2016}. In order to define an initial list of proteins targeted by the autoimmune reaction in APS1 patients, we implemented a protein microarray-specific pre-processing pipeline as well as performed differential analysis. 

We aimed at a more profound understanding of the mechanisms behind APS1 condition and autoimmunity in general. Therefore, protein targets identified in the previous publication were studied further \cite{Fishman2017}. We analysed multiple open databases and public protein datasets to determine common biological factors behind selected protein targets. We used a web server for the functional enrichment analysis to validate our results. 

A focal point of this PhD work is the protein microarray web-explorer (PAWER) -- an R-based web-tool, developed to enable semi-automatic protein microarray analysis \cite{Fishman2019}. PAWER incorporates all the relevant computational methods implemented in the previous papers. Its intuitive user interface and step-by-step workflow are designed to help perform protein microarray analysis with ease. 

Finally, in the fourth publication included in this thesis, we have explored the value of applying machine learning models along with classical statistical methods discussed in the previous publications. Here we analysed a case-control study of endometriosis \cite{Knific2019}. Prior statistical analysis had shown that no individual proteins are capable of distinguishing endometriosis patients from controls based on proteins found in the blood. We used several powerful machine learning metods to evaluate the predictive performance of combinations of proteins. In line with statistical test results, neither model achieved performance significantly better than the random chance. Therefore, machine learning results supported the hypothesis that neither measuring individual proteins nor in combination with others can predict endometriosis and thus help to diagnose the disease in the given samples.

\section{Main contributions of the thesis}
\begin{enumerate}
    \item Enabling a series of biological findings with a help of a custom all-around protein microarray data analysis pipeline, from protein microarray specific pre-processing and signal normalisation to differential and enrichment analysis.
    \item Development of the protein array web-explorer -- intuitive web-tool that incorporates computational methods relevant to protein microarrays and enables semi-automatic analysis of protein microarray data.
    \item Exploring the application of machine learning methods to protein concentration data, adding a new dimension to classical biomarker discovery practice.  
\end{enumerate}

%%%%%%%%%%%%%%%%%%%%%%%%%%%%%%%%%%%%%%%%%%%%%%%%%%%%%%%%%%%%%%%%%%%%%%%%%%%%%%%%%%%%%%%%%%%%%%%%%%%%%
% BIOLOGICAL BACKGROUND CHAPTER
%%%%%%%%%%%%%%%%%%%%%%%%%%%%%%%%%%%%%%%%%%%%%%%%%%%%%%%%%%%%%%%%%%%%%%%%%%%%%%%%%%%%%%%%%%%%%%%%%%%%%

\chapter{Introduction to proteins}

Life is beautiful in its complexity, and proteins are some of the most fundamental biological elements that enable this complexity. Often referred to as ``workhorses'' of the cell, proteins are responsible for almost every imaginable item on organisms' to-do list \cite{behave}. Proteins carry out tasks ranging from building tissue and replicating deoxyribonucleic acid (DNA) to enabling timely immune response and facilitating oxygen delivery, albeit different types of proteins are at work. The number of proteins present at any given moment is strictly regulated by cells \cite{Vogel2012}, as any significant deviation from the norm may cause malfunction and even disease. Therefore, information about protein abundance can offer valuable insight into the mechanisms of various diseases. In this work, we focused on quantifying protein abundance in a human body via protein microarray technology. The current chapter will present the biological context relevant to this thesis.

\section{From DNA to proteins and back}

One of the key principles behind the scientific approach is being open to new evidence that contradicts established doctrines. However, there is one scientific dogma that remained present in the discourse over the years -- the central dogma of molecular biology. Coined by Francis Crick in 1957 and published in 1958 \cite{centraldogma}, it states that DNA in the cell nucleus gets transcribed into ribonucleic acid (RNA), which in its turn is used to produce proteins. Although being proven wrong on a number of occasions, e.g., transcription factor proteins that regulate RNA production, the central dogma remains a useful approximation for the most important biological interactions in a cell.  

DNA is a long double-stranded molecule that consists of four nucleotides: adenine (A), thymine (T), guanine (G), and cytosine (C). Nucleotides form pair-wise bonds (A with T and C with G), helping to hold two strands of DNA together. Human DNA is made up of approximately 3.6 billion nucleotide pairs forming our complete genetic blueprint. Albeit an impressive number, only a fraction of DNA has been associated with relevant biological functions in the organism. These functional regions are called genes. The role of the majority of DNA remains largely unestablished. Because DNA is a static molecule that never leaves the cell nucleus, to execute biological functions, genes need to send instructions to the rest of the organism. They do it via the process of transcription, which transfers information stored in genes into a molecule called RNA. RNA travels to structures called ribosomes, where both take part in synthesising proteins.   
 
Proteins are complex molecules made up of 20 amino acids, each with its unique chemical properties \cite{Johnson2002}. Proteins vastly vary in length from about 200 to almost 27,000 amino acids \cite{Fulton1991}. Such variability naturally implies rich structural and functional diversity of resulting molecules. The total number of proteins in humans remains a subject of scientific debate, with estimates ranging from modest 20,000 proteins, if assumed that one gene is responsible for one protein (i.e., canonical proteins) to hundreds of thousands if the combinatorial nature of gene expression and alternative splicing is taken into account \cite{Ponomarenko2016}. While some proteins leave the native cell to operate elsewhere (e.g., pancreas produces insulin to help regulate blood sugar level), others remain to help facilitate domestic processes, including regulating the gene expression. Transcription factor proteins bind to DNA and either suppress or enhance RNA production of nearby genes, directly violating the basic premise of the central dogma of molecular biology. This creates a cycle of regulation, where genes create RNA that initiates the production of proteins, which in their turn regulates the gene expression. 

Proteins do most of the work in cells and tissues and are required for many critical processes in the body \cite{Johnson2002}. For example, the immune system employs special types of proteins -- antibodies to recognise foreign substances and fight infections. Most of the work presented in this thesis is focused on antibodies and their role in immune response, therefore the following section will dive into immunity.

\section{The immune system}

\textit{Despite continuous advances in the understanding of the immune system, the subject of the matter remains so vast and complex that we consider an in-depth discussion of immunity to be well beyond the scope of this thesis. The section below is meant to introduce readers to the most central concepts that are essential for the autoimmunity discussion that will follow.}

The term \textit{immunity} comes from Latin \textit{imm\=unit\=as}, which referred to legal protection offered to Roman senators during their time in the office \cite{abbas2011}. In biology, immunity is defined as the defense system of an organism from infections and other intruders. The immune system is a large network of cells, tissues, and organs that tirelessly work together to protect the organism from anything that is recognised as an `invader' or `foreign', for example, bacteria, virus, parasite, cancer cell, or toxin \cite{Marshall2018}. While not all invaders are necessarily harmful, those foreign substances that have disease-causing potential are called \textit{pathogens}. 

\begin{figure*}
    \centering
    \includegraphics[trim=0 0 5 0, clip, width=\columnwidth]{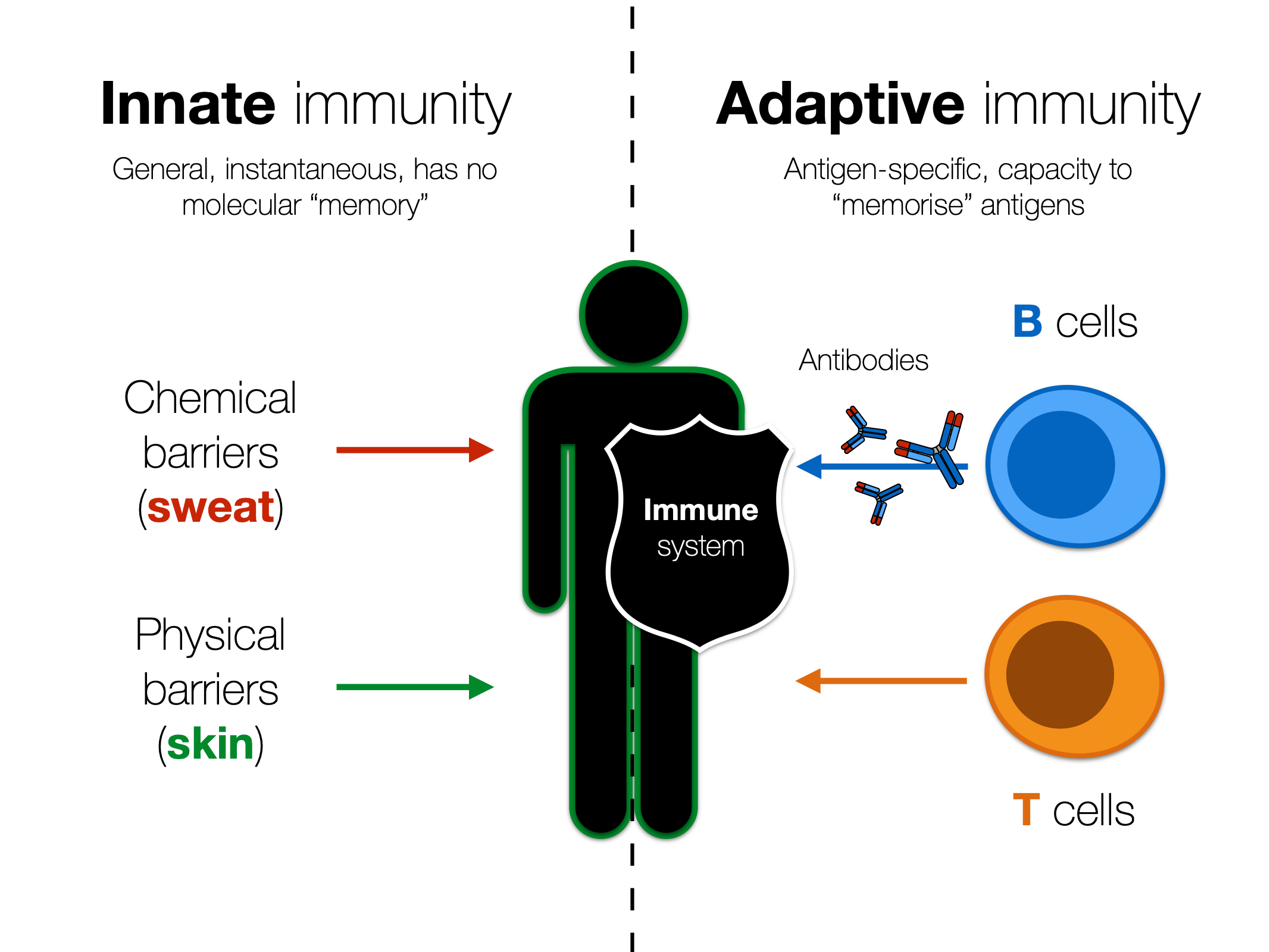}
    \caption{Immune system can be broadly divided into two subsystems: innate and adaptive immunity. Two systems supporting each other in their own ways to fight intruders. With the latter being slower and more specific, while the former is faster and more general.}
    \label{immunity}
\end{figure*}

The immune system can be broadly divided into two main subsystems: \textit{innate} and \textit{adaptive} immunity \cite{Marshall2018} (Figure \ref{immunity}).  Innate immunity is the first line of any organism's defence, it employes a wide set of strategies that are not directed against anyone pathogen in particular, but rather designed to protect against all possible threats. Skin is a prominent example and a key component of the innate immune system. Skin acts as a primary physical barrier between pathogens and the organism. The adaptive or acquired immunity supports innate immunity in defending against intruders. The adaptive immune system interacts with molecules (most often proteins), known as \textit{antigens}. Antigens are present on the surface of the pathogens and can guide the immune response \cite{Chaplin2010}. Unlike innate immunity, which has evolved to be antigen-independent, adaptive immunity is highly specific to \textit{antigens} in its response.

\begin{figure*}
    \centering
    \includegraphics[trim=0 0 5 0, clip, width=.9\textwidth]{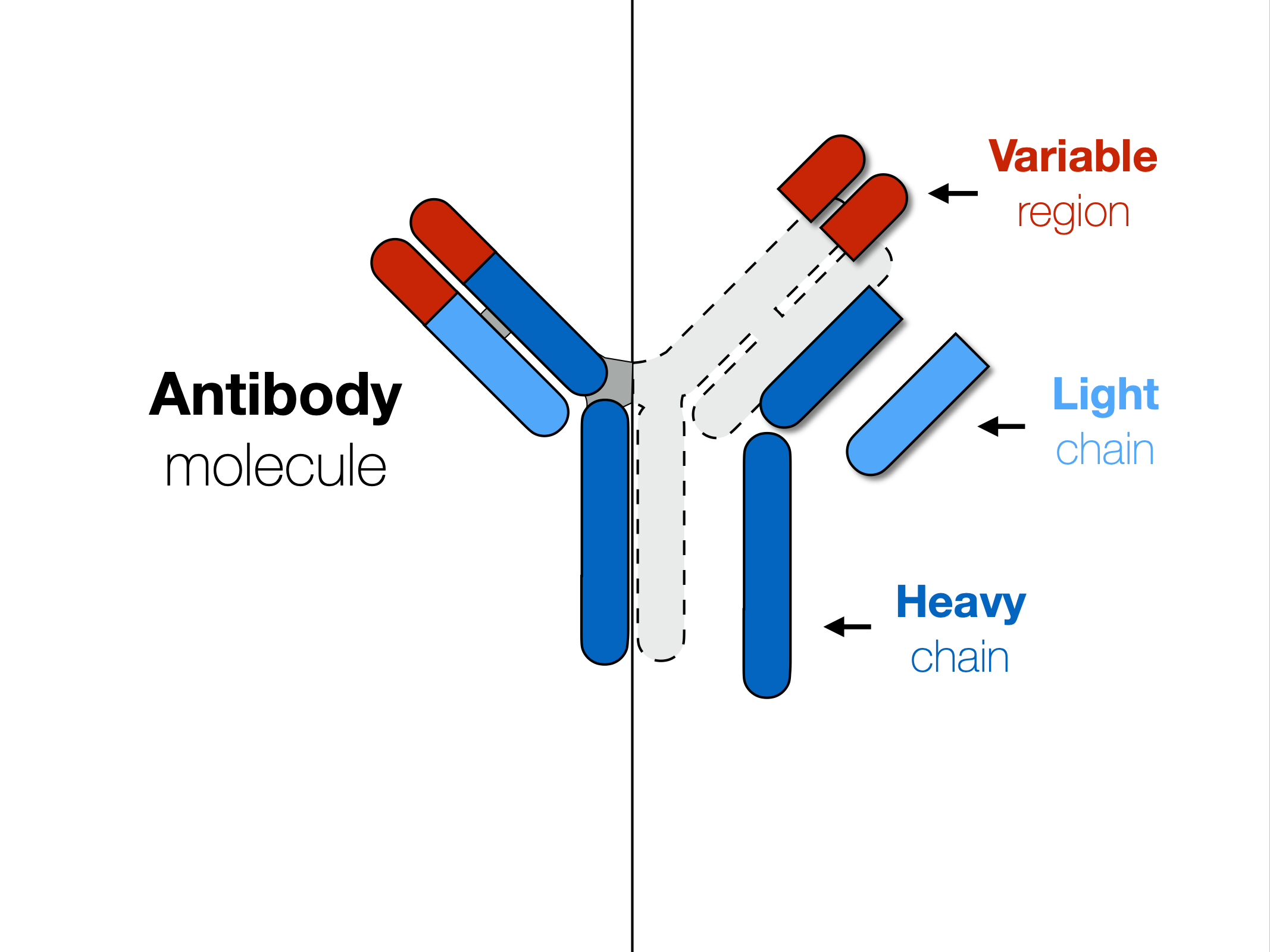}
    \caption{Structure of a typical antibody. Antibodies use their variable regions (coloured in red) as arms to bind to potential intruders. By binding to a molecule, antibodies send a signal to other immune cells to take action against the intruder.}
    \label{antibody}
\end{figure*}

The two main types of cells involved in the adaptive immune system are B and T cells. While the T cells target intracellular pathogens by directly triggering cell death mechanisms in the pathogen-infected cells, the B cells are mainly involved in targeting extracellular pathogens. After recognising foreign antigens on the surface of the pathogen, B cells differentiate into \textit{plasma B cells} that produce \textit{antibodies}. An antibody is a large Y-shaped protein that binds to a specific antigen. The structure of a typical antibody is presented in Figure \ref{antibody}. Produced antibodies bind to the cognate antigen which results in neutralization of this particular pathogen. Remaining antibodies are then circulating in the bloodstream where they can initiate the immune response.

Normally, the adaptive immune system produces T cells, B cells, and antibodies that are capable of living harmoniously together with the body's own cells, which are usually referred to as \textit{self} \cite{Klein2014}. Albeit, sometimes, due to various environmental and genetic factors, the immune system produces cells and proteins that can harm its own host, in a process called an \textit{autoimmune reaction} or \textit{autoimmunity}. Such cells are then called \textit{self-reactive} or \textit{auto-reactive}, as they attack domestic cells. These attacks may have a negligible effect if they occur rarely or are mild \cite{Romagnani2006}. However, more systematic failures accumulate and may lead to serious damage, causing various pathologies or even, in some circumstances -- premature death, and more so among women \cite{Walsh2000}. 

Typically, self-reactive B cells and T cells are safely removed or silenced by the immune system prior to any serious harm \cite{Elkon2008}. Two mechanisms are mainly responsible for eliminating malfunctioned immune cells: \textit{central} and \textit{peripheral} tolerance \cite{Romagnani2006}. Central tolerance normally occurs inside primary lymphoid organs: thymus and bone marrow and it targets self-reactive T cells and B cells in their infancy. Peripheral tolerance acts as a backup filter as it selects out self-reactive immune cells which central tolerance has failed to identify and neutralize. 

As part of the central tolerance, T cells undergo a two-stage selection procedure in the thymus \cite{Romagnani2006}. In the first stage (positive selection), immature T cells are tested for their ability to interact with special antigen-presenting cells in the thymus. T cells-to-be that show a lack of interest in such targets at this stage is eliminated. Later, in the second stage (negative selection), T cells are tested for the binding capacity to self. For this, a set of \textit{self-antigens} (i.e. antigens that belong to the body's own cells) is assembled with help of the autoimmune regulator gene (AIRE) and displayed to prospective T cells that successfully passed the first stage \cite{Kisand2011}. Only cells that ignore self-antigens are subject to further development \cite{Bettelli2006}. B cells are also subjects to central tolerance, although the exact details remain poorly understood \cite{Meyer2016}. Nevertheless, the tolerance in B cells is partly T cell-dependent, with defective T cells contributing to the production of autoreactive B cells \cite{Kinnunen2013, Meffre2008}. In healthy situations, B cells produce antibodies for our protection against pathogens but in case of bypassing defensive mechanisms of the immune system, self-reactive B cells produce antibodies that tag the organism's own cells and may trigger autoimmunity. Such self-reactive antibodies are called \textit{autoantibodies}. Thus, failure to recognise and mitigate self-reactive T and B cells may potentially lead to the accumulation of autoantibodies and result in autoimmune disorders \cite{Tsubata2017}.

\subsection{Autoimmune disorders}

Autoimmune diseases affect about 5\% to 7\% of the world population, with the majority of patients being women \cite{flaherty2011}. The most famous examples of autoimmune diseases are type 1 diabetes (T1D), celiac disease and multiple sclerosis \cite{Abel2014}. For instance, the onset of type 1 diabetes is caused by an autoimmune reaction against insulin-producing $\beta$ cells in the pancreas \cite{Pihoker2005}, exogenous gluten proteins are linked to the autoimmune reaction in celiac disease \cite{Sollid2013, Lauret2013} while patients with multiple sclerosis harbor autoantibodies against myelin i.e. fatty tissue in the brain and spinal cord that facilitates neurotransmission \cite{Fraussen2014}. 

One of the autoimmune disorders -- autoimmune polyendocrinopathy-\linebreak candidiasis-ectodermal dystrophy (APECED) or autoimmune polyendocrine syndrome type 1, is of particular importance for the researchers and doctors who study autoimmunity. APECED is a rare disorder caused by a small modification in genetic code. Mutations in the AIRE gene alter negative selection mechanisms that normally prevent the body from producing harmful autoantibodies. As a result, a wide range of autoantibodies is released into the bloodstream, causing various types of damage to the organism's own cells and tissues. Many of these autoantibodies that are produced in APECED are shared with other diseases such as autoantibodies against $\beta$ cells in T1D \cite{Fishman2017}. Precisely due to such high diversity of self-reactive antibodies, APECED is considered an important disease model that helps to understand the processes that drive autoimmunity in general \cite{PETERSON2004}. The study of APECED is central for two publications included in this thesis, which we will examine in later chapters. 

Autoantibodies have been shown to play an important role in many other diseases: various cancers, neurodegenerative diseases, cardiovascular and infectious disorders. However, the direction of the association between autoantibodies and disease onset is not always clear -- do autoantibodies cause disease or whether autoimmunity is merely a side effect \cite{Aziz2019}? Nevertheless, it has been demonstrated that the presence of autoantibodies in blood, may suggest the development of a disease and provide information about its nature and intensity \cite{Leslie2001}. Studies show that the information about the quantity of disease-specific autoantibodies may provide decisive diagnostic information \cite{Pihoker2005, encyclopedia, Nagele2011, Huang2017, Abel2014, Gupta2017}. Therefore, detecting and characterising autoantibodies present in the organism may shed light on the disease development. 

One of the earliest attempts to experimentally detect the presence of an antibody engaged in a reaction against a patient's own tissue was successfully carried out in 1955 by Dr. Henry Kunkel \cite{Holman2011}. Dr. Kunkel used antibodies tagged with a fluorescent marker (also known as secondary antibodies) to detect autoantibodies in lupus erythematosus cells extracted from the serum of patients with systemic lupus erythematosus disease \cite{Hepburn2001}. Later, a number of simpler and more accurate methods have been developed: radiobinding assay, western blot, and enzyme-linked immunosorbent assays (ELISA). These methods allowed for the detection of antibodies associated with a pre-defined antigen \cite{Rosenberg2015}. A need for an \textit{a priori} hypothesis about the antigen was a significant limiting factor, preventing the discovery of autoantibodies against new previously deemed unrelated antigens. Therefore, in order to discover novel autoantibodies, researchers needed a way to screen a much wider range of potential candidate molecules in a high-throughput manner \cite{Abel2014}. 

\section{DNA microarrays}

DNA microarray technology was developed in the 1990s by the American researcher Patric O. Brown and has revolutionised the analysis of biological systems \cite{Sobek2006, Rosenberg2015}. DNA microarray is a collection of microscopic DNA fragments, short sections of genes printed on a solid surface  \cite{Rosenberg2015}. A biological sample containing fluorescently labeled complementary DNA or RNA molecules can be then applied to each array enabling researchers to quantify the expression of each gene i.e. gene productivity. DNA microarrays were the first high-throughput technology enabling quantification of gene expression in a parallel fashion using minimal sample input requirements \cite{Rosenberg2015}. Soon after the first DNA microarrays appeared it was demonstrated that similar technology involving protein binding molecules can be used to estimate the number of proteins, including autoantibodies from patients' blood \cite{Rosenberg2015}. Therefore DNA microarrays have played a pivotal role in the emergence and the development of protein microarrays.

\section{Protein microarrays}

Similar to DNA microarrays, protein microarrays (or protein chips) contain a large collection of individually isolated (purified) molecules, densely printed on the solid glass-based surface \cite{Huang2017}. Based on the type of molecule incubated on the slide, protein microarrays can be categorised into three broad groups: \textit{functional}, \textit{analytical}, and \textit{reverse-phase} \cite{Hall2007, Moore2016}. Functional protein microarrays are produced by printing full-length proteins on the glass surface. Printing the entire protein helps to preserve its original structure and as a result, also function. Functional protein microarrays detect autoantibodies that bind to the proteins on the slide. This type of arrays gained a lot of popularity in the last decade, with the number of clinical applications steadily growing \cite{Zhu2012}. In contrast to functional microarrays, analytical or capture arrays utilize panels of antibodies attached to the slide to detect and measure proteins from the sample \cite{Moore2016}. Instead of printing an arbitrary set of antibodies or proteins on the slide, in reverse-phase protein microarrays (RPPA or also known as lysate arrays), all proteins from a specific cell interior (lysate) are printed and the antibody binding from the sample is detected \cite{Moore2016, Neagu2019}. In this thesis, we are going to focus solely on functional protein microarrays that are used to detect and measure the abundance of autoantibodies in the blood \cite{Rosenberg2015}.

\subsection{Functional protein microarrays} 

\begin{figure*}
\begin{center}
\includegraphics[trim=0 0 5 0, clip, width=.9\textwidth]{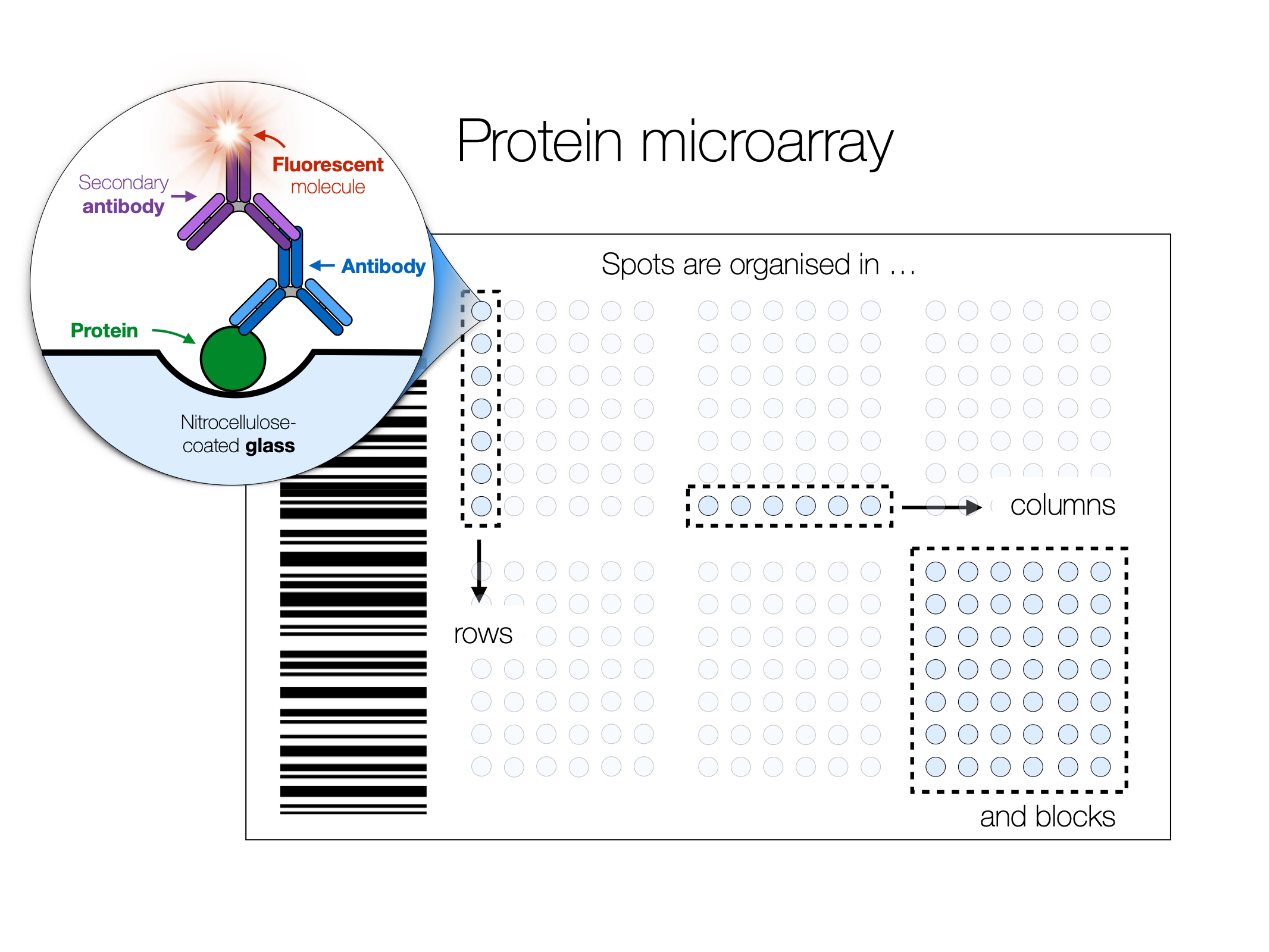}
\end{center}
\caption{Schema of the exemplary protein microarray. In functional protein microarrays, proteins and protein fragments are printed inside spots (blue circles) arranged in rows, columns and blocks.}
\label{arraystructure}
\end{figure*}

In order to use functional protein microarrays to accurately quantify the amounts of autoantibodies, at first, a sample in the form of plasma, serum or other solution from a human subject is applied to protein microarray slides. Autoantibodies from human sample bind to proteins immobilized on microarray surface in spots or tiny cavities in the glass. Array slides are then washed and dried to get rid of all molecules that failed to bind and remained on the surface. Later, autoantibodies that have genuinely bound to spotted proteins are detected by applying fluorescently labeled secondary antibodies. Microarrays are then again washed and dried. Fluorescent signals coming from each well are acquired with a microarray scanner and later analysed using computer software \cite{Huang2017}. The amount of light coming from each well is associated with the levels of autoantibody binding to the specific protein (from this well). Schema of a typical protein microarray is presented in Figure \ref{arraystructure}.

Fabrication and further handling of functional protein microarrays is a complex process. It involves multiple consecutive steps including printing and immobilization of proteins on the slide surface, incubation with a sample, repeated washing, and drying and scanning of arrays. Hence, a substantial number of technical factors influence the quality of the resulting autoantibody binding signal \cite{Duarte2017}. Uncalibrated printing machinery may result in uneven distribution of proteins in spots or in proteins being carried over to neighbouring sites by the printing pin. Irregular washing and drying of the slides can also cause the variation of the actual protein content \cite{zhu2006}. Several studies have reported a cross-talk between neighbouring spots that resulted in unlikely highly correlated signals between neighbouring spots \cite{Landegren2016, Fishman2017}. Technical variability introduced by mechanical liabilities can mask the true underlying biological signal \cite{sboner}. Therefore, sufficient care must be taken in design, fabrication, and subsequent analysis to account for possible technical biases.

\subsubsection{ProtoArray human protein microarray} 
One of the most popular examples of wide coverage commercial protein microarrays is ProtoArray\textsuperscript{\textregistered}. Originally designed and manufactured by Invitrogen company, ProtoArray\textsuperscript{\textregistered} became a part of the Life Technologies brand in 2008 that was later acquired by Thermo Fisher \linebreak Scientific in 2014. ProtoArray\textsuperscript{\textregistered} includes more than 9,000 full-length human proteins as well as several thousand control proteins. All proteins are spotted twice on the array to enable quality control.  More than 6,100 proteins that are included on the chip are potential drug targets and thus, relevant to disease processes. Data from ProtoArray\textsuperscript{\textregistered} experiments can be analysed by a number of publicly available tools including the manufacturer's own -- ProtoArray Prospector. ProtoArray\textsuperscript{\textregistered} has a broad spectrum of potential applications, including discovering novel disease biomarkers via analysing autoimmune reactions, discovery, validation, and development of novel drug targets \cite{Duarte2017}. Nevertheless, the majority of studies that used the technology have focused on discovering and characterizing autoimmune targets from the blood in a specific disease \cite{Nagele2011, Long2016, Huang2017, Meyer2016, Fishman2017, Abel2014}. 

Experimental data produced via ProtoArray\textsuperscript{\textregistered} platform has been shown to suffer from a multitude of technical errors \cite{Duarte2017}. Printing and contamination artifacts, non-specific protein binding, and high background signal all were shown to contribute to the distortion of biological findings, and reduced reproducibility of the experiments \cite{Duarte2017}. In later chapters, we discuss some methodological ways to tackle these challenges. Perhaps in part due to the aforementioned technical biases in 2018, Thermo Fisher Scientific has discontinued all the services related to ProtoArray\textsuperscript{\textregistered}. Despite this, a large number of experiments involving ProtoArray\textsuperscript{\textregistered} had been performed, and much of this data was made available through data repositories like Gene Expression Omnibus \cite{geo} and ArrayExpress \cite{arrayEx}. This publicly available data remains valuable for many researchers who plan to either reproduce old results or make their own analysis of protein microarray data.  

\subsubsection{HuProt human protein microarray} HuProt is an actively maintained alternative platform to ProtoArray\textsuperscript{\textregistered}. Initially developed by CDI laboratory at John Hopkins University in 2012 \cite{hu2012, Duarte2017}, HuProt contained 16,368 full-length functional proteins, representing 12,586 protein-coding genes \cite{Duarte2017, jeong2012}. At the time of writing, the most recent Human Protein Microarray v4.0 in total contains more than 21,000 unique human proteins and protein variants, covering more than 81\% of the canonical human proteome. Similar to ProtoArray\textsuperscript{\textregistered}, HuProt has mostly been used for detecting and evaluating autoimmune reaction in patients across various disorders, including: primary biliary cirrhosis \cite{hu2012}, ovarian \cite{Jung2014} and gastric cancers \cite{Yang2015}, neuropsychiatric lupus \cite{Hu2015} and Behcet's syndrome \cite{Hu2016}. As a data generation platform, HuProt was shown to be susceptible to similar biases as  ProtoArray\textsuperscript{\textregistered}, including non-specific binding and printing contamination \cite{Duarte2017}.

\subsubsection{Custom protein microarrays}
Both equipment and constituents necessary for creating custom protein microarrays are commercially available from private and public vendors. Also, fully assembled protein microarrays are available on the market. Coverage of commercial microarray products ranges from arrays with few dozens of carefully selected proteins to vast collections that include almost the entire known proteome \cite{Rosenberg2015}.

In preparation for this thesis, mostly data from ProtoArray\textsuperscript{\textregistered} and HuProt protein microarrays were used.

%%%%%%%%%%%%%%%%%%%%%%%%%%%%%%%%%%%%%%%%%%%%%%%%%%%%%%%%%%%%%%%%%%%%%%%%%%%%%%%%%%%%%%%%%%%%%%%%%%%%%
% METHODS CHAPTER
%%%%%%%%%%%%%%%%%%%%%%%%%%%%%%%%%%%%%%%%%%%%%%%%%%%%%%%%%%%%%%%%%%%%%%%%%%%%%%%%%%%%%%%%%%%%%%%%%%%%%

\chapter{Protein microarray data analysis}

Protein microarray chips are high-throughput platforms capable of measuring thousands of protein interactions in parallel across multiple samples \cite{DeLuca2011}. Computational analysis of a typical protein microarray study starts with acquiring high-resolution images of stained protein arrays, using the special microarray scanner (e.g. GenePix Microarray scanner). Signal information about each spot on the array is then extracted into a file by segmentation and registration software (e.g. GenePix Pro 7). Each array usually results in one file. Data from such files is then used to assemble a data matrix, which serves as an input to the analysis pipeline. Each column in such a matrix represents an individual sample while a row is associated with a protein. This initial matrix is called raw data, as it has not been ``purified'' by pre-processing methods. Techniques such as background correction, signal transformation, outlier detection, and normalisation are essential for further statistical analysis as they help to remove or at least minimise the effects of technical noise and thus, enable a fair comparison between samples. Normalised and pre-processed data can be visualised and explored further. If a study follows a case-control design \cite{Zheng2018} (including Publications I, II, and IV in this thesis), it is possible to compare protein signals in patients with those in healthy individuals. This enables researchers to pinpoint proteins in which concentration levels can reliably differentiate patients from controls. Such differential proteins can both be used as clinical biomarkers as well as reveal important insights about the mechanisms of the disease.

Protein microarray analysis has notably benefited from the set of analytical methods developed for DNA microarrays, as both technologies enable measurement of numerous molecules immobilised on the slide surface, e.g. the array scanning approach \cite{sboner}. But as it soon turned out, not all statistical methods designed for the DNA microarrays may directly be applied to the protein microarrays, as the latter is grounded on different biological assumptions. Namely, DNA microarrays assume comparable levels of gene expression across individuals regardless of their clinical condition. While this may be true for genes, the number of autoantibodies present in the blood of a healthy person and a patient is vastly different \cite{sboner, Duarte2017, DeLuca2011}. Such discrepancy between assumptions has motivated the use of a different normalisation strategy from the one used in DNA microarrays, which will be discussed in the sections below. 

\begin{figure*}
\begin{center}
\includegraphics[trim=0 0 5 0, clip, width=\columnwidth]{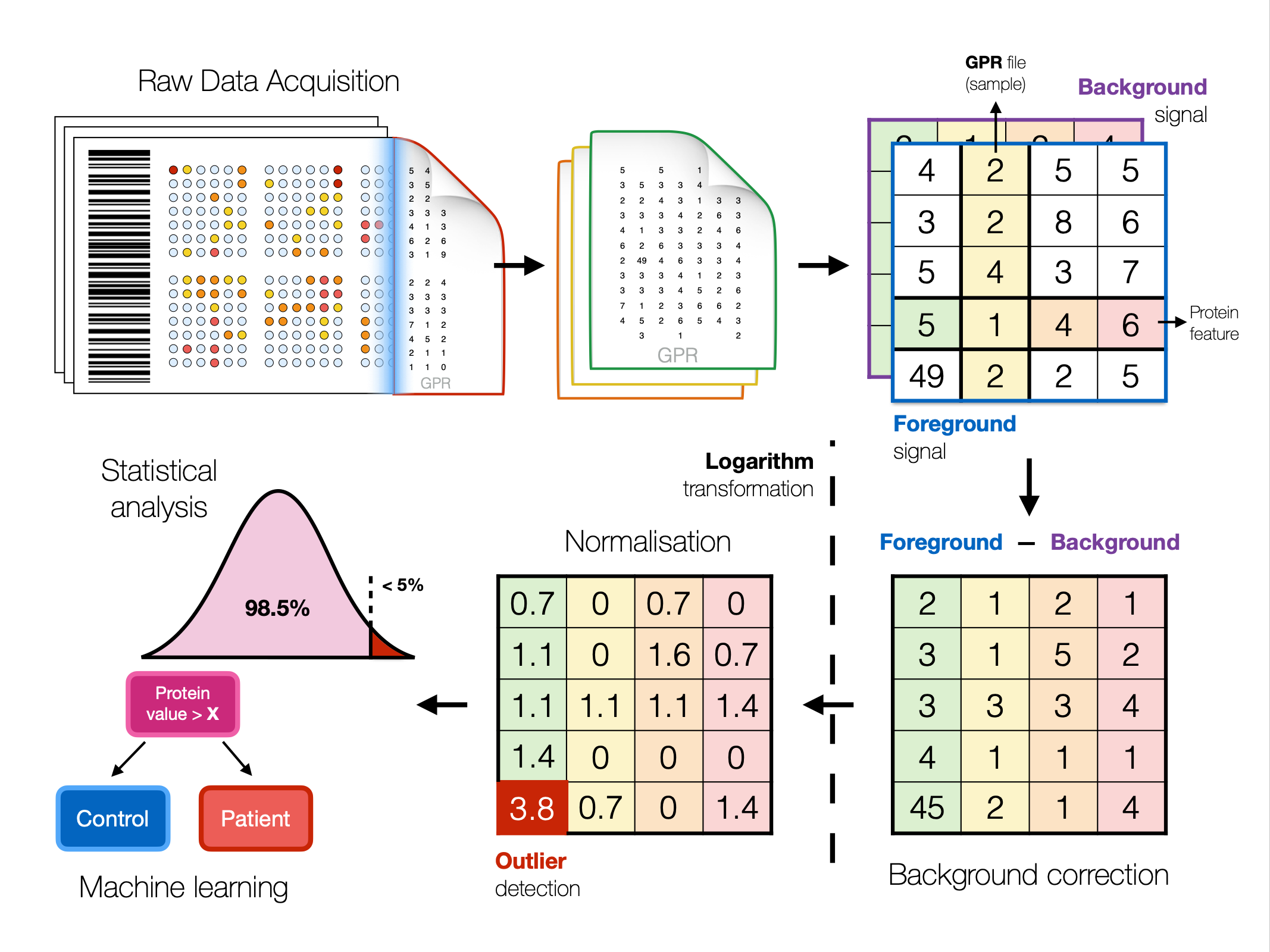}
\end{center}
\caption{A rough outline of data analysis methods used for protein microarray experiments. Raw data after the acquisition is assembled into a large matrix. This matrix passes through a number of steps, including background correction, signal transformation, outlier detection, and normalisation before being used as an input for statistical analysis and machine learning. Next, results of statistical analysis and machine learning modelling are interpreted in the context of the existing body of knowledge.}
\label{pipeline}
\end{figure*}

In this thesis, we describe a set of computational methods used to analyse functional protein microarray experiments. These techniques are bundled together into a general data analysis pipeline (Figure \ref{pipeline}), which treats raw GPR files as an input while providing normalised data, a list of differential proteins, and the results of the enrichment analysis as an output. The pipeline was first designed as means to analyse data in Publication I, it was later expanded for Publication II, and finally packaged and released to the public as a web-tool (PAWER) in Publication III. Finally, in Publication IV we have explored the utility of using machine learning methods alongside classical statistical algorithms described in previous publications with a goal to increase the choice of methods available for analysis. Below, we describe each essential part of our pipeline in detail.

\section{Raw data acquisition}

The process of data acquisition starts with scanning incubated arrays with a special microarray scanner that produces high-resolution 16-bit images. These images are saved in Tag Image File Format (TIFF) data format and later processed by the image segmentation and registration software (e.g. GenePix Pro 7). This software accurately detects each spot and quantifies its signal intensity with respect to the local background. Then the software uses information about each spot's location and contents to link estimated intensity values with corresponding proteins that were printed on the microarray. Data about array design and spatial location of each protein is stored in an auxiliary GenePix Array List (GAL) file and can be added separately to the analysis. Finally, estimated intensities from an individual array are saved into GenePix Result (GPR) file -- a \textit{de facto} standard format for storing protein microarray data \cite{Abel2014}. Each collected sample normally corresponds to one GPR file. Typical protein microarray studies collect dozens or even hundreds of samples, resulting in a corresponding number of GPR files. 

GPR files are text files in disguise, hence, they are tab-delimited text files that can be read by most popular spreadsheet programs such as Microsoft Excel. GPR files contain a header with relevant meta-information about the experiment and the data matrix, which contains raw fluorescent intensity values of each spot on the chip. If several fluorescent molecules with different wavelengths were used in the experiment, foreground and background signals are measured and reported for each. Different types of arrays may have vastly different contents of both meta-information and intensity matrix. In this thesis, mostly GPR files from ProtoArray and HuProt platforms were analysed, thus we will focus on them. 

\section{Data pre-processing}

At the beginning of the protein microarray analysis, researchers extract individual raw signal values from GPR files and combine them into a large matrix of raw data. Multiple studies have shown that raw protein microarray data should not be used directly in the computational analysis \cite{sboner, Abel2014, Duarte2017}. Various technical issues discussed previously can introduce a significant amount of noise into the output signal \cite{Abel2014, sboner, Duarte2017}. This noise can hinder the analysis by masking the true signal, rendering the final results indecisive. Therefore, raw data must be carefully pre-processed prior to any further analysis. Pre-processing helps to identify and get rid of technical noise at the same time preserving valuable biological signal. Below we discuss a set of common pre-processing strategies, which can be applied in various orders depending on the experimental setup. 

\subsection{Background correction and signal transformation}

\begin{figure*}
\begin{center}
\includegraphics[trim=0 0 5 0, clip, width=\columnwidth]{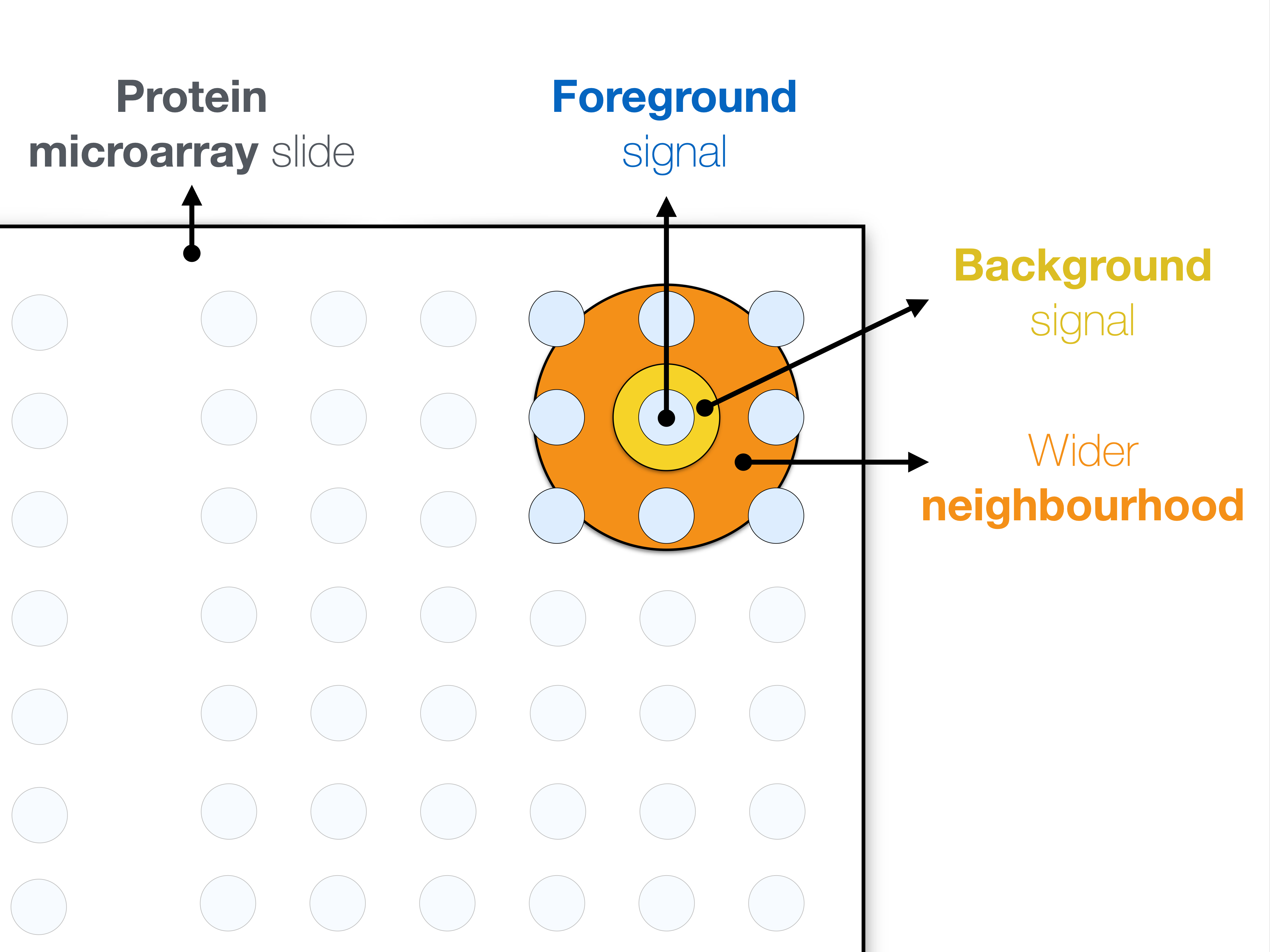}
\end{center}
\caption{Foreground signal, background signal and wider neighbourhood of the spot. The fluorescent signal of the protein is determined by subtracting the median of the immediate background of the spot from the median of its foreground signal. Alternatively, background value can be calculated taking a wider neighbourhood of the spot into consideration.}
\label{forebackground}
\end{figure*}

In functional protein microarrays, proteins of interest are immobilized in rows and columns on the glass surface forming a grid of spots \cite{Dez2012}. After a slide is incubated with a sample solution (usually blood), autoantibodies bind to immobilized proteins. One of the technical challenges related to correctly quantifying the fluorescent signal emitted by the labeling antibody is to discriminate signal produced by the genuine biological reaction and local residuary background light \cite{Dez2012, Duarte2017}. The true signal is usually derived by subtracting the median background intensity of the spot, i.e. \textit{background signal}, from the amount of fluorescent signal registered within the spot, i.e. \textit{foreground signal} (Figure \ref{forebackground}) \cite{Abel2014}. However, several other more elaborate background corrections methods have been proposed in the past \cite{DaGamaDuarte2018, Ritchie2007, zhu2006}. For example, it has been suggested that instead of using an immediate local background intensity it is beneficial to consider a wider neighbourhood of the spot when calculating the median background signal \cite{zhu2006}.

Another commonplace practice in biomedical research is to apply one of the data transform techniques. For example, log-transformation is known to make fold changes symmetric around zero, reduce the skew in the data, and provide a good approximation for the normal distribution -- desirable property for many methods \cite{Feng2014, Abel2014}, including protein microarray specific normalisation strategy that we are going to discuss in a later section. In spite of some researchers revealing negative effects of logarithm-based data transformation \cite{Feng2014}, it remains popular and was used in a number of recently published works related to protein microarray analysis \cite{Long2016, Landegren2016}, including Publications I and II included into this thesis \cite{Fishman2017, Meyer2016}.

\subsection{Outlier detection}

In order to satisfy assumptions imposed by most of the statistical methods, protein expression levels registered by a panel of protein microarrays should follow a normal distribution: most of the values being close to the average signal with few very low and high values in the tails. However, in practice, due to a multitude of technical factors e.g. inattentive handling of microarray slides, sample quality, or manufacturing errors, proteins can exhibit expression levels vastly inconsistent with the rest of the data. Such proteins are known as \textit{outliers} or ``anomalous data points''. In protein microarray experiments, both individual proteins and entire microarray slides may exhibit abnormal expression levels and thus considered anomalous. The presence of outliers can be unfavourable for the downstream analysis and resulting conclusions \cite{Pearson}. Hence, once detected, such values usually are either removed completely or substituted with a reasonable approximation e.g. average or median of the corresponding protein.

Several methods exist to automatically detect outliers. One of the most popular and suitable for data that follows normal distribution is to label as outliers all data points that fall outside three standard deviations from the corresponding mean (\textit{i.e. three-sigma rule} or \textit{empirical rule}). The common assumption is that such extreme values are unlikely to be generated by the same biological process as the rest of the data. This reasoning is based on the definition of the normal distribution, for which 95.45\% of its data lie within two standard deviations from its mean, while 99.73\% within three standard deviations. If the value is either larger or smaller than the aforementioned threshold of three standard deviations, it has only 0.27\% of the chance to come from the same distribution as other values. This line of thought is valid only in case data follows the normal distribution. In other circumstances, the above calculations may not apply. In practice, it has been shown that protein expression profiles vary a lot between individuals rarely resulting in signal values that follow the normal distribution \cite{Kisand2010, Puel2010, Meyer2016, Fishman2017}. Therefore data from protein microarray experiments must be transformed (e.g. using log-transform) if the three-sigma rule is to be applied.

Another popular approach for identifying outliers is using boxplots \cite{mcgill1978variations}. Boxplots are graphical structures, that show how the data points are spread out. Boxplots have at least two relevant merits. Firstly, they offer a natural way to visualise the data, and secondly, they can be used to detect outliers in a way that is indifferent to the underlying data distribution. Boxplot summarises data using five quantitative measures: minimum, three quartiles, and maximum. While the second quartile ($Q_2$) corresponds to a median (middle value), the first ($Q_1$), and the third ($Q_3$) quartiles enclose the first 25\% and 75\% of data distribution respectively. The distance between the first and third quartiles is called the interquartile range ($IQR$) and given by $Q_3 - Q_1$. Genuine data points must be larger than $Q_1 - 1,5*IQR$ and smaller than $Q_3 + 1,5*IQR$. Data points outside this range are considered to be outliers. Boxplots are usually rendered as rectangles (hence the name ``boxplot'') with a fixed width, and length equal to $IQR$, with outliers, visualised as circles outside of the box either at the top or bottom of the figure. Although boxplots can be plotted side by side to compare distributions of multiple features, they are not suitable for identifying outliers from multivariate data (i.e. data points characterized by more than one feature).

\begin{figure*}
\begin{center}
\includegraphics[trim=0 0 5 0, clip, width=5in]{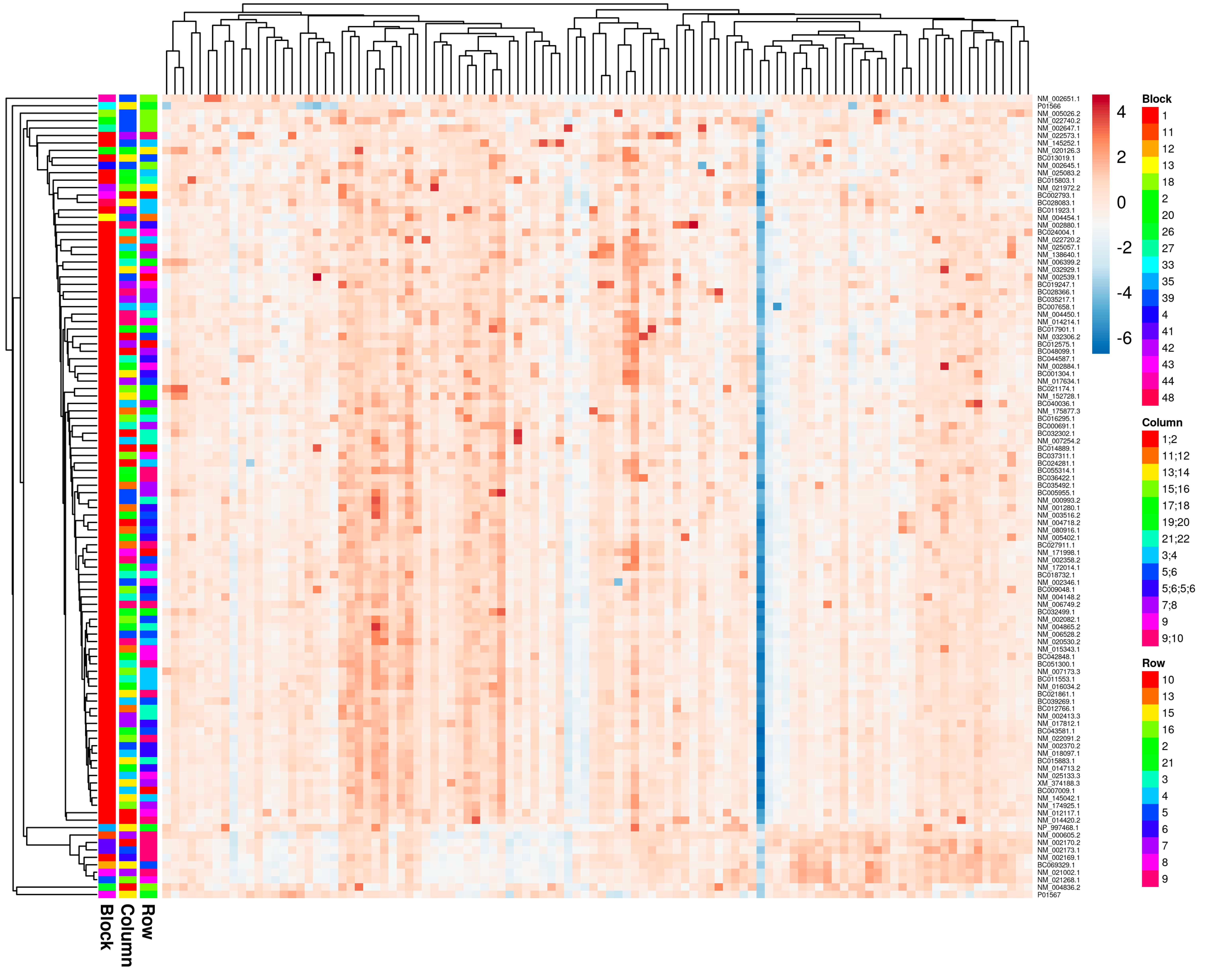}
\end{center}
\caption{An example of a heatmap obtained from a protein microarray experiment using ClustVis tool. Each cell of the heatmap is coloured based on the signal level of the corresponding protein (in rows) across all samples (in columns). Additional meta-information available about samples and proteins is visualised using extra colour legends. Hierarchical clustering results are presented in the form of dendrograms at the top and on the left of the heatmap. }
\label{heatmap}
\end{figure*}

Clustering techniques in combination with various visualisation strategies can be used to recognise outliers in multivariate data, such as protein microarray read-outs. Hierarchical clustering is an algorithm that recursively groups data into clusters based on a predefined distance metric e.g. Euclidean distance \cite{ZepedaMendoza2013}. The result of hierarchical clustering is a dendrogram. Dendrogram visually shows the arrangement of clusters produced by the algorithm. Anomalous samples or proteins will stand out far from the rest of the clusters on the dendrogram, making them easy to spot and remove. Dendrograms are often coupled with another visualisation approach named heatmaps (Figure \ref{heatmap}). Heatmaps use colour to represent the magnitude of individual signals. Heatmaps supplement dendrograms with an additional context about single expression values. Despite enabling rich visualisations, hierarchical clustering does not label samples as outliers automatically. One of several metrics can be used on top of hierarchical clustering results to detect clusters and therefore identify outliers (e.g. elbow method and silhouette score). Density-based spatial clustering of applications with noise (DBSCAN) is another clustering method that uses the spatial density of points as a factor for creating clusters \cite{dbscan}. Data points from the low-density regions, far from established clusters are considered to be outliers or noise. Therefore, unlike hierarchical clustering, DBSCAN can detect outliers without human intervention. However, DBSCAN requires several key parameters to be fixed to work. Although DBSCAN is the only technique mentioned in this section that was not explicitly applied in publications included in this thesis, we consider it to be an important addition, potentially valuable for the readers that may decide to use it in their work.   

\subsection{Signal normalisation}

One of the primary goals of protein microarrays is to compare the amount of binding between individual samples or groups of samples (e.g. healthy and controls). For this process to yield realistic results, the protein binding signal measured from multiple arrays must be comparable. This can be problematic due to the potential difference in the number of proteins printed on the slides and other technical factors that can introduce systematic biases \cite{zhu2006}. Such biases may significantly distort or shift signal distribution for some or all proteins on one or several protein microarrays. Therefore, unlike the above-mentioned data pre-processing approaches, signal normalisation acts globally combining information about signal variation from all arrays and proteins to successfully eliminate non-biological differences. The most commonly used approaches for protein signal normalisation were adopted from the DNA microarray context. These techniques, make strong assumptions about underlying signal distribution, which are not always in line with biological mechanisms at work in protein microarrays \cite{sboner, Abel2014, DaGamaDuarte2018}. Protein microarray-specific normalisation strategy based on robust linear model \cite{sboner} makes use of control proteins printed on each array and block. Control proteins can be positive or negative, but in either case, they are assumed to exhibit constant signal levels across all samples. Any differences in signal values of these proteins are considered to be technogenic and thus, corrected for. Below we present some of the most popular approaches for signal normalisation implemented in the computational tools used for protein microarray analysis \cite{Duarte2017, PAA, Fishman2019}.

\subsubsection{Global scaling} One of the standard methods for most DNA microarrays (e.g. Affymetrix platform) that was also applied in protein microarrays is the global scaling approach \cite{Bolstad2003, Wu2005, sboner}. In short, the signal levels of each array are divided by the median signal of the corresponding array. Namely, for each array $S$, normalised signal $S_n$ would be calculated as $$S_n = S/median(S)$$ \cite{Wu2005, sboner}. This ensures that the median signal is the same across all arrays. The global scaling method assumes that the total amount of signal is the same in all arrays. Although this assumption may hold for DNA microarrays, where approximately the same number of genes is expressed regardless of the phenotype, it may not be true for protein microarrays \cite{Duarte2017}. For example blood from patients with an autoimmune disease is expected to contain more autoantibodies and therefore produce a higher total signal comparing to serum from healthy individuals.

\subsubsection{Quantile normalisation} Quantile normalisation substitutes the largest value in each array with a median (or mean) of the largest values across arrays, all second largest values with a median of the second-largest values, etc. \cite{sboner, Bolstad2003}. This algorithm assumes that signal distribution for all arrays is nearly the same while major differences between samples are mainly of technical, not the biological origin, which can be the case for DNA microarrays \cite{Duarte2017}. However, as discussed previously, the autoimmune profile has been shown to be very heterogeneous \cite{Kisand2010, Puel2010, Meyer2016, Fishman2017} with a subset of protein features demonstrating a signal very different from the rest of the platform. Therefore, samples can produce distinct distributions due to genuine biological differences. Quantile normalisation thus eliminates such biologically legible differences, by equalising the underlying distributions.

\subsubsection{Cyclic loess} Cyclic loess normalization is performed for a pair of microarrays, and its main intuition is usually described using the so-called $M$ versus $A$ plot (MA plot) \cite{Ballman2004}. Here $M$ is the difference of $log_2$ expression values, and $A$ is the average of $log_2$ values. More formally, for a pair of arrays $i$ and $j$ and protein $p$, $$ M_p = log_2{(x_{pi})} - log_2{(x_{pj})} = log_2{(x_{pi}/x_{pj})}$$ and $$A_p = \frac{1}{2}(log_2{(x_{pi})} + log_2{(x_{pj})}) = \frac{1}{2}log_2{(x_{pi}*x_{pj})}$$ \cite{Bolstad2003}.  Thus, the MA plot for any pair of microarrays can be illustrated as a scatter plot with $M_p$ on the y-axis and $A_p$ on the x-axis. Similar to quantile, cyclic loess normalization assumes that the expression of the vast majority of genes (in the case of DNA microarrays) does not change between the conditions, therefore two perfectly normalized arrays would result in a MA plot in which points are scattered around $M = 0$ line \cite{Ballman2004}. Locally estimated scatterplot smoothing curve, which is also referred to as \textit{loess} is computed for the given MA plot to estimate the deviation from the ideal $M = 0$ line \cite{Cleveland1979}. A correction factor is then applied to individual signals of both arrays to achieve convergence of the loess curve and the ideal line. If there are multiple arrays in the experiment, the above procedure is applied until all possible pairs have been compared and normalised. Typically, several cycles of the algorithm are required for the final convergence \cite{Ballman2004}. However, if the number of arrays is large, a substantial amount of time is needed to make sure all arrays are normalised. 

\subsubsection{Robust linear model} Robust linear model (RLM) \cite{sboner} makes use of special sets of proteins -- controls that are often built into protein microarrays to enable normalisation of the signal. Controls can be either positive, that are guaranteed to have a high signal regardless of the blood content, or negative that should not react with serum under any circumstances, for example on microarray slides it is common to use empty spots as negative controls. Controls are present on every array as well as in every block of proteins on each array. Any significant deviation in the control signal may indicate the presence of unwanted noise or bias. Hence, RLM is employing controls to quantify and control for potential biases associated with arrays and blocks. 

To estimate such biases, observed signal $s_{ijkr}$ from the array $i$, block $j$, control protein $k$ and probe $r$ is modelled as a linear combination of the following coefficients: $\alpha_i$ from an array, $\beta_j$ from a block, $\tau_k$ from a protein feature and random noise $\epsilon_r$ using the following formula \cite{sboner}:

\begin{equation}
\label{rlm}
s_{ijkr} = \alpha_i + \beta_j + \tau_k + \epsilon_r
\end{equation}

\begin{figure*}
\begin{center}
\includegraphics[trim=0 0 5 0, clip, width=\columnwidth]{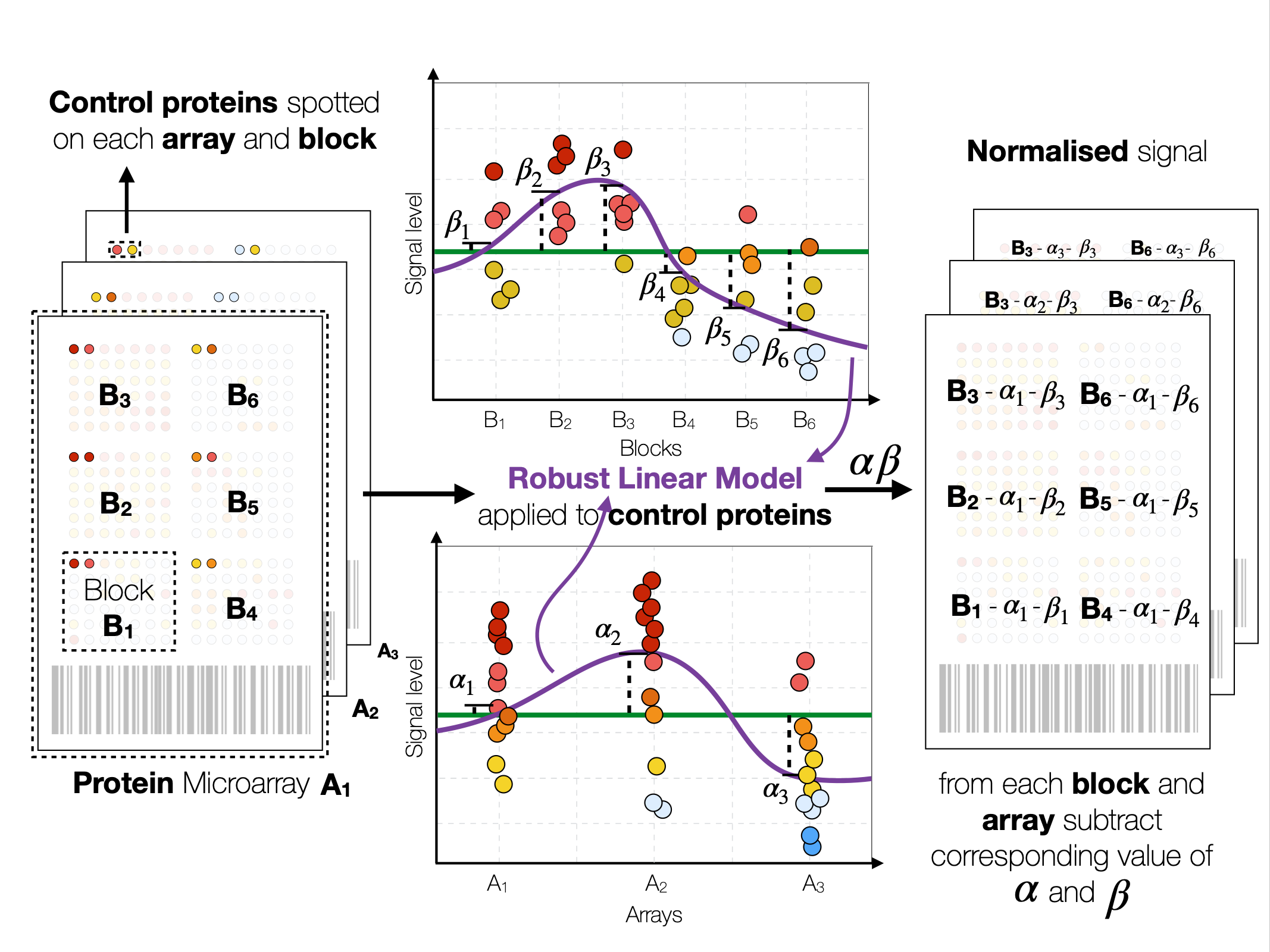}
\end{center}
\caption{Normalising protein microarray signal using the robust linear model and control proteins. First, intensity values of the control proteins are modelled as a linear combination of the array ($\alpha$), block ($\beta$), type of control protein, and noise using a robust linear model (purple line). The resulting coefficients for arrays and blocks are considered to be associated with unwanted technical biases. Then the normalised signal is obtained by subtracting the corresponding coefficients associated with a specific array and block.}
\label{rlmfigure}
\end{figure*}

The linear model presented by \ref{rlm} is built iteratively via a re-weighted least-squares algorithm that assigns weights to individual observations depending on their distance to the fitted curve. In order to make the algorithm more robust to outliers, the median is used instead of the more conventional sum of least squares to drive the optimization process \cite{sboner}. Once the model is trained, the coefficients associated with each array and block are calculated ($\alpha$ and $\beta$ in \ref{rlm}). The values of these coefficients describe how much signal of control proteins on a particular array or block deviates from the average. These deviations are considered to be of technogenic origin and thus, the normalised signal is calculated by subtracting corresponding coefficients from the signal of each spot as follows: 

$$
s'_{ijkr} = s_{ijkr} - (\alpha_i + \beta_j)
$$
for all possible $i$ and $j$ values. Figure \ref{rlmfigure} describes the normalisation process using RLM. In publications presented in this thesis, RLM normalisation was implemented in R, using functions from MASS \cite{mass} and limma \cite{limma} packages.

There is another, likely more familiar way to represent the linear model presented equation by \ref{rlm} using matrix and vector notation:

\begin{equation}
\label{lm}
{\bf y} = {\bf Xw} + {\bf \epsilon}
\end{equation} 

Here, we will discuss how the latter equation (\ref{lm}) can be translated into the former (\ref{rlm}) using an artificial example. Variable ${\bf w}$ from the latter equation is a vector of all coefficients of the linear model, namely: $\{w_0, w_1, ..., w_n\}$, where $n$ is the total number of coefficients included into the linear model. These coefficients reflect contributions from arrays, blocks and protein features. If we decide to call coefficients that represent contributions from arrays $\alpha$, from blocks $\beta$ and from protein types $\tau$, vector ${\bf w}$ will transform into \linebreak $\{\alpha_0, \alpha_1, ..., \alpha_{n_a}, \beta_0, \beta_1, ..., \beta_{n_b}, \tau_0, \tau_1, ..., \tau_{n_t}\}$, where $n_a$, $n_b$ and $n_t$ represent the total number of arrays, blocks and types of control proteins respectively, such that $n_a + n_b + n_t = n$. Matrix ${\bf X}$ in equation \ref{lm} is of size $S \times n$, where $S$ is the total number of control signals in the experiment, including all possible copies. For example, the total number of control signals ($S$) in the experiment with two arrays, two blocks on each array, and three types of controls is 12 ($2 \times 2 \times 3$), provided that each control protein is present in each block and each array. At the same time, the number of coefficients $n$ in ${\bf w}$ for the same example is 7 (2 arrays + 2 blocks + 3 control types). Each row in ${\bf X}$ encodes the location of one control signal. In the same imaginary protein microarray experiment with two arrays, two blocks and three control proteins, the first row of matrix ${\bf X}$ might look as follows:  $\{1, 0, 0, 1, 0, 1, 0\}$. This control protein therefore comes from the first array (first 1), second block (second 1 at fourth position) and happens to be the second type of control proteins (third 1 at sixth position). The dot product between the first row of matrix ${\bf X}$ and vector ${\bf w}$ will produce the following result: $\alpha_0*1 + \alpha_1*0 + \beta_0*0 + \beta_1*1 + \tau_0*0 + \tau_1*1 + \tau_2*0$. After a straightforward simplification, we get $\alpha_0 + \beta_1 + \tau_1$. Therefore, ${\bf X_0w} = \alpha_0 + \beta_1 + \tau_1$. If we insert this result into the linear model equation above, we get $ y_0 = \alpha_0 + \beta_1 + \tau_1 + {\bf \epsilon}$, where $y_0$ is modelled signal of the corresponding control protein. Random noise ${\bf \epsilon}$ is sampled from the normal distribution for each control protein independently. All in all, in a generic case we get $ y_{ijkr} = \alpha_i + \beta_j + \tau_k + \epsilon_r$, where $i$, $j$, $k$ and $r$ are indexes of corresponding array, block, protein type and protein signal. This final equation is equivalent to \ref{rlm}.

RLM is considered a preferred normalization strategy for protein microarrays as it exploits protein microarray specific control proteins and does not assume the near equal signal distribution across arrays \cite{sboner}. On the other hand, RLM assumes a normal distribution of the underlying signal to work well. Logarithm transformation that was discussed earlier, can be applied to the raw protein microarray data to approximate the normal distribution.

Unfortunately, a source code of the RLM method was not available at a time when work that laid the foundation of this thesis was performed. RLM was implemented as part of Prospector software -- a standard analysis tool provided by the manufacturer of ProtoArray. At the same time, Prospector had a prohibitive limit on the number of input samples, rendering it futile for larger studies as ours \cite{Turewicz2013}. Later, RLM was introduced in a protein array analyser (PAA) -- an R package developed by Michael Turewicz \cite{PAA}. However, for a significant stretch of time RLM was nowhere to be found and thus, we had to create our own implementation of the RLM normalisation module for publications I, II, and later released it to the research community as part of the web-tool (publication III of this thesis). The absence of a source code, as well as a pseudo-code of the RLM method in the original publication \cite{sboner}, made re-implementation of the RLM one of the most challenging part of the protein microarray analysis pipeline built in this thesis.

\section{Statistical analysis}

After protein data was properly pre-processed, relevant statistical methods can be applied. Statistical analysis is a vast field with a large number of techniques available at researchers' disposal. Characteristics of data and the research question determine the choice of the statistical method. For example, often researchers are looking for proteins that are capable of reliably distinguishing between two (or more) groups of samples, e.g. disease versus controls. Such protein features, in which intensity levels are sufficiently different between studied conditions are considered significantly differential and can be used as important biological markers suggesting the presence or absence of the condition in question often called \textit{outcome variable}. To assess the relationship between the intensity of a single protein and the outcome variable, univariate analysis tools are used. If simple univariate analysis yields no results or there is a good reason to believe that multiple proteins in combination can be predictive of the sample's outcome, multivariate analysis can be performed. Finally, once influential proteins are identified with either multivariate or univariate techniques, the enrichment analysis can be used to discover their common properties. The following sections focus on statistical methods used in this thesis, while adjacent methods are described only briefly.

Although statistical tests that we are going to talk about further, work slightly differently, some basic notions remain universal. The common starting point in hypothesis testing is defining a baseline or a \textit{null hypothesis} ($H_0$) -- a general statement about the absence of the assumed phenomenon. For example, $H_0$ may be formulated as observing no difference between means of protein intensity signals of two groups of samples. Namely $H_0: \mu_1 = \mu_2$, where $\mu_1$ and $\mu_2$ are means of the group 1 and group 2 respectively. An opposite to $H_0$, \textit{alternative hypothesis} ($H_1$) thus can be formulated as $H_1: \mu_1 \neq \mu_2$. The statistical test usually results in either rejecting the $H_0$ and thus favouring the alternative hypothesis or failing to reject the null hypothesis. In either case, employing statistical tests inevitably entails a number of background \textit{assumptions}, which are made either about the data or about the ways how the data has been gathered. An example of a data-based assumption could be that protein intensity signals follow a normal distribution. Invalid assumptions may lead to invalid test results, therefore it is of utmost importance to establish correct assumptions and choose an appropriate test. 

Statistical tests use a numeric quantity derived from data to perform the hypothesis test. This quantity is commonly referred to as \textit{test statistic}. The observed value of the test statistic can be calculated from the data at hand and compared with a known theoretical distribution of the test statistic under the null hypothesis (null distribution). If the observed value of the test statistic is at the far ends of the distribution i.e. either much larger or smaller than most of the values in the distribution, it is considered to be sufficient evidence for rejecting the null hypothesis. However, instead of relying on vague notions such as ``far ends'' or ``much larger'', researchers compute a probability of the observed value of the test statistic to be sampled from the null distribution - a \textit{p-value}. If the p-value is less than some pre-defined threshold value, the corresponding $H_0$ should be rejected. Common threshold values are 5\% and 1\%, more about this in the following sections.

Finally, most of the statistical tests can generally be divided into two large categories: one-sample and two-sample tests. A one-sample test explores the possibility of the mean of the sample being statistically different from the known population mean. Two-sample tests are used to assess the significance of the difference between means of two groups (e.g. patients and controls). Two-sample tests can be paired or unpaired (or independent). The unpaired two-sample test assumes no overlap between tested groups (e.g. two independent groups of mice). As it follows from the name, a two-sample test can compare only two groups. If there is a need to estimate the significance of the difference between three or more groups, an analysis of variance can be performed (also known as ANOVA). The sizes of groups to be compared and corresponding variances also influence the choice of the test. In this thesis, we make use of both one-sample and independent two-sample tests, which are discussed further in more detail. 

\subsection{Differential analysis}

Many studies involving protein microarrays follow case-control study design (including Publications I and II in this thesis), where protein concentrations in patients can be compared to those in healthy individuals \cite{Zheng2018}. Proteins that can reliably differentiate between patients and controls are referred to as differential and process of identifying such proteins -- differential analysis \cite{Abel2014, DeLuca2011}. More formally differential proteins are the proteins for which the probability to observe the corresponding value of the test statistic to be sampled from the null distribution is below the acceptable significance threshold leading to rejection of the null hypothesis. 

Differential proteins can be used to explore the mechanisms of the disease (such as in Publications I and II) or as a screening tool -- measuring autoantibody reaction to these proteins in the general population can ideally reveal individuals at risk (Publication IV). These individuals could be treated early and less aggressively, increasing their chances for long-term well-being. Below we discuss different statistical methods used in this thesis to detect differential proteins. 

\subsubsection{Z-score analysis}

A substantial number of protein microarray-based studies (including Publications I and II in this thesis) have used an approach referred to as \textit{Z-score analysis} or \textit{Z-test} to define differential proteins \cite{Meyer2016, Fishman2017, Dez2012, DeLuca2011, Landegren2016}.

Classical Z-test is used to either evaluate the difference between two groups of samples or a group and a known population. The null hypothesis for the latter case can be formulated as $H_0: \mu = \mu_0$ in two-sided version or as either $H_0: \mu >= \mu_0$ and $H_0: \mu <= \mu_0$ for one-sided version, where $\mu$, $\mu_0$ are group and population mean respectively. Z-test uses \textit{Z-scores} (or standard scores) as a test statistic, which can be defined as
\begin{equation}
\label{zscore}
z = \frac{\mu - \mu_0}{\sigma}
\end{equation}

where $z$ is the Z-score for a given group of samples with a mean $\mu$, while $\mu_0$ and $\sigma$ are population mean and standard deviation respectively. The distribution of Z-scores under the null hypothesis is well-known and can be approximated by a corresponding normal distribution. The group with unusually high or low Z-score value is considered to have a mean value different from the one of null-distribution. The relevant p-value can be estimated as the percentage of the null distribution falling above or below the observed Z-score.

However, unlike the classical Z-test described above, in protein microarray analysis it is common to use individual spot's concentration values $x$ in a place of group mean $\mu$ in \ref{zscore} \cite{DeLuca2011}, leading to:

\begin{equation}
z = \frac{x - \mu_0}{\sigma}
\end{equation}

This approach is similar to the \textit{three-sigma rule} described in the outlier detection section, as protein concentrations with a Z-score of more than 3 or less than -3 in one or more samples are considered to be significantly differential \cite{Meyer2016, Fishman2017, DeLuca2011}. However, due to the potentially high number of tests, such a strategy may lead to a substantial number of proteins deemed differential by mistake (see a section on multiple testing correction). Such risk can still be justified in the case the goal is not to identify proteins that are consistently differential across all patients, but proteins that have abnormally higher concentration values only in a handful of patients. This is the case for APECED patients, who develop heterogeneous sets of autoantibodies that may differ from patient to patient and thus target vastly different sets of proteins spotted on protein microarray slides.

Two main assumptions should be met in order for Z-test to be applicable: Z-scores should follow a normal distribution and population parameters i.e. mean and standard deviation should be known in advance, which is not always possible. Often population mean and standard deviation can be estimated using the sample mean and standard deviation, which transforms Z-test into a t-test.

\subsubsection{Student's t-test}

The Student's t-test (or a t-test) is one of the most popular statistical approaches for hypothesis testing. The test was named after William Sealy Gosset who published the method under the alias ``Student''. 

In this thesis we have employed a two-sample version of the t-test, which explores the difference between two groups of protein microarray slides (patients and controls), $H_0$ for such test has the following familiar formulation $H_0: \mu_1 = \mu_2$, where $\mu_1$ and $\mu_2$ are means of the group 1 and group 2 respectively. An alternative hypothesis ($H_1$) is therefore $H_1: \mu_1 \neq \mu_2$. Similar to Z-score analysis which is relying on Z-scores, the t-test computes t-values (denoted as $t$), which are used to decide the outcome of the test. T-value is a test statistic for the t-test and calculated using the following formula:

\begin{equation}
\label{tvalue}
t = \frac{\mu_1 - \mu_2}{s_p \sqrt{\frac{1}{n_1} + \frac{1}{n_2}}}
\end{equation} 

where  
\begin{equation}
\label{sp}
s_p = \sqrt{\frac{s^2_1}{2} + \frac{s^2_2}{2}}
\end{equation}

Above, $s_p$ is a pooled standard deviation (\ref{sp}), $n_1$ and $n_2$ represent the number of samples in group 1 and group 2 respectively, while $s_1$ and $s_2$ are the standard deviations of these two groups. The formula for $t$ (\ref{tvalue}) is valid as long as there is a good reason to believe that groups have similar sizes and are sampled from the populations with equal variances. In other circumstances, slightly different formulas for $t$ and $s_p$ must be applied. From the definition, it follows that $t$ is the distance between group means in units of pooled standard deviation. If the null hypothesis is true the value of $t$ should be close to 0, suggesting no difference between the two means. However, the larger the value of $t$, the less likely is $H_0$. Thus, using computed $t$-value it is possible to quantify the p-value by comparing $t$ to a null-distribution.  If the p-value is less than a predefined threshold, the null hypothesis is considered to be false and can be rejected. It is common to use 0.05 as a threshold imposed on p-values. Rejection of $H_0$ under 0.05 threshold can be interpreted as that there is less than 5\% chance that the observed difference between groups is due to random chance. 

Multiple assumptions should be satisfied for the above equations and reasoning to work. The above-mentioned equality of group sizes and variances is one of such assumptions. The other assumptions are described below. T-test should be applied only to continuous data. Also, sample means from populations being compared should follow the normal distribution, which makes the t-test a member of \textit{parametric tests} family, i.e. tests that rely on a specific probability distribution.  Compared groups should be independent of each other (no overlap, unless paired t-test is used). Lastly, samples i.e. patients and controls should be independent of each other. The aforementioned assumptions are by default assumed to be satisfied, therefore it is the responsibility of the researcher to make sure that data is suitable, otherwise, results produced by the t-test may not be sensible. The normality assumption is especially hard to satisfy for the researchers working with highly heterogeneous protein microarray data. In such cases, more powerful alternatives to classical t-test can be used, such as moderated t-test \cite{Smyth2004} or Mann–Whitney U test \cite{Mann1947} discussed in the next section. 

The t-test can be expressed in terms of linear models discussed in the previous sections and can be formulated as ${\bf y} = {\bf Xw} + {\bf \epsilon}$ (\ref{lm}). Here we will pay no attention to $\epsilon$ term as it is independent for each sample and cannot be accounted for. In our case, ${\bf X}$ is an indicator of whether a sample was drawn from the first or second group and thus can be written simply as $x_i$. The above equation (\ref{lm}) can be reformulated as follows: $y_i = x_i{\bf w}$, where $y_i$ is predicted signal value of $i$-th sample. This equation can be further simplified $y_i = w_0 + w_1*x_i$. If an $i$-th sample is drawn from the first group, $x_i$ becomes 0 and the whole equation transforms into $y_i = w_0 + w_1*0$ or simply $y_i = w_0$. Hence, $w_0$ is a predicted signal for the samples in the first group. Since the best way to summarise a set of points is via their mean, $w_0$ represents a mean signal of the first group. When sample is from the second group, the $x_i$ equals to 1 and thus the core equation changes to $y_i = w_0 + w_1$. With this, we model the second group by adding $w_1$ to the mean of the first group $w_0$, and therefore $w_1$ is the difference between the means of the two groups. The null hypothesis can be formulated accordingly as $H_0: w_1 = 0$. Not only the linear model formulation of the t-test can help to understand the procedure better, but it also facilitates the implementation using programming languages. Various statistical software packages e.g. \textit{limma} in R, uses linear model formulation as a basis for t-test implementation. T-test-based expression analysis performed and presented in Publication II of this thesis was implemented in \textit{limma} and formulated in terms of the linear model. 

\subsubsection{Mann–Whitney U test} Mann-Whitney U test (also known as the Wilcoxon rank-sum test) is an alternative to the two-sample equal variance t-test discussed before \cite{Mann1947}. Contrary to the t-test, the Mann-Whitney U test belongs to the family of \textit{non-parametric tests} that do not rely on any particular parameterized distribution for hypothesis testing. Therefore, it is applicable to data that does not necessarily follow the normal distribution as is often the case in biology. Under null hypothesis $H_0$ the two compared distributions should be considered equal. More formally, the probability of an observation from the first group to be larger (or smaller) than an observation from the second group is not consistently different from the probability of the opposite, namely, that observation from the second group being larger (or smaller) than an observation from the first group. Thus, the Mann-Whitney U test assumes that observations are comparable, i.e. it is possible to say if one is bigger than the other. In linear model formulation, Mann–Whitney U test is very similar to the standard t-test, except the model is built on ranks of $x$ and $y$ instead of actual values: $rank(y_i) = w_0 + w_1*rank(x_i)$. In this thesis, the Mann-Whitney U test has been used in an attempt to identify differential cytokines (Publication IV).

\subsubsection{Permutation test}
Another way to compare two independent groups of samples without assuming a particular distribution is called a permutation test (or randomization test) \cite{Edgington2011}. It starts with calculating a predefined test statistic on the original data. In the case of two independent groups, we may decide to calculate the difference between two means $\mu_1$ and $\mu_2$ of two groups with $n_1$ and $n_2$ samples respectively. Hence, $df_o = \mu_1 - \mu_2$ is considered an observed value of the test statistic for the original data. In order to obtain a distribution of the test statistic under a null hypothesis $H_0: \mu_1 = \mu_2$, the permutation test performs the following steps. First, it randomly shuffles all the data and assigns $n_1$ first observations to the new first group. The remaining $n_2$ samples are assigned to the second group. Next, the difference between means of randomly created groups $df_r$ is calculated. If obtained value $df_r$ is larger than $df_o$, the pre-initialized counter $i$ is increased by 1. Later, the data is reshuffled again and all the same, steps are repeated a large number of times (e.g. 10,000), each time a new $df_r$ is computed. To estimate the corresponding p-value, the observed value of test statistic $df_o$ should be compared to the distribution of test statistic under the null hypothesis, thus the distribution of $df_r$. This can be done by dividing the resulting value of the counter $i$ by the number of repetitions that were performed. For example, if after 10,000 repetitions only on seven occasions $df_r$ was larger than $df_o$, the probability to observe $df_r$ as extreme as $df_o$ under the null hypothesis is 0.0007, which is less than a classical significance threshold of 0.05, and therefore small enough to reject the null hypothesis.

There is no need to calculate all possible permutations of the original data, as this number can be extremely large (e.g. two groups with 30 observations in each will result in $1.18*10^{17}$ possible permutations) \cite{howell2010statistical}. Instead, a large enough random sample of all possible combinations would be sufficient. The larger the sample, the more precise estimate it will generate. Such a sampling procedure is usually referred to as the Monte Carlo approach. Modern software tools, as well as processing hardware, enable researchers to shuffle their data enough times to obtain sufficiently precise estimates in almost no time, making randomization tests a practical solution to hypothesis testing.

In the research presented in this thesis (in Publication II) we used a permutation test to test a hypothesis that proteins targeted by the autoantibodies in the blood of APECED-positive patients originate from genetically more conservative (i.e. those that accumulate fewer mutations over time) regions of the DNA.

\subsection{Enrichment analysis}

The identified set of significantly differential proteins (i.e. proteins with signal levels significantly different between conditions) can be interpreted with respect to the existing body of knowledge. Such interpretation can be the key to the understanding of biological processes, e.g. mechanisms of the disease. Autoimmune disorders such as APS1, discussed in earlier chapters, are caused by the genetic mutations that undermine the immune system's native ability to prevent self-targeting antibodies from entering the bloodstream. Hence, APS1 patients' blood is filled with a large number of aggressive autoantibodies. Researchers analyze a pool of proteins that are targeted by released autoantibodies, trying to identify properties and functions that are common among the targets. Pinpointing these properties and functions might shed some light on autoantibodies and the autoimmune process, for example, it may provide clues as to autoantibodies' origin. In general, the process of determining properties that are over-represented in a group of proteins or genes is usually referred to as \textit{enrichment analysis}. Enrichment analysis is often performed by quantifying the size of the overlap between a group of proteins with a known biological property, e.g. proteins expressed in lymphoid cells, and a group in question. A statistical test is then used, e.g. hypergeometric test, to estimate the probability that this overlap or larger was observed by random chance. If such probability is deemed sufficiently low (< 5\%), the overlap between groups is considered significant and therefore, genuine. In this case, the group in question is said to share the same biological property as a group with which it was compared. A large number of public databases, such as Gene Ontology \cite{geo}, KEGG \cite{kegg}, Reactome \cite{Reactome}, Human Phenotype Ontology \cite{hpo} and Human Protein Atlas \cite{hpa} are available with protein and gene groups characterized with various biological properties and functions. Hence, in practice, enrichment analysis means comparing the obtained group of target proteins to hundreds or even thousands of groups stored in public databases. A number of potential databases and datasets that can be searched to find relevant terms has long become prohibitively large for humans to manually work through. Thus many software tools, e.g. g:Profiler \cite{Raudvere2019} were developed to automate the enrichment analysis, saving dozens of researchers' work hours. In this thesis, specifically in publication II, we have used g:Profiler as well as a hypergeometric test to identify enriched terms in the group of targeted proteins.

\subsubsection{Hypergeometric test} 

We can explain the idea of the hypergeometric test with the following example: we are drawing balls from the urn which contains balls of two colours: white and black. We took a fixed number of balls from this urn. Some of the balls turned out to be black and some white. The hypergeometric test attempts to answer the following question -- what is the probability of observing as many white balls as we have or more if the balls were drawn from this urn at random. In the context of this thesis, it is possible to reformulate this example as follows: the urn is a protein microarray platform (e.g. ProtoArray), the balls drawn from the urn represent the list of differential proteins we extracted, white balls are the proteins that represent a specific biological function and black balls are all the remaining proteins. The null hypothesis here is that the overlap between the list of extracted proteins and proteins associated with some biological process is of the same size as it would be expected if we drew the proteins from the platform at random. Therefore, we ought to find a probability to observe as many proteins that belong to a biological process as we do in our list if we drew these proteins from the protein microarray platform at random. More formally, the probability to draw $k$ proteins that are associated with a biological process of interest in our list by chance $p(k)$ can be calculated as follows:

\begin{equation}
\label{hypertest}
p(k) = \frac{{K \choose k}{N - K \choose n - k}}{{N \choose n}}
\end{equation}
where $K$ is the number of such proteins in total on the platform. The number of proteins in our list is $n$ and $N$ is the number of proteins overall on the platform. By inserting all possible values of $k$ (from 0 to $n$) into \ref{hypertest} it is possible to obtain corresponding hypergeometric distribution. By accumulating parts of the distribution that correspond to actual $k_{actual}$ observed from the data, it is possible to estimate the probability to observe $k$ as extreme as $k_{actual}$ or larger by chance. If this probability is low enough (less than 0.05) we reject the null hypothesis and assume that there are more proteins associated with a biological property among extracted proteins than what we can expect at random. Although we have used g:Profiler to determine biological processes common among targeted proteins, we still applied the hypergeometric test in Publication II to be able to perform enrichment analysis on datasets that were not part of the g:Profiler tool and also to cross-check our findings.

\subsection{Multiple testing correction}

Let us recall that the p-value implies the probability to observe the current outcome of the experiment or even more extreme under the null hypothesis. If this probability is less than 5\% it is commonly accepted that the null hypothesis is unlikely to be true and therefore can be safely rejected. Here, ``unlikely'' does not mean ``impossible'', thus it is still imaginable to obtain a p-value less than 5\% under the null hypothesis by chance and the probability of this event is the same 5\%. Although 5\% does not seem to be a very high value, consider an example, which involves 20 simultaneous tests with the same p-value threshold of 5\% \cite{Goldman2008}. Let us calculate the probability of having at least one test out of 20 to generate a p-value of 5\% or less by pure luck. This is equivalent to asking for the probability to obtain at least one head by tossing a biased coin (which has a 95\% chance of coming tails) 20 times. This probability can be calculated as follows:

\begin{equation}
\begin{split}
p(\textrm{at least 1 significant test}) & = 1 - p(\textrm{no significant tests})^{20} \\
 & = 1 - (1 - 0.05)^{20} \\
 & = 1 - 0.36 \\
 & = 0.64. 
\end{split}
\end{equation}

There is a 64\% chance to obtain at least one falsely significant result out of 20 independent tests. Differential analysis for a protein microarray implies running statistical tests described in the previous section for each of thousands of proteins seeded on the platform. For example, analysing a typical HuProt protein microarray experiment would mean performing about 20,000 tests simultaneously. For each test, a p-value will be produced, and the null hypothesis is rejected if the corresponding p-value is less than 0.05. If we assume the absence of true differential proteins on the platform, 5\%, namely 1000 (0.05*20,000) proteins will be considered significant simply by chance (because the number of tests is so large). For example, for the ProtoArray study discussed in Publication II, this would mean that half of the proteins in the positive group are false. Therefore, the number of tests needs to be taken into account when performing statistical analysis. A number of methods have been proposed to adjust the classical statistical significance threshold of 5\%.

\subsubsection{Bonferroni correction}
The simplest multiple testing correction approach is called the Bonferroni correction \cite{bonferroni1936}. This method adjusts the p-value threshold by dividing it by the number of experiments. In the above example of HuProt array with $\sim20000$ proteins and threshold of 0.05, the new significance threshold becomes $0.05/20000 = 2.5*10^{-6}$. Applied to $n$ tests with a p-value threshold of $\alpha$ Bonferroni correction ensures that the probability of observing at least one false significant result is $\alpha$ \cite{Noble2009}, while it is usually sufficient to optimize for some acceptable proportion of false predictions. Thus this procedure was commonly regarded as overly strict for most of the practical applications \cite{Noble2009, Goldman2008}.

\subsubsection{Benjamini and Hochberg correction} 
The method by Benjamini and Hochberg (also known as FDR correction) attempts to keep the number of falsely significant results at a certain predefined level (e.g. 5\%) \cite{bhmethod}. The proportion of falsely admitted associations (false positives) among all significant results is called a false discovery rate. The method starts with ordering all $m$ unadjusted p-values in a descending order \cite{Goldman2008}. Then for the $i$-th p-value $p_i$, the algorithm checks if this value is less or equal to $(i/m)*\alpha$, where $\alpha$ is acceptable level of FDR. As soon as such $p_i$ was found consider it to be a significant threshold. We used the method by Benjamini and Hochberg to correct unadjusted p-values for multiple testing in publications II and IV.

\section{Machine learning modelling}

While classical statistical methods help to analyse the significance of each protein feature separately \cite{Zumbo2014}, more sophisticated methods, such as machine learning algorithms, are needed to assess the predictive performance of multiple proteins combined. Machine learning (sometimes also can be referred to as artificial intelligence) is a field of computer science that develops algorithms capable of learning valuable relationships from data without being explicitly programmed. Such relationships can then be used further by the same machine learning models to accurately predict the value of the outcome variable in new unseen data. Machine learning models are frequently used in biology in an attempt to build diagnostic tools for certain diseases (e.g. publication IV of this thesis). Depending on outcome variable type (discrete or continuous) machine learning models can be broadly divided into classification and regression algorithms. Myriads of machine learning algorithms have been developed for each of these categories \cite{Tarca2007}. Some of the most popular machine learning methods capable of working with both continuous and discrete outcome variables are decision trees \cite{Quinlan1986}, and random forest \cite{Breiman2001}. 

\subsubsection{The decision tree algorithm} 
\begin{figure*}
\begin{center}
\includegraphics[trim=0 0 5 0, clip, width=.9\textwidth]{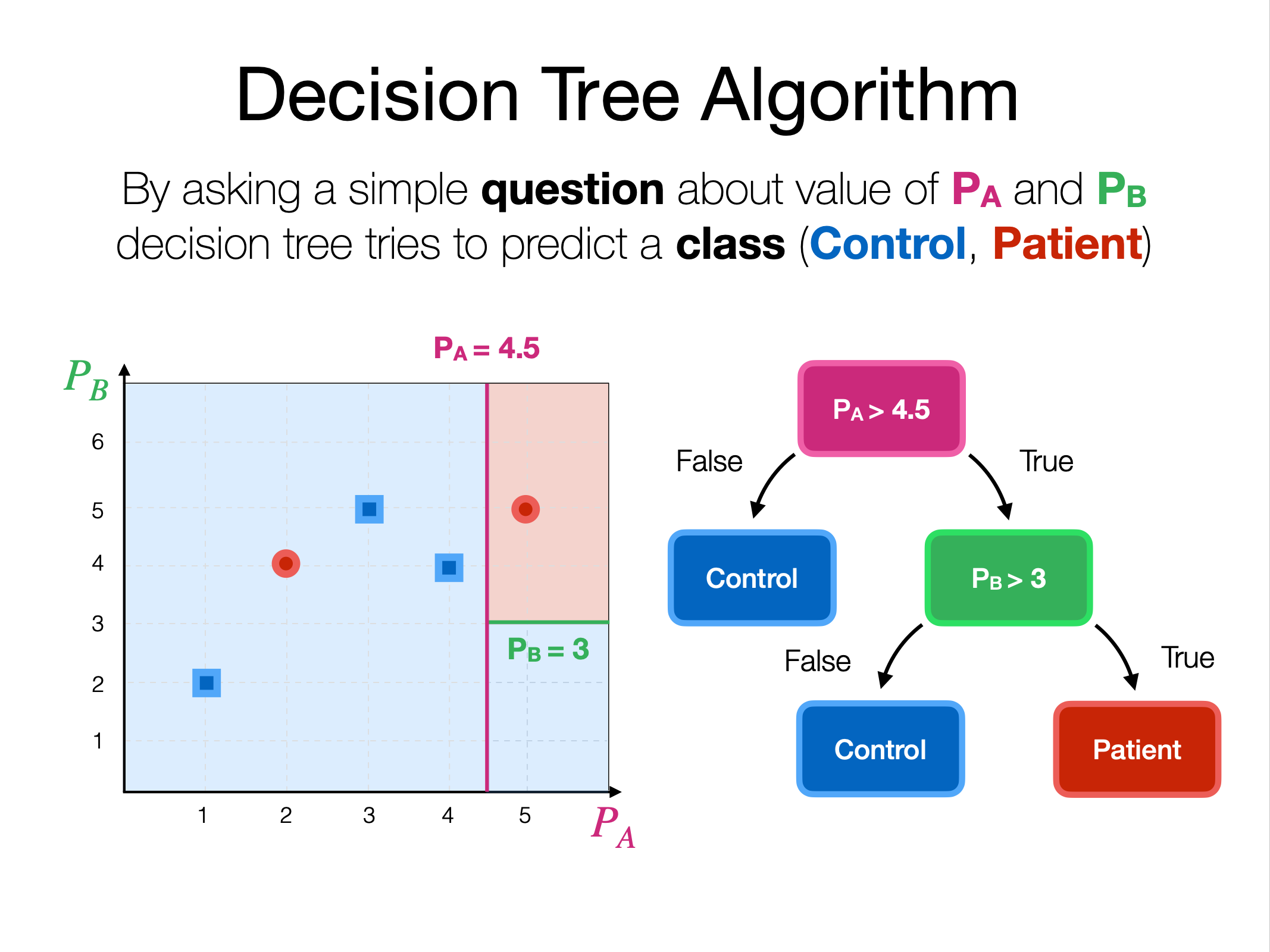}
\end{center}
\caption{Decision tree algorithm uses information about individual protein intensities ($P_A$ and $P_B$) in order to explain the outcome variable. For example, the decision tree algorithm may assign a class ''control`` to a sample for which protein A ($P_A$) has an intensity value less than $4.5$.}
\label{dt}
\end{figure*}

Decision tree algorithm \cite{Breiman2017} uses values of input features (e.g. protein intensities) to infer the value of the outcome variable. A decision tree has a recursive structure with a root at its origin and leaves at the bottom. Each node can potentially have two children nodes. All nodes except leaves encode conditions in a form of questions. For example one of the nodes may inquire if the normalised intensity of protein A has a value of more than 4.5 in order to decide on a value of the outcome variable (Figure \ref{dt}). The tree starts with checking if the input data satisfies the initial condition at the root and descends down the tree depending on the outcome. Leaf nodes of the tree determine the outcome of the algorithm: class in the case of classification or continuous value for regression. To build the tree the decision tree algorithm uses values of all available features. 

\subsubsection{The random forest algorithm} 
\begin{figure*}
\begin{center}
\includegraphics[trim=0 0 5 0, clip, width=.9\textwidth]{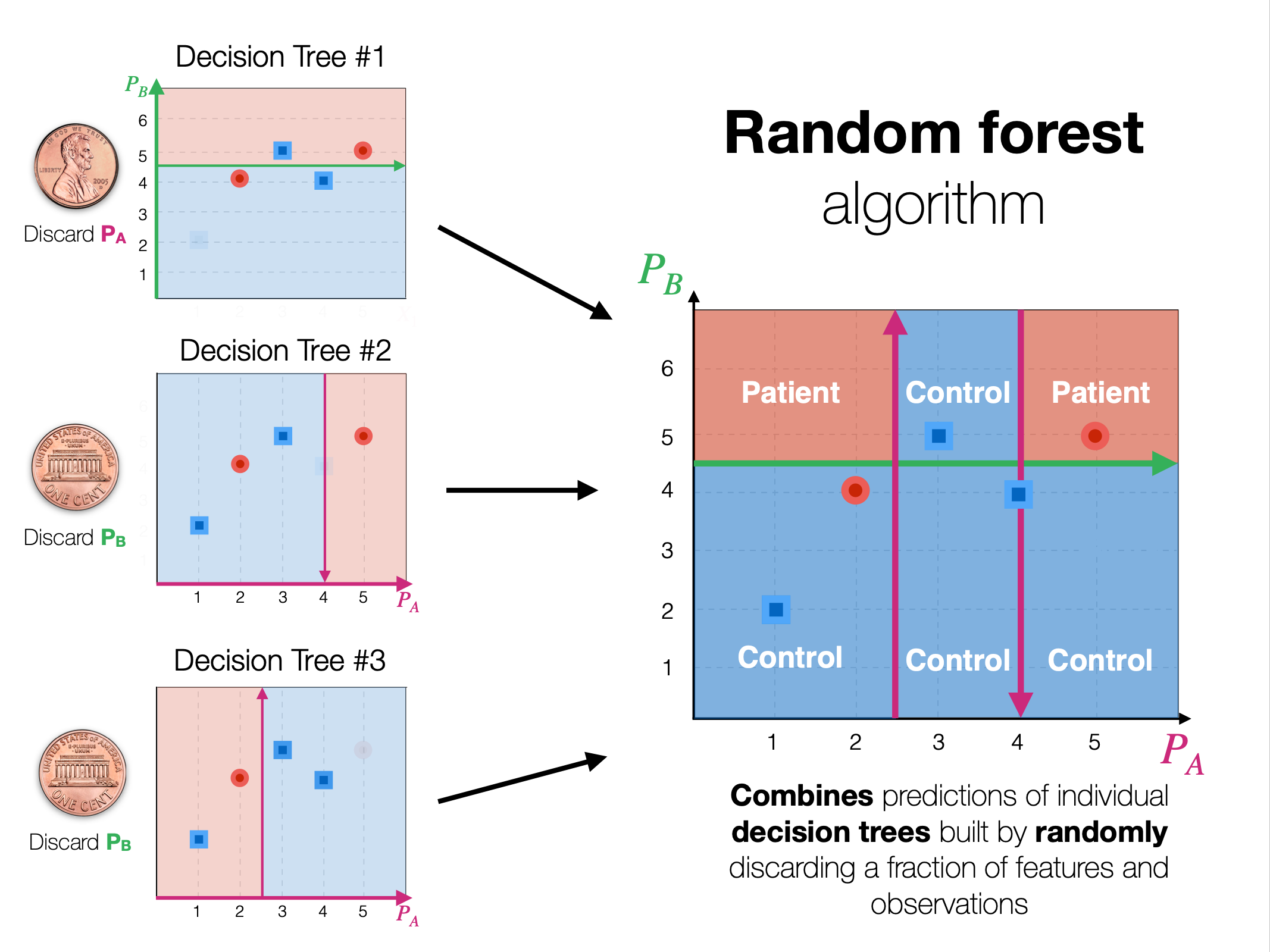}
\end{center}
\caption{Random forest algorithm uses predictions produced by individual decision trees in order to predict an outcome variable. Decision trees that are part of a random forest ensemble are built by randomly discarding a pre-defined number of features and observations. For example, decision tree 1 has been built after discarding protein A intensity and the first control sample.}
\label{RF}
\end{figure*}

The random forest algorithm \cite{Breiman2001} can be considered an extension of the decision tree algorithm discussed above. Instead of building one tree and inferring the predictions from that tree, the random forest algorithm creates several trees, which are used to predict the value of the outcome variable in parallel. Predictions from the individual trees are then combined together to obtain the final joint prediction. Such approach is often called \textit{ensemble learning} or \textit{ensembling}. The random forest algorithm has another important difference from the decision tree algorithm. The individual trees are constructed using a random subset of input features and input data points (e.g. proteins). For example, it is common to use random 80\% of samples for each tree in the forest and when building a tree, nodes are optimized using only random 80\% of the features (Figure \ref{RF}). This has been shown to make random forest models extremely robust to noise and highly generalizable to unseen data.   

\subsubsection{Evaluating machine learning models} Creating a fully functional machine learning model (e.g. for diagnosis), starts with exposing it to data that closely resembles the data on which it is expected to perform well in the future. This initial dataset is called \textit{training data}. Most machine learning algorithms (including random forest algorithm) have various parameters that can be tweaked in the hope to obtain a better working model. By tweaking such parameters, machine learning models can be made powerful enough to completely memorize all possible aspects of the training data. Counterintuitively, this can be harmful to the model's performance on new data, if its distribution exhibits even the slightest deviation from the distribution of the training set. Therefore, correctly estimating the performance when training machine learning algorithms and tweaking their parameters is a very important step in creating a viable data analysis pipeline. The performance of a model is usually evaluated using independent parts of data (referred to as validation or test set). This approach implies detaching a substantial part of data from the training set, which may not be used for model training. However, if the number of samples is limited, which is usually the case in biological research, creating a separate validation set can be prohibitively data-expensive. Another way to evaluate the performance of the model is called the cross-validation (CV) algorithm. It starts by randomly splitting the initial data set into a finite number of chunks (folds). The total number of folds depends on the size of the data set, often ranging between 3 and 6. At every iteration of the CV algorithm, the machine learning model is trained on all folds except one, the remaining fold is used as a validation set, allowing to take a snapshot of the model's performance. Part of the data that serves as a validation set is changed at every iteration, allowing to obtain several performance estimates using one data set. The cross-validation algorithm provides a safe way to estimate the unbiased performance of the model under different sets of parameters without letting the model memorize training data. Once the perfect combination of parameters was found and its performance estimated, the model can be trained on the entire dataset. The cross-validation approach was used in the publication IV of this thesis as well as in our previous work \cite{Ottas2017} to efficiently evaluate machine learning models.

At present, it is impossible to imagine a section on machine learning that would ignore a connection between machine learning and deep learning techniques. Deep learning algorithms are machine learning methods that use a popular type of machine learning models -- neural networks to tackle the most challenging problems, often previously unsolvable by humans. Since the success of the AlexNet neural network in 2010 \cite{alexnet} these methods have been considered to be the most advanced form of machine learning. Although deep learning approaches are considered to be state-of-the-art in many areas, including biology \cite{Angermueller2016, Jones2017}, due to several substantial limitations of neural networks (e.g. low transparency and data-intensive nature of training), here we have focused on less complex, yet still powerful machine learning methods that have been used in publication IV of this thesis.

%%%%%%%%%%%%%%%%%%%%%%%%%%%%%%%%%%%%%%%%%%%%%%%%%%%%%%%%%%%%%%%%%%%%%%%%%%%%%%%%%%%%%%%%%%%%%%%%%%%%%
% PUBLICATION I CHAPTER
%%%%%%%%%%%%%%%%%%%%%%%%%%%%%%%%%%%%%%%%%%%%%%%%%%%%%%%%%%%%%%%%%%%%%%%%%%%%%%%%%%%%%%%%%%%%%%%%%%%%%

\chapter{Protein microarray analysis in the search for high-affinity ameliorating autoantibodies (Publication I)}

The AIRE gene has an important role in central tolerance as it is responsible for assembling self-antigens used in T cell maturation \cite{Kisand2011, Kont2008, Nagamine1997}. These antigens are presented to T cells during the so-called \textit{negative selection phase}. Binding between prospective T cells and self-antigens indicates the potentially self-reactive tendency of T cells. As a result, such T cells are deemed dangerous for the organism and normally are removed from the pool of potential immune cells.

Occasional genetic mutations can alter or even cease the function of AIRE, jeopardising central T cell tolerance \cite{Nagamine1997} and thus resulting in the accumulation of self-reactive immune cells in the bloodstream. T cells have been linked with activation of  B cells \cite{Coutinho1984} that produce autoantibodies. The theory emerged that genetically deficient AIRE gene may distort not only the selection of T cells but also may indirectly create autoimmune B cells and as a result -- disease-causing autoantibodies. Despite such an important role, precise molecular mechanisms of AIRE have remained poorly understood \cite{PETERSON2004}.  Genetic alternations of AIRE cause the APECED/APS1 autoimmune condition mentioned previously. This disease is characterised by large numbers of autoantibodies against self-antigens expressed in the peripheral organs present in the patients' blood \cite{Kluger2012}. Although this condition is rather rare in the general population (depending on the country it can range from 1 in 25,000 to 1 in 1,000,000), it is often considered as a model disease for human autoimmunity. We, therefore, reasoned that studying autoantibodies from a sufficiently large set of APS1/APECED patients may help to gain a better understanding of the interplay between AIRE and autoimmunity, for example, collect evidence for the hypothesis that specific protein features may be an indirect cause of B cell autoimmunity \cite{PETERSON2004, Meyer2016, Fishman2017}.

Overall, eighty-one APS1/APECED patients of five distinct geographical origins were recruited into the study that was published in 2016 \cite{Meyer2016}. Some of the patients were sampled several times over the course of the study, resulting in a total of ninety-seven patient samples. Control samples were extracted from nine first-degree relatives and twelve healthy volunteers. In total, data from a hundred eighteen samples were analysed. Protein microarray chips from Fisher Scientific (ProtoArray) were used to quantitatively measure the presence of autoantibodies in these samples. 

% Results and our contribution (perhaps should be expanded)

According to a standard data acquisition pipeline, ProtoArray chips were scanned with a GenePix scanner, resulting in hundred eighteen GPR files. Signal was acquired from each GPR file using \textit{readMAimages} function from \textit{limma} package in R. Data was pre-processed and normalised using robust linear model \cite{sboner}. Later, in order to identify proteins that showed higher levels of autoimmune reaction, the signal from each spot was transformed into Z-score by subtracting the mean and dividing by the standard deviation of the combined healthy and first-degree relatives group. We considered a protein to be a positive hit if the corresponding Z-score had a value of 3 or larger in at least one sample. This resulted in a high number of positive targets (3,731) jointly recognised by the group of patients and 406 proteins recognized by controls. This observation was in concordance with a widely held view that APS1/APECED patients are very heterogeneous in the nature of their autoimmune response with only a handful of proteins (such as type I interferon family) being recognised by the majority of patients \cite{Kisand2010, Puel2010}, while other constituting a private set.

To conclude, this paper had at least two major findings. Firstly, our statistical analysis showed that APS1/APECED patients as a group develop a unique set of autoantibodies that recognise approximately a hundred body's own proteins. Almost all samples contained high concentrations of autoantibodies against a small set of proteins (about 10), such as type I interferon or interleukin-22 (IL22). The remaining proteins were collectively recognised by autoantibodies present only in a handful of individuals. Thus, it was observed that blood from all 81 patients collectively contained autoantibodies against more than 3,700 human proteins (which is about 14\% of the canonical proteome). Secondly and perhaps even more importantly, the presence of autoantibodies against type I interferons had a surprising negative correlation with type I diabetes. In this publication, our main contribution is of two folds: implementation of a protein microarray specific data pre-processing pipeline, including re-implementation of the robust linear model for normalisation and identification of positive hits using Z-scores. A more comprehensive analysis of autoimmune targets was performed in Publication II, which will be discussed in the following chapter.

An important comment was brought to our attention several years after the publication. A group of independent researchers pointed out that the mean and standard deviation of combined control and healthy relatives group that we have used to calculate Z-scores is inherently small. Thus, such a procedure is bound to produce higher numbers of proteins with Z-scores above the aforementioned threshold of 3 in patients \cite{Landegren2019}. Hence, they concluded that it is likely that a large part of reported in the paper 3,731 positive proteins are false positives -- a side effect of employing an imperfect statistical method. The authors of the comment suggested that using classical statistical methods for comparing two distributions would alleviate the problem and produce more trustworthy results. In our response, published along with the original criticism \cite{response2019}, we emphasized the goal of the study -- broadly characterize the nature of auto-reactivity in APS1/APECED patients. We showed that this goal implied maintaining a false-negative rate as low as possible. Moreover, we stressed an important clinical aspect of APS1/APECED, namely that each patient is highly individual in the range of symptoms, which manifests in diverse sets of autoantibodies present in the blood of patients \cite{response2019}. In this vein, classical statistical tests such as Fisher's exact test \cite{Fisher1922} or moderated t-test \cite{limma} may not have been applicable as they would filter out proteins recognised only by few individuals, as insignificant. While we have not denied the fact this approach could have led to the higher number of false positives, we nevertheless used it in our analysis, as true targets were to be verified by independent lab experiments and follow-up studies \cite{Meyer2016, Rodero2017, Fremond2017, Fishman2017, Sng2019}. This argument is relevant to the analysis performed for the second publication.

The comment published in eLife \cite{Landegren2019} and our subsequent response \cite{response2019} enabled us to appreciate the complexity involved in devising a precise rule for identifying true-positive autoantibody targets in protein microarray experiments. Where the classical statistical methods are deemed unsuitable for the task due to the stochastic nature of the autoimmunogenesis, the ad hoc solutions may lack the desirable precision. Arguably the most common way to respond to such a challenge is to employ several orthogonal assays to confirm the findings. Hence, there seems to be a need for the non-parametric technique that would enable robust yet non-restrictive analysis of inherently variable signal intensities often exhibited by the protein microarrays.

%%%%%%%%%%%%%%%%%%%%%%%%%%%%%%%%%%%%%%%%%%%%%%%%%%%%%%%%%%%%%%%%%%%%%%%%%%%%%%%%%%%%%%%%%%%%%%%%%%%%%
% PUBLICATION II CHAPTER
%%%%%%%%%%%%%%%%%%%%%%%%%%%%%%%%%%%%%%%%%%%%%%%%%%%%%%%%%%%%%%%%%%%%%%%%%%%%%%%%%%%%%%%%%%%%%%%%%%%%%

\chapter{Characterising autoimmune targets further (Publication II)}

Autoantibodies have been shown to play an important role in the onset and progression of various autoimmune diseases \cite{Bratland2011, Herold2013}. Yet, while recently a lot more has been discovered about cellular and genetic factors that contribute to the emergence of autoimmunity, our understanding of properties of involved autoantibodies remains limited \cite{Fishman2017}. We made an initial attempt to characterise autoantibodies and their targets in Publication I \cite{Meyer2016}, where we showed that patients' blood contains high concentrations of autoantibodies against a set of well-known proteins. Some of these proteins such as type I interferons (especially IFN-$\alpha$), became diagnostic markers for APECED \cite{Kisand2011}. Autoantibodies against other proteins, like IL17A, IL17F and IL22 were shown to contribute to the onset of chronic mucocutaneous candidiasis -- another distinctive feature of APECED \cite{Okada2016,Fishman2017}. Previous studies suggested that there are autoantibodies that are shared between APECED and other complex autoimmune diseases, such as Addison's disease \cite{Bratland2011} and T1D \cite{Herold2013}. But the number of such commonly targeted and widely known proteins, is small, comparing to the total number of proteins collectively targeted by autoantibodies in all patients (3,731 in total). Hence, in Publication II we focused on an in-depth analysis of autoimmune targets identified in Publication I. 

Protein data from the same samples as in Publication I was pre-processed using an earlier developed pipeline with few minor modifications. As before, the background-subtracted signal (we used basic median local background subtraction) was log-transformed before being normalised using RLM. The resulting values were standardized using the mean and standard deviation of control samples. After several studies that have employed ProtoArrays highlighted a danger of cross-contamination between neighbouring protein spots \cite{Landegren2016} we decided to add another quality control step into our pipeline. Thus, unlike the workflow of the first publication, here, we removed 31 proteins with unexpectedly highly correlated signals (with Pearson's coefficient of 0.6 or higher) with the expression of nearby proteins or well-known protein targets (e.g. IFN-$\alpha$). The remaining proteins with a standardized score of 3 or higher in three or more patients were considered to be true positive targets \cite{Fishman2017}. Additional filtering criteria of three patients were introduced in this work to reduce the number of possible false positives that could distort the analysis. This requirement narrowed down the list of positive proteins from an initial 3,731 to 963, which we later referred to simply as the ``positive group''. Although the positive group was significantly decreased in size, it remained big enough to include proteins recognised ``privately'' i.e. by fewer patients, and therefore capable of providing details on the mechanisms of the autoimmunity. 

Most of the further analysis was centered around characterisation of the positive group, by matching our protein list with various public databases and analysing available meta-information about samples. In the course of this research, we looked for biological processes that are over-represented in the positive group. We quantified the number of single nucleotide polymorphisms (SNPs) and APECED related mutations as well as the level of evolutionary conservation of relevant gene regions. We also performed differential and clustering analysis with respect to associated clinical conditions, which however revealed few notable results. Finally, we ran a longitudinal analysis of protein expression patterns in the positive group. Performing all these experiments involved a number of technical challenges, solutions to which, are our main contributions to this publication. 

First and foremost, in order to be able to use public databases in our analysis we needed a way to unambiguously compare entities stored in these databases (usually proteins or genes) with our positive group. Most of the tools and datasets, employed in this work, operated with Ensembl gene IDs (ENSG). While for a few others, symbolic gene names must have been used. Analysing overlaps with these databases meant converting all native Protoarray names used by the manufacturer (approximately nine thousand Reference Sequence IDs), into ENSGs and gene names. We used g:Covert web-tool \cite{Raudvere2019} available at \url{https://biit.cs.ut.ee/gprofiler/convert} to obtain initial results. But a substantial number of missing and duplicated gene ids in the results presented themselves as a serious problem for further analysis. On one hand, due to factors such as alternative splicing, when a gene can be associated with multiple proteins, some number of duplicated IDs was expected. However, a lot of protein IDs were not converted by g:Convert into ENSGs at all. To impute as many missing gene IDs as possible, we
parsed the official ProtoArray content file, applied g:Convert tool, and manually searched NCBI \cite{Agarwala2017} and Ensembl databases \cite{Yates2019}. Eventually, the number of proteins that could not have been translated was reduced to 324, which is below 4\% of all proteins on the platform.

%   Contribution II

Next, data from all relevant databases must have been acquired and unified prior to further analysis. Often datasets that we required for the analysis were redistributed over multiple files and stored in different formats, using conflicting or inconsistent notations. Sometimes, we had to prepare a custom dataset based on public records. For example, to compare the mutation rate of the positive group with the overall platform, we extracted information about all mutations in the human genome and then programmatically searched for SNPs associated with relevant genes and gene regions (introns, exons, etc.). 

Finally, we proceeded to analyse the autoimmune targets by comparing them to data from public databases \cite{hpa,kegg, Reactome, hpo}. Some databases contained only gene lists associated with a certain biological class (e.g. genes expressed only in some tissue), while others supplemented genes with quantitative measures (e.g. level of evolutionary conservation). Therefore, the third major technical challenge we faced in this work can be broadly described as building a multi-headed statistical pipeline to characterise various biological properties of our positive group. We used a number of statistical tests to accomplish this, each in a specific context. The hypergeometric test was used to assess the significance of the overlap between the positive group and various biological processes, permutation test to compare quantitative characteristics, various univariate tests to check for statistical differences between distributions. The resulting p-values were adjusted using the Benjamini-Hochberg method \cite{bhmethod} to account for a large number of tests executed in parallel. Finally, we applied a classical significance threshold of 0.05 to adjusted p-values to find statistically relevant properties of the positive group. To cross-check the results of our statistical pipeline, we submitted the list of positive targets to g:Profiler web-tool \cite{Raudvere2019}. We used an unordered query with all ProtoArray proteins as a statistical background \cite{Fishman2017}. 

Described contributions helped us to discover a number of features shared by the autoimmune targets. For example, we have shown that our positive group was on average significantly more evolutionary conservative (i.e. had fewer mutations) comparing to other proteins from the platform (with an adjusted p-value of 0.0162). A significant proportion of the proteins from the positive group are found in the cell nucleus or cytosol. We also reported a significant association between the number of recognised proteins in each sample and the three most common APS1 mutations. Lastly, based on our results we hypothesised that  APECED ``autoimmunome'' is comprised of two distinct groups of autoantibodies, one of which is likely to be traced back to the AIRE gene discussed in the previous chapter, while the other originates from lymphoid tissue \cite{Fishman2017}. These observations expanded our understanding of the biological properties of autoimmune targets in APECED. Protein microarray data used as a basis for Publications I and II can be found online via the accession number ``E-MTAB-5369'' on ArrayExpress \href{https://www.ebi.ac.uk/arrayexpress/experiments/E-MTAB-5369/}{website}.

To summarise, in this publication, we performed an analysis of autoimmune protein targets in APECED \cite{Fishman2017}. The main engineering and data analysis challenges include lossless conversion of ProtoArray protein IDs into ENSGs and gene names, the transformation of the data from more than fifteen public databases into a format usable for further analysis with a multitude of statistical approaches. The substantial complexity of the performed analysis presented a need for a user-friendly tool that would automate the most laborious parts of the protein microarray analysis. We went on to build such a tool that will be presented separately in the next chapter.

%%%%%%%%%%%%%%%%%%%%%%%%%%%%%%%%%%%%%%%%%%%%%%%%%%%%%%%%%%%%%%%%%%%%%%%%%%%%%%%%%%%%%%%%%%%%%%%%%%%%%
% PUBLICATION III CHAPTER
%%%%%%%%%%%%%%%%%%%%%%%%%%%%%%%%%%%%%%%%%%%%%%%%%%%%%%%%%%%%%%%%%%%%%%%%%%%%%%%%%%%%%%%%%%%%%%%%%%%%%

\chapter{Automating protein microarray analysis with Protein Array Web ExploreR (Publication III)}

Building a complete protein microarray analysis pipeline presented in the previous two sections proved to be time-consuming as well as required skills in statistics and programming. People with such background are not always available in biological labs -- primary sources of protein microarray data. At the same time, the existing analysis tools are either outdated (e.g. manufacturer's own tool -- Prospector) or require familiarity with programming to work with \cite{PAA} or hard to use. As the resulting protein microarray analysis became a real challenge for the practitioners. Therefore, we decided to develop a user-friendly web-tool -- Protein Array Web ExploreR that enables biologists, who produce data, to carry out protein microarray analysis independently using publicly available web service. PAWER is based on paweR -- the R package that was first developed as part of this work. 

\section{PAWER pipeline} 

To start the analysis, PAWER expects fluorescent signal array readings -- GenePix Results files as an input. Files are read in and assembled into a single data matrix using \textit{limma} package in R \cite{limma}. Then the protein signal is estimated by subtracting the background signals from foreground intensities. Depending on the platform background subtraction is done using either default values (for ProtoArray or HuProt) or user-defined feature names. The resulting values are first log-transformed and then normalised via the robust linear model algorithm \cite{sboner}. RLM models signal of control proteins that by default should exhibit no variation between conditions, using protein location (array and block) and the protein type. Later non-zero coefficients related to individual arrays and blocks are subtracted from the corresponding signal intensities to correct for existing biases. Control proteins that are used as a basis for the linear model can also be chosen manually by the user, through a convenient search interface. Otherwise, several reasonable default options are provided. The obtained normalised data matrix can be downloaded as a separate file. This file can then be used as an input to other tools to obtain additional insights. To enable principal component and clustering analysis of the normalised protein microarray data, PAWER is explicitly linked to ClustVis tool \cite{Metsalu2015}. The user only needs to click a button and upload the file with normalised data to run the additional analysis.

Next, the user can provide meta-information about samples in order to identify proteins in which intensity levels significantly differ between conditions. Metadata can be either uploaded as a separate file or entered manually by clicking on a corresponding radio button for each GPR file in the input. Once the metadata has been successfully uploaded, PAWER performs differential protein analysis. In this work, we used a moderated t-test, implemented in \textit{limma} package \cite{limma}. In order to perform a moderated t-test, the number of samples must be larger than the number of conditions (at least by one). Therefore, since currently, PAWER supports only binary conditions, it requires at least three samples (in total) to perform the differential analysis. As in publication II, the Benjamini-Hochberg method \cite{bhmethod} has been used to adjust computed p-values for multiple tests and thus, greatly reduce the number of false-positive proteins. Proteins with adjusted p-values of less than 0.05 are displayed in the results table along with boxplots that illustrate the distribution of the signal of each protein in the table. The content of the table can be changed by selecting only a subset of proteins. Both the table and boxplot visualization can be downloaded separately.

\begin{figure*}
\begin{center}
\includegraphics[trim=0 0 5 0, clip, width=.9\textwidth]{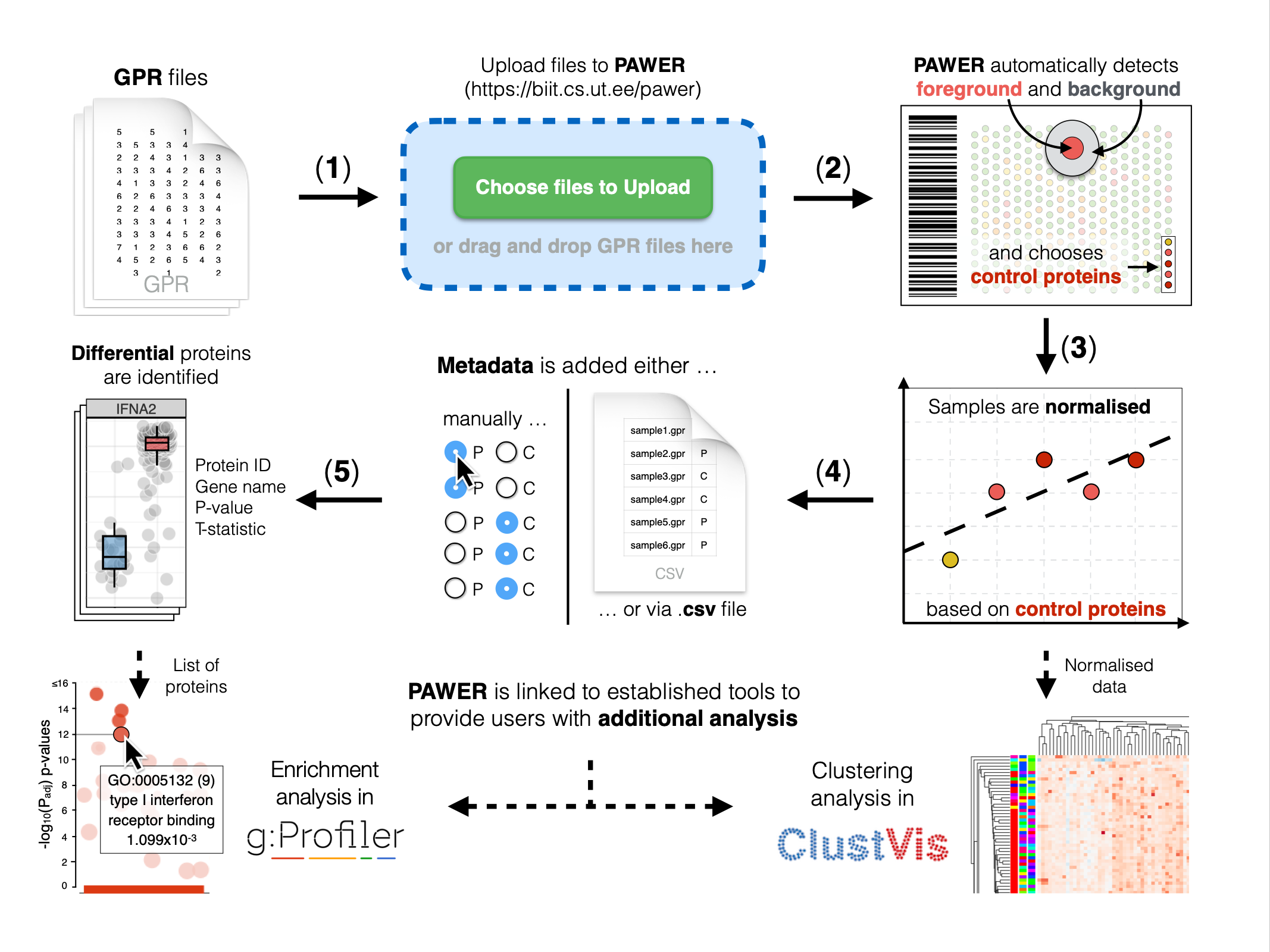}
\end{center}
\caption{PAWER pipeline. Raw GPR files are uploaded to PAWER (\textbf{1}), then the system proceeds to identify foreground and background intensities and a panel of control proteins that can be used for normalisation (\textbf{2}). The robust linear model is then used to estimate and remove the technical artifacts associated with each array and array block (\textbf{3}). Normalised data is then combined with sample metadata (\textbf{4}) to produce a list of differentially expressed proteins (\textbf{5}). PAWER is linked with two other tools (g:Profiler and ClustVis) to enable additional analysis, namely: protein enrichment analysis and cluster analysis of normalised expression values.}
\label{PAWER}
\end{figure*}

Proteins that exhibit differential levels of the signal across studied conditions are then characterised using \textit{gprofiler2} R package, which is an interface to g:Profiler web service \cite{Raudvere2019}. g:Profiler runs the enrichment analysis using differential proteins as a query. The top six most significant enrichment terms are visualised in a form of a bar plot that can also be downloaded. More visualisations and the remaining list of relevant enrichment terms can be accessed directly via g:Profiler by pressing ``Open in g:Profiler'' button in PAWER. The overall pipeline is presented in Figure \ref{PAWER}.

One of our focuses for PAWER was designing a user-friendly interface, which would enable people with no special computer science background to carry out independent protein microarray analysis. To fulfill this vision, we added a comprehensive help page that helps new users to get started with PAWER. We recognised that preparing the data in the right format can be a problem for adapting bioinformatics tools as PAWER. Therefore, a sample data set with a corresponding metadata file is available for download from the main page as an example of the file format that PAWER expects. Also, demo results are linked from the home page. Initially, we developed PAWER for GPR files generated from ProtoArray and HuProt \cite{huprot} platforms. Later, through additional customisation, PAWER has been made in principle compatible with any protein microarray system as long as it produces text files with consistent headers and a few key parameters (e.g. names of foreground and background intensity features) are specified. 

\section{Implementation}

PAWER consists of two major parts: R package paweR at its computational core and the web interface. The heart of PAWER -- paweR was written in R version 3.4.2 and uses functions from the following packages: \textit{limma} \cite{limma}, \textit{reshape2} \cite{reshape2}, \textit{MASS} \cite{mass} and \textit{gprofiler2} \cite{Raudvere2019}. The web interface was implemented as a single-page application using React.js and Redux architecture on the client-side and node.js on the server-side. The figures are created and rendered with a help of D3.js \cite{d3} and DataTables libraries. Both the R package and the webserver code are freely available under the GNU GPL v2. license.

\section{Comparison to other tools}

To the best of our knowledge there are four major tools available for protein microarray analysis (besides PAWER): Prospector, ProtoArray Analyser (PAA) \cite{PAA}, protein microarray analyser (PMA)  \cite{DaGamaDuarte2018} and an online tool available as part of the protein microarray database (PMD)  \cite{Xu2016}. When we started working on the first publication, the only option to analyse ProtoArray data was through Prospector software. Prospector was software developed and distributed by the manufacturer of the ProtoArray platform -- Invitrogen, which later has been acquired by Thermo Fisher Scientific. The initial version of the Prospector did not allow us to analyse more than ~65 samples at the same time \cite{Turewicz2013}, rendering it impractical for larger studies (e.g. publications I and II of this thesis). Although in later versions this limitation has been removed, the software still remained closed sourced and compatible only with outdated operating systems -- Windows XP and Windows 7. Later, ProtoArray analyser -- open-source R package was introduced \cite{PAA}. PAA has become a standard tool for a lot of bioinformaticians that worked with protein microarray data. Most of the data analysis pipeline described in the previous sections  (except enrichment analysis) has been implemented and thus accessible via PAA. Although, being very powerful, PAA still required substantial programming skills in order to be used efficiently. Another tool for analyzing protein microarray data -- PMA, was developed and publicly published in 2018 \cite{DaGamaDuarte2018}. PMA is a multiplatform Java application, that can be used by anyone through the point and click interface. Notably, PMA has implemented a lot of state-of-the-art computational methods for protein microarray analysis \cite{DaGamaDuarte2018}. However, the lack of supporting documentation and maintenance, a large number of confusing parameters, and unclear input format, make this application hard to use. Also, PMA is not a stand-alone tool, after the signal normalisation is performed with PMA it requires other resources to be employed for the downstream analysis. Lastly, a tool that resides on the PMD website is the only web-based solution developed prior to current work \cite{Xu2016}. According to the original publication, PMD offers an all-encompassing analysis of protein microarrays, including enrichment and differential analysis. However, when thoroughly tested using existing GPR files, the tool failed to produce results. Error messages displayed to the user, come straight from R and generally are not easy to understand to non-specialists. Moreover, no documentation file linked from the PMD tool page was accessible at the time of writing. Details of the implemented algorithms could be found in neither the original publication nor on the website. In response to the above-mentioned challenges, we developed PAWER -- a freely available web-tool as a user-friendly way to analyse protein microarray data. PAWER attempts to combine the key features of the above-listed tools: it has an online interface like PMD, there is a rigid R core and separate R package (similar to PAA), it provides all-around analysis using state-of-the-art computational methods like PMA. The comparison between PAWER and other tools is presented in Table \ref{Tab:comparison}.

\begin{table}[h!]
\caption{Comparison between currently available protein microarray analysis tools: Prospector, PAA, PMA, PMD, and PAWER. The presence or absence of relevant features (in columns) are shown as pluses highlighted in green (present features) or minuses in red (absent features). We were not able to obtain results using the PMD tool, thus all the relevant entries are based on the claims made in the original publication and highlighted in gray.}
\centering
\begin{adjustbox}{width=1\textwidth}
\begin{tabular}{ | c | c | c | c | c | c | c | c | c |}

\hline
    Tool name &
    License  & 
    \makecell{Last \\ updated} & 
    Platform 
    & GUI 
    & Normalisation 
    & \makecell{Biomarker \\ identification} & 
    \makecell{Functional \\ annotation} & 
    \makecell{Downloadable \\ visuals} 
    \\ 
\hline
    \textbf{Prospector} & 
    \makecell{No license \\ specified} &
    2015 & 
    \makecell{Windows 7 \\ desktop application} & 
    \cellcolor{green}+ & 
    \cellcolor{green}+ & 
    \cellcolor{green}+ & 
    \cellcolor{red}- & 
    \cellcolor{red}- 
    \\
\hline
    \textbf{PAA} & 
    BSD 3 & 
    2019  & 
    R package & 
    \cellcolor{red}- & 
    \cellcolor{green}+ & 
    \cellcolor{green}+ & 
    \cellcolor{red}- & 
    \cellcolor{green}+
    \\ 
\hline
    \textbf{PMA} & 
    \makecell{No license \\ specified}  & 
    2018 & 
    \makecell{Java \\ desktop application} & 
    \cellcolor{green}+ & 
    \cellcolor{green}+ & 
    \cellcolor{red}- & 
    \cellcolor{red}- & 
    \cellcolor{red}-
    \\ 
\hline
    \textbf{PMD} & 
    \makecell{No license \\ specified}  & 
    2020 & 
    \makecell{Web server \\ R code} & 
    \cellcolor{green}+ & 
    \cellcolor{gray}+ & 
    \cellcolor{gray}+ & 
    \cellcolor{gray}+ & 
    \cellcolor{gray}+
    \\ 
\hline
\textbf{PAWER} & 
GNU GPL V2. & 
2020 & 
\makecell{Web server, \\ R package} & 
\cellcolor{green}+ & 
\cellcolor{green}+ & 
\cellcolor{green}+ & 
\cellcolor{green}+ & 
\cellcolor{green}+ \\ 
\hline
\end{tabular}
\label{Tab:comparison}
\end{adjustbox}
\end{table}

\section{Summary}

% Update % 

PAWER is the only web-based solution solely focusing on providing state-of-the-art analysis methods for protein microarray experiments data. All the data and visualisations are freely downloadable from the website. User-friendly interface, comprehensive help page, pre-loaded results page, and sample data set -- all make PAWER very easy to get started with and execute one's own analysis. Code for both: R package and web interface is publicly available.

%%%%%%%%%%%%%%%%%%%%%%%%%%%%%%%%%%%%%%%%%%%%%%%%%%%%%%%%%%%%%%%%%%%%%%%%%%%%%%%%%%%%%%%%%%%%%%%%%%%%%
% PUBLICATION IV CHAPTER
%%%%%%%%%%%%%%%%%%%%%%%%%%%%%%%%%%%%%%%%%%%%%%%%%%%%%%%%%%%%%%%%%%%%%%%%%%%%%%%%%%%%%%%%%%%%%%%%%%%%%

\chapter{Validating results of statistical tests using machine learning models on protein expression data (Publication IV)}

In this final publication, a research group from the Faculty of Medicine at the University of Ljubljana (Slovenia) in collaboration with Medical University Vienna (Austria) measured the protein expression from patients’ blood in an attempt to identify early signs of the common gynecological condition -- endometriosis. Endometriosis is a benign gynecological disease that results in endometriotic lesions found outside the uterine cavity \cite{Knific2019}. Endometriosis is an inflammatory disease, which often manifests itself in pelvic pain and results in infertility. As many inflammatory processes are regulated by a set of secreted proteins called cytokines \cite{Zhang2007}, researchers hypothesised that levels of cytokines in patient blood can be predictive of disease status and can potentially be used as biomarkers in clinic. 

The aforementioned problem definition presented itself as a suitable scenario for the case-control study. Such a study would normally employ univariate statistical tests to identify a handful of cytokines with concentration levels sufficiently different between endometriosis positive (cases) and negative (controls) subjects. While we would expect complex interactions between molecules to be the most decisive in biology, univariate tests will only spot protein features that can sufficiently explain the output variable in isolation from other inputs \cite{HuynhThu2012}. Hence, in order to discover more relevant biomarkers, more sophisticated analysis, capable of exploring multivariate associations between protein features must be performed. Additionally, case-control studies have been shown to be prone to various biases \cite{Mann2003, Melamed2018}. It has been suggested that any evidence obtained from case-control studies must be carefully scrutinised and reviewed from multiple angles \cite{Melamed2018}. In this publication, we employed machine learning methods as a means to explore multivariate relationships between protein features and confirm observations made by the univariate tests with respect to single variables. We have already applied such a strategy in our own work on modelling severity score of Psoriasis \cite{Ottas2017}. 

In this publication, unlike previous papers, presented in this thesis, quantification of protein expression was performed using the xMAP Luminex platform, not protein microarrays. xMAP Luminex technology is a flow cytometric method based on colour-coded microspheres (also known as beads) \cite{Kang2012}. Microspheres coupled with target-specific antibodies are used to capture target molecules from the sample. Colour coding enables the system to unambiguously identify each type of beads, hence, accurately recognise the presence of target molecules. \cite{Kellar2002}. According to the manufacturer's website, up to 500 different microspheres can be designed targeting as many molecules. Unlike protein microarrays, where thousands of proteins are incubated on the glass surface, and one sample essentially corresponds to one array, the xMAP platform allows to measure one sample per well, making this approach much more affordable if the narrower set of proteins of interest is known in advance. In this work, xMAP Luminex platform was used to measure the concentration levels of 40 cytokines, mostly chemokines that have not been previously studied in the context of diagnosing endometriosis \cite{Knific2019}.  

After samples were collected  (210 samples in total), the expression of target proteins was measured using xMAP Luminex technology in accordance with relevant guidelines and protocols. Initial pre-processing of the raw data was done by the proprietary software tool -- Bio-Plex\texttrademark Manager Software. Clinical and demographic data about subjects were included in the analysis. 

We have first analysed individual protein features using a two-sided Wilcoxon rank-sum test (Mann-Whitney U test). The resulting p-values were corrected using the Bonferroni method to account for multiple tests (as many as there are protein features) being performed at the same time. The protein feature was considered to have significantly different expression levels between conditions if the corresponding corrected p-value was less than 0.05. In this work, none of the analysed cytokines and cytokine ratios showed statistically significant differences. Hence, univariate statistics suggested an absence of a meaningful association between cytokine levels in the blood and endometriosis. Next, we hypothesized that a more complex multivariate relationship can be present. To validate this hypothesis, we employed machine learning. 

In terms of machine learning, we were dealing with a supervised learning problem as the outcome variable (diagnosis) was collected and available in advance. The binary nature of the outcome variable (endometriosis positive or negative) suggested a need to employ classification algorithms. A lot of classification models have been created and made readily available for researchers as part of software libraries (e.g. \textit{caret} in R or \textit{scikit-learn} in Python). These models differ in many aspects. As the goal of our research was to build a diagnostic model that could be used in clinical practice, the model's explainability was an important factor. Hence, we decided to use the decision tree algorithm (implemented in \textit{rpart} package \cite{Strobl2009}). To validate the results of the decision tree algorithm, we also trained three other classification algorithms: random forest \cite{Breiman2001}, generalised linear model \cite{Nelder1972} and weighted k-nearest neighbour algorithm \cite{Gou2011}. Selected models (all but decision tree and random forest) represented intrinsically different families of machine learning methods, at the same being powerful enough to capture complex non-linear relationships between input features and the outcome \cite{Knific2019}. Therefore, all four models were further trained on cytokines' concentrations in an attempt to predict the diagnosis.

Powerful machine learning models (such as the ones we have selected), trained on a fairly small set of observations, are fully capable of memorizing the signal distribution \cite{Walsh2020}. Evaluated on familiar examples, such models eagerly report highly optimistic performance that does not necessarily reflect a true differentiating power. This problem is known in the machine learning community as \textit{overfitting}. In order to obtain realistic results, it is a common practice to divide the data into a few independent non-overlapping parts. These usually referred to as training, validation, and test (or hold out) sets. Models are then trained on the training set, higher-order parameters are optimised using validation data and the final performance is assessed using the test set. For small datasets this approach can result in poorly performing models, as a sizeable part of data is not used for training. To utilise data more effectively, we applied repeated 4-fold cross-validation (CV) algorithm. This algorithm shuffles and then divides data into four equal parts, three-quarters of which are used for training while the last set is treated as a nominal validation set. This was done to generate unbiased performance estimates for each selected model without the need for a separate test set. The median of obtained estimates was thus considered an approximation of the final performance. In this work, none of the models showed performance sufficiently different from what can be expected at random, thus, confirming results obtained by univariate tests.

To summarise, in this collaborative work, we looked for an association between cytokines expression in blood with the presence or absence of endometriosis. Neither univariate tests nor machine learning algorithms found protein features that could alone or in combination accurately explain the value of the outcome variable. Our main contribution in this publication is two-fold: firstly, we used four different machine learning algorithms to support and validate the results of the univariate statistics, and secondly, we employed repeated 4-fold cross-validation instead of explicit validation and test sets to estimate the performance of selected machine learning models. The first contribution improved the trustworthiness of reported results, while the second enabled us to train models efficiently without fear of overfitting to available data. Although this work is based on data generated by other than the protein microarray platform, we feel that it fits nicely into the overall narrative of analysing protein expression data using means of modern data science. 

This work has become pivotal for the contribution that the authors of this thesis have made to the DOME recommendations for supervised machine learning validation in biology which was accepted for publication in Nature Methods \cite{Walsh2020}. DOME presents a set of community-wide recommendations aiming at establishing standards for describing key aspects of machine learning pipelines: data, optimization strategies, models, and evaluation. These recommendations should improve the quality as well as increase the reproducibility of biological papers that employ machine learning models.

\chapter{Conclusions}

Accurate information about protein levels in the organism has shown to be a valuable asset in the understanding of human biology. The presence of certain types of proteins is associated with a threat to health and well-being, while the abundance of others can be life-saving. One of the ways to estimate protein quantities is through protein microarrays. Although, in many ways similar to DNA microarrays, protein microarrays are subject to distinct biological assumptions, rendering computational methods designed for DNA microarrays inadequate. An overarching theme of this thesis is exploring computational methods needed for all-around analysis of protein microarrays. To fulfill this vision, we have employed various approaches from statistics, data science, and machine learning.

In Publication I, we performed extensive data pre-processing including data normalisation and filtering in order to explore the protein profile of the autoimmune disorder APS1. We expanded this analysis further in Publication II as we discovered a group of proteins that were more likely to exhibit high-intensity levels in patients comparing to controls. We thoroughly characterised this group of positive proteins, using a vast number of publicly available datasets and tools for the enrichment analysis. Having recognised the amount of work and complexity such thorough analysis demands, we decided to develop an intuitive web-based tool for protein microarray analysis. Protein microarray web explorer has been developed and presented in Publication III. PAWER was built for researchers without a programming background. In publication IV we have explored the usefulness of machine learning methods for the analysis of protein concentrations. In collaboration with clinical partners from Slovenia and Austria, we have employed statistical tests as well as more sophisticated machine learning methods to differentiate between endometrium cases and controls, based on protein data. Here, we used a more data-efficient cross-validation approach to estimate the performance of the machine learning models. Both machine learning and classical statistical analysis have failed to tell the difference between samples, suggesting an absence of reliable biomarkers among cytokines measured in the study. Although data for this publication, unlike other papers was obtained using xMAP Luminex technology, we believe that similar methods can be applied to protein microarray experiments.

The capstone and the most important contribution of this thesis is the web-tool PAWER. PAWER encompasses all the important parts of the protein microarray analysis implemented in other publications, namely: the entire pre-processing pipeline including the normalisation strategy using robust linear model, differential, and enrichment analysis. The source code of the tool is available online.

To conclude, the work included in this thesis has explored a set of computational methods available for the protein microarray analysis as well as included practical recommendations both of which could be useful for those who plan to carry out their own analysis of protein expression data. The most essential methods presented here were included in the web-tool PAWER, allowing researchers to perform semi-automated analysis online in a drag-and-drop and point-and-click manner. A number of research projects have also benefited from work presented in the thesis, including BioEndoCar (\url{https://bioendocar.eu/}) -- an international consortium with an aim to combine protein microarray data with information about blood metabolites to identify diagnostic and prognostic markers for endometrial cancer. The first-hand experience of applying machine learning methods in a biological context has inspired a contribution to the first set of recommendations for validating machine learning methods in biological studies \cite{Walsh2020}. These guidelines if adopted widely by the community may increase trust in machine learning research as well as improve the reproducibility of published findings, accelerating the progress in the field.

% == bibliography ==

\bibliography{aticollection}

\begin{thebibliography}{100}

\bibitem{Meyer2016}
Steffen Meyer, Martin Woodward, Christina Hertel, Philip Vlaicu, Yasmin Haque,
  Jaanika K\"{a}rner, Annalisa Macagno, Shimobi~C. Onuoha, Dmytro Fishman, Hedi
  Peterson, Kaja Metsk\"{u}la, Raivo Uibo, Kirsi J\"{a}ntti, Kati Hokynar,
  Anette~S.B. Wolff, Kai Krohn, Annamari Ranki, P\"{a}rt Peterson, Kai Kisand,
  Adrian Hayday, Antonella Meloni, Nicolas Kluger, Eystein~S. Husebye,
  Katarina~Trebusak Podkrajsek, Tadej Battelino, Nina Bratanic, and Aleksandr
  Peet.
\newblock {AIRE}-deficient patients harbor unique high-affinity
  disease-ameliorating autoantibodies.
\newblock {\em Cell}, 166(3):582--595, July 2016.

\bibitem{Fishman2017}
Dmytro Fishman, Kai Kisand, Christina Hertel, Mike Rothe, Anu Remm, Maire
  Pihlap, Priit Adler, Jaak Vilo, Aleksandr Peet, Antonella Meloni,
  Katarina~Trebusak Podkrajsek, Tadej Battelino, {\O}yvind Bruserud, Anette
  S.~B. Wolff, Eystein~S. Husebye, Nicolas Kluger, Kai Krohn, Annamari Ranki,
  Hedi Peterson, Adrian Hayday, and P\"{a}rt Peterson.
\newblock Autoantibody repertoire in {APECED} patients targets two distinct
  subgroups of proteins.
\newblock {\em Frontiers in Immunology}, 8, August 2017.

\bibitem{Fishman2019}
Dmytro Fishman, Ivan Kuzmin, Jaak Vilo, and Hedi Peterson.
\newblock {PAWER}: Protein array web {ExploreR}.
\newblock {\em BMC Bioinformatics}, July 2019.

\bibitem{Knific2019}
Tamara Knific, Dmytro Fishman, Andrej Vogler, Manuela Gst\"{o}ttner, Ren{\'{e}}
  Wenzl, Hedi Peterson, and Tea~Lani{\v{s}}nik Ri{\v{z}}ner.
\newblock Multiplex analysis of 40 cytokines do not allow separation between
  endometriosis patients and controls.
\newblock {\em Scientific Reports}, 9(1), November 2019.

\bibitem{behave}
Robert~M. Sapolsky.
\newblock {\em Behave: The Biology of Humans at Our Best and Worst}.
\newblock Penguin Press, 2017.

\bibitem{Vogel2012}
Christine Vogel and Edward~M. Marcotte.
\newblock Insights into the regulation of protein abundance from proteomic and
  transcriptomic analyses.
\newblock {\em Nature Reviews Genetics}, 13(4):227--232, March 2012.

\bibitem{centraldogma}
Francis Harry~Compton Crick.
\newblock {{O}n protein synthesis}.
\newblock {\em Symp Soc Exp Biol}, 12:138--163, 1958.

\bibitem{Johnson2002}
Bruce Alberts, Alexander Johnson, Julian Lewis, Martin Raff, Keith Roberts, and
  Peter Walter.
\newblock {\em Molecular Biology of the Cell}.
\newblock Garland Science, 2002.

\bibitem{Fulton1991}
Alice~B. Fulton and William~B. Isaacs.
\newblock Titin, a huge, elastic sarcomeric protein with a probable role in
  morphogenesis.
\newblock {\em {BioEssays}}, 13(4):157--161, April 1991.

\bibitem{Ponomarenko2016}
Elena~A. Ponomarenko, Ekaterina~V. Poverennaya, Ekaterina~V. Ilgisonis,
  Mikhail~A. Pyatnitskiy, Arthur~T. Kopylov, Victor~G. Zgoda, Andrey~V.
  Lisitsa, and Alexander~I. Archakov.
\newblock The size of the human proteome: The width and depth.
\newblock {\em International Journal of Analytical Chemistry}, 2016:1--6, 2016.

\bibitem{abbas2011}
Abul~K. Abbas, Andrew H.~H. Lichtman, and Shiv Pillai.
\newblock {\em Cellular and Molecular Immunology: with STUDENT CONSULT Online
  Access (Abbas, Cellular and Molecular Immunology)}.
\newblock Saunders, 2011.

\bibitem{Marshall2018}
Jean~S. Marshall, Richard Warrington, Wade Watson, and Harold~L. Kim.
\newblock An introduction to immunology and immunopathology.
\newblock {\em Allergy, Asthma {\&} Clinical Immunology}, 14(S2), September
  2018.

\bibitem{Chaplin2010}
David~D. Chaplin.
\newblock Overview of the immune response.
\newblock {\em Journal of Allergy and Clinical Immunology}, 125(2):S3--S23,
  February 2010.

\bibitem{Klein2014}
Ludger Klein, Bruno Kyewski, Paul~M. Allen, and Kristin~A. Hogquist.
\newblock Positive and negative selection of the t cell repertoire: what
  thymocytes see (and don't see).
\newblock {\em Nature Reviews Immunology}, 14(6):377--391, May 2014.

\bibitem{Romagnani2006}
Sergio Romagnani.
\newblock Immunological tolerance and autoimmunity.
\newblock {\em Internal and Emergency Medicine}, 1(3):187--196, September 2006.

\bibitem{Walsh2000}
S~J Walsh and L~M Rau.
\newblock Autoimmune diseases: a leading cause of death among young and
  middle-aged women in the united states.
\newblock {\em American Journal of Public Health}, 90(9):1463--1466, September
  2000.

\bibitem{Elkon2008}
Keith Elkon and Paolo Casali.
\newblock Nature and functions of autoantibodies.
\newblock {\em Nature Clinical Practice Rheumatology}, 4(9):491--498, September
  2008.

\bibitem{Kisand2011}
Kai Kisand and P\"{a}rt Peterson.
\newblock Autoimmune polyendocrinopathy candidiasis ectodermal dystrophy: known
  and novel aspects of the syndrome.
\newblock {\em Annals of the New York Academy of Sciences}, 1246(1):77--91,
  December 2011.

\bibitem{Bettelli2006}
Estelle Bettelli, Yijun Carrier, Wenda Gao, Thomas Korn, Terry~B. Strom,
  Mohamed Oukka, Howard~L. Weiner, and Vijay~K. Kuchroo.
\newblock Reciprocal developmental pathways for the generation of pathogenic
  effector {TH}17 and regulatory t cells.
\newblock {\em Nature}, 441(7090):235--238, April 2006.

\bibitem{Kinnunen2013}
Tuure Kinnunen, Nicolas Chamberlain, Henner Morbach, Jinyoung Choi, Sangtaek
  Kim, Joseph Craft, Lloyd Mayer, Caterina Cancrini, Laura Passerini, Rosa
  Bacchetta, Hans~D. Ochs, Troy~R. Torgerson, and Eric Meffre.
\newblock Accumulation of peripheral autoreactive b cells in the absence of
  functional human regulatory t cells.
\newblock {\em Blood}, 121(9):1595--1603, February 2013.

\bibitem{Meffre2008}
Eric Meffre and Hedda Wardemann.
\newblock B-cell tolerance checkpoints in health and autoimmunity.
\newblock {\em Current Opinion in Immunology}, 20(6):632--638, December 2008.

\bibitem{Tsubata2017}
Takeshi Tsubata.
\newblock B-cell tolerance and autoimmunity.
\newblock {\em F1000Research}, 6:391, March 2017.

\bibitem{flaherty2011}
Dennis Flaherty.
\newblock {\em Immunology for Pharmacy - Elsevier eBook on VitalSource (Retail
  Access Card)}.
\newblock Mosby, 2011.

\bibitem{Abel2014}
Laura Abel, Simone Kutschki, Michael Turewicz, Martin Eisenacher, Jale
  Stoutjesdijk, Helmut~E. Meyer, Dirk Woitalla, and Caroline May.
\newblock Autoimmune profiling with protein microarrays in clinical
  applications.
\newblock {\em Biochimica et Biophysica Acta ({BBA}) - Proteins and
  Proteomics}, 1844(5):977--987, May 2014.

\bibitem{Pihoker2005}
C.~Pihoker, L.~K. Gilliam, C.~S. Hampe, and A.~Lernmark.
\newblock Autoantibodies in diabetes.
\newblock {\em Diabetes}, 54(Supplement 2):S52--S61, November 2005.

\bibitem{Sollid2013}
Ludvig~M. Sollid and Bana Jabri.
\newblock Triggers and drivers of autoimmunity: lessons from coeliac disease.
\newblock {\em Nature Reviews Immunology}, 13(4):294--302, March 2013.

\bibitem{Lauret2013}
Eugenia Lauret and Luis Rodrigo.
\newblock Celiac disease and autoimmune-associated conditions.
\newblock {\em {BioMed} Research International}, 2013:1--17, 2013.

\bibitem{Fraussen2014}
Judith Fraussen, Nele Claes, Laura de~Bock, and Veerle Somers.
\newblock Targets of the humoral autoimmune response in multiple sclerosis.
\newblock {\em Autoimmunity Reviews}, 13(11):1126--1137, November 2014.

\bibitem{PETERSON2004}
P.~Peterson, J.~Pitkanen, N.~Sillanpaa, and K.~Krohn.
\newblock Autoimmune polyendocrinopathy candidiasis ectodermal dystrophy
  ({APECED}): a model disease to study molecular aspects of endocrine
  autoimmunity.
\newblock {\em Clinical and Experimental Immunology}, 135(3):348--357, March
  2004.

\bibitem{Aziz2019}
Farhanah Aziz, Muneera Smith, and Jonathan~M Blackburn.
\newblock Autoantibody-based diagnostic biomarkers: Technological approaches to
  discovery and validation.
\newblock In {\em Autoantibodies and Cytokines}. {IntechOpen}, April 2019.

\bibitem{Leslie2001}
David Leslie, Peter Lipsky, and Abner~Louis Notkins.
\newblock Autoantibodies as predictors of disease.
\newblock {\em Journal of Clinical Investigation}, 108(10):1417--1422, November
  2001.

\bibitem{encyclopedia}
Peter Delves.
\newblock {\em Encyclopedia of immunology}.
\newblock Academic Press, San Diego, 1998.

\bibitem{Nagele2011}
Eric Nagele, Min Han, Cassandra DeMarshall, Benjamin Belinka, and Robert
  Nagele.
\newblock Diagnosis of alzheimer's disease based on disease-specific
  autoantibody profiles in human sera.
\newblock {\em {PLoS} {ONE}}, 6(8):e23112, August 2011.

\bibitem{Huang2017}
Yi~Huang and Heng Zhu.
\newblock Protein array-based approaches for biomarker discovery in cancer.
\newblock {\em Genomics, Proteomics {\&} Bioinformatics}, 15(2):73--81, April
  2017.

\bibitem{Gupta2017}
Shabarni Gupta, K.~P. Manubhai, Shuvolina Mukherjee, and Sanjeeva Srivastava.
\newblock Serum profiling for identification of autoantibody signatures in
  diseases using protein microarrays.
\newblock In {\em Methods in Molecular Biology}, pages 303--315. Springer New
  York, 2017.

\bibitem{Holman2011}
H~Holman.
\newblock The discovery of autoantibody to deoxyribonucleic acid.
\newblock {\em Lupus}, 20(5):441--442, February 2011.

\bibitem{Hepburn2001}
A.~L. Hepburn.
\newblock The {LE} cell.
\newblock {\em Rheumatology}, 40(7):826--827, July 2001.

\bibitem{Rosenberg2015}
Jacob~M. Rosenberg and Paul~J. Utz.
\newblock Protein microarrays: A new tool for the study of autoantibodies in
  immunodeficiency.
\newblock {\em Frontiers in Immunology}, 6, April 2015.

\bibitem{Sobek2006}
Jens Sobek, Kerstin Bartscherer, Anette Jacob, Jvrg Hoheisel, and Philipp
  Angenendt.
\newblock Microarray technology as a universal tool for high-throughput
  analysis of biological systems.
\newblock {\em Combinatorial Chemistry {\&} High Throughput Screening},
  9(5):365--380, June 2006.

\bibitem{Hall2007}
David~A. Hall, Jason Ptacek, and Michael Snyder.
\newblock Protein microarray technology.
\newblock {\em Mechanisms of Ageing and Development}, 128(1):161--167, January
  2007.

\bibitem{Moore2016}
Cedric~D Moore, Olutobi~Z Ajala, and Heng Zhu.
\newblock Applications in high-content functional protein microarrays.
\newblock {\em Current Opinion in Chemical Biology}, 30:21--27, February 2016.

\bibitem{Zhu2012}
Heng Zhu and Jiang Qian.
\newblock Applications of functional protein microarrays in basic and clinical
  research.
\newblock In {\em Advances in Genetics Volume 79}, pages 123--155. Elsevier,
  2012.

\bibitem{Neagu2019}
Monica Neagu, Marinela Bostan, and Carolina Constantin.
\newblock Protein microarray technology: Assisting personalized medicine in
  oncology (review).
\newblock {\em World Academy of Sciences Journal}, June 2019.

\bibitem{Duarte2017}
Jessica~G. Duarte and Jonathan~M. Blackburn.
\newblock Advances in the development of human protein microarrays.
\newblock {\em Expert Review of Proteomics}, 14(7):627--641, July 2017.

\bibitem{zhu2006}
Xiaowei Zhu, Mark Gerstein, and Michael Snyder.
\newblock Procat: a data analysis approach for protein microarrays.
\newblock {\em Genome Biology}, 7(11):R110, 2006.

\bibitem{Landegren2016}
Nils Landegren, Donald Sharon, Eva Freyhult, {\AA}sa Hallgren, Daniel Eriksson,
  Per-Henrik Edqvist, Sophie Bensing, Jeanette Wahlberg, Lawrence~M. Nelson,
  Jan Gustafsson, Eystein~S. Husebye, Mark~S. Anderson, Michael Snyder, and
  Olle K\"{a}mpe.
\newblock Proteome-wide survey of the autoimmune target repertoire in
  autoimmune polyendocrine syndrome type 1.
\newblock {\em Scientific Reports}, 6(1), February 2016.

\bibitem{sboner}
A.~Sboner, A.~Karpikov, G.~Chen, M.~Smith, D.~Mattoon, M.~Dawn,
  L.~Freeman-Cook, B.~Schweitzer, and M.~B. Gerstein.
\newblock {{R}obust-linear-model normalization to reduce technical variability
  in functional protein microarrays}.
\newblock {\em J. Proteome Res.}, 8(12):5451--5464, Dec 2009.
\newblock [PubMed:\href{http://www.ncbi.nlm.nih.gov/pubmed/19817483}{19817483}]
  [doi:\href{https://doi.org/10.1021/pr900412k}].

\bibitem{Long2016}
Jintao Long, Genhua Pan, Emmanuel Ifeachor, Robert Belshaw, and Xinzhong Li.
\newblock Discovery of novel biomarkers for alzheimer's disease from blood.
\newblock {\em Disease Markers}, 2016:1--9, 2016.

\bibitem{geo}
T.~Barrett, S.~E. Wilhite, P.~Ledoux, C.~Evangelista, I.~F. Kim,
  M.~Tomashevsky, K.~A. Marshall, K.~H. Phillippy, P.~M. Sherman, M.~Holko,
  A.~Yefanov, H.~Lee, N.~Zhang, C.~L. Robertson, N.~Serova, S.~Davis, and
  A.~Soboleva.
\newblock {{N}{C}{B}{I} {G}{E}{O}: archive for functional genomics data
  sets--update}.
\newblock {\em Nucleic Acids Res.}, 41(Database issue):D991--995, Jan 2013.
\newblock [PubMed:\href{http://www.ncbi.nlm.nih.gov/pubmed/23193258}{23193258}]
  [doi:\href{https://doi.org/10.1093/nar/gks1193}{10.1093/nar/gks1193}]
  [PubMedCentral:
  \href{https://www.ncbi.nlm.nih.gov/pmc/articles/PMC3531084/}{PMC3531084}].

\bibitem{arrayEx}
Awais Athar, Anja Füllgrabe, Nancy George, Haider Iqbal, Laura Huerta, Ahmed
  Ali, Catherine Snow, Nuno~A Fonseca, Robert Petryszak, Irene Papatheodorou,
  Ugis Sarkans, and Alvis Brazma.
\newblock {ArrayExpress update – from bulk to single-cell expression data}.
\newblock {\em Nucleic Acids Research}, 47(D1):D711--D715, 10 2018.

\bibitem{hu2012}
Chao-Jun Hu, Guang Song, Wei Huang, Guo-Zhen Liu, Chui-Wen Deng, Hai-Pan Zeng,
  Li~Wang, Feng-Chun Zhang, Xuan Zhang, Jun~Seop Jeong, Seth Blackshaw, Li-Zhi
  Jiang, Heng Zhu, Lin Wu, and Yong-Zhe Li.
\newblock Identification of new autoantigens for primary biliary cirrhosis
  using human proteome microarrays.
\newblock {\em Molecular {\&} Cellular Proteomics}, 11(9):669--680, May 2012.

\bibitem{jeong2012}
Jun~Seop Jeong, Lizhi Jiang, Edisa Albino, Josean Marrero, Hee~Sool Rho,
  Jianfei Hu, Shaohui Hu, Carlos Vera, Diane Bayron-Poueymiroy, Zully~Ann
  Rivera-Pacheco, Leonardo Ramos, Cecil Torres-Castro, Jiang Qian, Joseph
  Bonaventura, Jef~D. Boeke, Wendy~Y. Yap, Ignacio Pino, Daniel~J. Eichinger,
  Heng Zhu, and Seth Blackshaw.
\newblock Rapid identification of monospecific monoclonal antibodies using a
  human proteome microarray.
\newblock {\em Molecular {\&} Cellular Proteomics}, 11(6):O111.016253, February
  2012.

\bibitem{Jung2014}
Jin-Gyoung Jung, Alexander Stoeck, Bin Guan, Ren-Chin Wu, Heng Zhu, Seth
  Blackshaw, Ie-Ming Shih, and Tian-Li Wang.
\newblock Notch3 interactome analysis identified {WWP}2 as a negative regulator
  of notch3 signaling in ovarian cancer.
\newblock {\em {PLoS} Genetics}, 10(10):e1004751, October 2014.

\bibitem{Yang2015}
Lina Yang, Jingfang Wang, Jianfang Li, Hainan Zhang, Shujuan Guo, Min Yan,
  Zhenggang Zhu, Bin Lan, Youcheng Ding, Ming Xu, Wei Li, Xiaonian Gu, Chong
  Qi, Heng Zhu, Zhifeng Shao, Bingya Liu, and Sheng-Ce Tao.
\newblock Identification of serum biomarkers for gastric cancer diagnosis using
  a human proteome microarray.
\newblock {\em Molecular {\&} Cellular Proteomics}, 15(2):614--623, November
  2015.

\bibitem{Hu2015}
Chaojun Hu, Wei Huang, Hua Chen, Guang Song, Ping Li, Qiang Shan, Xuan Zhang,
  Fengchun Zhang, Heng Zhu, Lin Wu, and Yongzhe Li.
\newblock Autoantibody profiling on human proteome microarray for biomarker
  discovery in cerebrospinal fluid and sera of neuropsychiatric lupus.
\newblock {\em {PLOS} {ONE}}, 10(5):e0126643, May 2015.

\bibitem{Hu2016}
Chao-Jun Hu, Jian-Bo Pan, Guang Song, Xiao-Ting Wen, Zi-Yan Wu, Si~Chen,
  Wen-Xiu Mo, Feng-Chun Zhang, Jiang Qian, Heng Zhu, and Yong-Zhe Li.
\newblock Identification of novel biomarkers for behcet disease diagnosis using
  human proteome microarray approach.
\newblock {\em Molecular {\&} Cellular Proteomics}, 16(2):147--156, October
  2016.

\bibitem{DeLuca2011}
David~S. DeLuca, Ovidiu Marina, Surajit Ray, Guang~Lan Zhang, Catherine~J. Wu,
  and Vladimir Brusic.
\newblock Data processing and analysis for protein microarrays.
\newblock In {\em Protein Microarray for Disease Analysis}, pages 337--347.
  Humana Press, 2011.

\bibitem{Zheng2018}
Yingye Zheng.
\newblock Study design considerations for cancer biomarker discoveries.
\newblock {\em The Journal of Applied Laboratory Medicine}, 3(2):282--289, May
  2018.

\bibitem{Dez2012}
Paula D{\'{\i}}ez, Noelia Dasilva, Mar{\'{\i}}a Gonz{\'{a}}lez-Gonz{\'{a}}lez,
  Sergio Matarraz, Juan Casado-Vela, Alberto Orfao, and Manuel Fuentes.
\newblock Data analysis strategies for protein microarrays.
\newblock {\em Microarrays}, 1(2):64--83, August 2012.

\bibitem{DaGamaDuarte2018}
Jessica Da~Gama Duarte, Ryan~W. Goosen, Peter~J. Lawry, and Jonathan~M.
  Blackburn.
\newblock {PMA}: Protein microarray analyser, a user-friendly tool for data
  processing and normalization.
\newblock {\em {BMC} Research Notes}, 11(1), February 2018.

\bibitem{Ritchie2007}
M.~E. Ritchie, J.~Silver, A.~Oshlack, M.~Holmes, D.~Diyagama, A.~Holloway, and
  G.~K. Smyth.
\newblock A comparison of background correction methods for two-colour
  microarrays.
\newblock {\em Bioinformatics}, 23(20):2700--2707, August 2007.

\bibitem{Feng2014}
Changyong Feng, Hongyue Wang, Naiji Lu, Tian Chen, Hua He, Ying Lu, and Xin~M.
  Tu.
\newblock Log-transformation and its implications for data analysis.
\newblock {\em Shanghai Archives of Psychiatry}, 26(2):105--109, April 2014.

\bibitem{Pearson}
Ronald~K. Pearson, Gregory~E. Gonye, and James~S. Schwaber.
\newblock Outliers in microarray data analysis.
\newblock In {\em Methods of Microarray Data Analysis {III}}, pages 41--55.
  Kluwer Academic Publishers, 2003.

\bibitem{Kisand2010}
Kai Kisand, Anette S.~B{\o}e Wolff, Katarina~Trebu{\v{s}}ak Podkraj{\v{s}}ek,
  Liina Tserel, Maire Link, Kalle~V. Kisand, Elisabeth Ersvaer, Jaakko
  Perheentupa, Martina~Moter Erichsen, Nina Bratanic, Antonella Meloni,
  Filomena Cetani, Roberto Perniola, Berrin Ergun-Longmire, Noel Maclaren, Kai
  J.~E. Krohn, Mikul{\'{a}}{\v{s}} Pura, Berthold Schalke, Philipp Str\"{o}bel,
  Maria~Isabel Leite, Tadej Battelino, Eystein~S. Husebye, P\"{a}rt Peterson,
  Nick Willcox, and Anthony Meager.
\newblock Chronic mucocutaneous candidiasis in {APECED} or thymoma patients
  correlates with autoimmunity to th17-associated cytokines.
\newblock {\em The Journal of Experimental Medicine}, 207(2):299--308, February
  2010.

\bibitem{Puel2010}
Anne Puel, Rainer D\"{o}ffinger, Angels Natividad, Maya Chrabieh, Gabriela
  Barcenas-Morales, Capucine Picard, Aur{\'{e}}lie Cobat, Marie
  Ouach{\'{e}}e-Chardin, Antoine Toulon, Jacinta Bustamante, Saleh Al-Muhsen,
  Mohammed Al-Owain, Peter~D. Arkwright, Colm Costigan, Vivienne McConnell,
  Andrew~J. Cant, Mario Abinun, Michel Polak, Pierre-Fran{\c{c}}ois
  Bougn{\`{e}}res, Dinakantha Kumararatne, L{\'{a}}szl{\'{o}} Marodi, Amit
  Nahum, Chaim Roifman, St{\'{e}}phane Blanche, Alain Fischer, Christine
  Bodemer, Laurent Abel, Desa Lilic, and Jean-Laurent Casanova.
\newblock Autoantibodies against {IL}-17a, {IL}-17f, and {IL}-22 in patients
  with chronic mucocutaneous candidiasis and autoimmune polyendocrine syndrome
  type i.
\newblock {\em The Journal of Experimental Medicine}, 207(2):291--297, February
  2010.

\bibitem{mcgill1978variations}
Robert McGill, John~W Tukey, and Wayne~A Larsen.
\newblock Variations of box plots.
\newblock {\em The American Statistician}, 32(1):12--16, 1978.

\bibitem{ZepedaMendoza2013}
Marie~Lisandra Zepeda-Mendoza and Osbaldo Resendis-Antonio.
\newblock Hierarchical agglomerative clustering.
\newblock In {\em Encyclopedia of Systems Biology}, pages 886--887. Springer
  New York, 2013.

\bibitem{dbscan}
Martin Ester, Hans-Peter Kriegel, J\"{o}rg Sander, and Xiaowei Xu.
\newblock A density-based algorithm for discovering clusters in large spatial
  databases with noise.
\newblock In {\em Proceedings of the Second International Conference on
  Knowledge Discovery and Data Mining}, KDD’96, page 226–231. AAAI Press,
  1996.

\bibitem{PAA}
Michael Turewicz, Maike Ahrens, Caroline May, Katrin Marcus, and Martin
  Eisenacher.
\newblock {PAA}: an r/bioconductor package for biomarker discovery with protein
  microarrays.
\newblock {\em Bioinformatics}, 32(10):1577--1579, January 2016.

\bibitem{Bolstad2003}
B.M. Bolstad, R.A Irizarry, M.~Astrand, and T.P. Speed.
\newblock A comparison of normalization methods for high density
  oligonucleotide array data based on variance and bias.
\newblock {\em Bioinformatics}, 19(2):185--193, January 2003.

\bibitem{Wu2005}
Wei Wu, Nilesh Dave, GeorgeC Tseng, Thomas Richards, EricP Xing, and Naftali
  Kaminski.
\newblock Comparison of normalization methods for codelink bioarray data.
\newblock {\em {BMC} Bioinformatics}, 6(1):309, 2005.

\bibitem{Ballman2004}
K.~V. Ballman, D.~E. Grill, A.~L. Oberg, and T.~M. Therneau.
\newblock Faster cyclic loess: normalizing {RNA} arrays via linear models.
\newblock {\em Bioinformatics}, 20(16):2778--2786, May 2004.

\bibitem{Cleveland1979}
William~S. Cleveland.
\newblock Robust locally weighted regression and smoothing scatterplots.
\newblock {\em Journal of the American Statistical Association},
  74(368):829--836, December 1979.

\bibitem{mass}
William~N Venables and Brian~D Ripley.
\newblock {\em Modern applied statistics with S-PLUS}.
\newblock Springer Science \& Business Media, 2013.

\bibitem{limma}
G.~K. Smyth.
\newblock {{L}inear models and empirical bayes methods for assessing
  differential expression in microarray experiments}.
\newblock {\em Stat Appl Genet Mol Biol}, 3:Article3, 2004.
\newblock [PubMed:\href{http://www.ncbi.nlm.nih.gov/pubmed/16646809}{16646809}]
  [doi:\href{https://doi.org/10.2202/1544-6115.1027}].

\bibitem{Turewicz2013}
Michael Turewicz, Caroline May, Maike Ahrens, Dirk Woitalla, Ralf Gold,
  Swaantje Casjens, Beate Pesch, Thomas Br\"{u}ning, Helmut~E. Meyer, Eckhard
  Nordhoff, Miriam B\"{o}ckmann, Christian Stephan, and Martin Eisenacher.
\newblock Improving the default data analysis workflow for large autoimmune
  biomarker discovery studies with {ProtoArrays}.
\newblock {\em {PROTEOMICS}}, 13(14):2083--2087, June 2013.

\bibitem{Smyth2004}
Gordon~K Smyth.
\newblock Linear models and empirical bayes methods for assessing differential
  expression in microarray experiments.
\newblock {\em Statistical Applications in Genetics and Molecular Biology},
  3(1):1--25, January 2004.

\bibitem{Mann1947}
H.~B. Mann and D.~R. Whitney.
\newblock On a test of whether one of two random variables is stochastically
  larger than the other.
\newblock {\em The Annals of Mathematical Statistics}, 18(1):50--60, March
  1947.

\bibitem{Edgington2011}
Eugene~S. Edgington.
\newblock Randomization tests.
\newblock In {\em International Encyclopedia of Statistical Science}, pages
  1182--1183. Springer Berlin Heidelberg, 2011.

\bibitem{howell2010statistical}
David Howell.
\newblock {\em Statistical methods for psychology}.
\newblock Thomson Wadsworth, Australia Belmont, CA, 2010.

\bibitem{kegg}
Minoru Kanehisa, Miho Furumichi, Mao Tanabe, Yoko Sato, and Kanae Morishima.
\newblock {KEGG}: new perspectives on genomes, pathways, diseases and drugs.
\newblock {\em Nucleic Acids Research}, 45(D1):D353--D361, November 2016.

\bibitem{Reactome}
Antonio Fabregat, Steven Jupe, Lisa Matthews, Konstantinos Sidiropoulos, Marc
  Gillespie, Phani Garapati, Robin Haw, Bijay Jassal, Florian Korninger, Bruce
  May, Marija Milacic, Corina~Duenas Roca, Karen Rothfels, Cristoffer Sevilla,
  Veronica Shamovsky, Solomon Shorser, Thawfeek Varusai, Guilherme Viteri, Joel
  Weiser, Guanming Wu, Lincoln Stein, Henning Hermjakob, and Peter D'Eustachio.
\newblock The reactome pathway knowledgebase.
\newblock {\em Nucleic Acids Research}, 46(D1):D649--D655, November 2017.

\bibitem{hpo}
Sebastian K\"{o}hler, Leigh Carmody, Nicole Vasilevsky, Julius O~B Jacobsen,
  Daniel Danis, Jean-Philippe Gourdine, Michael Gargano, Nomi~L Harris, Nicolas
  Matentzoglu, Julie~A McMurry, David Osumi-Sutherland, Valentina Cipriani,
  James~P Balhoff, Tom Conlin, Hannah Blau, Gareth Baynam, Richard Palmer,
  Dylan Gratian, Hugh Dawkins, Michael Segal, Anna~C Jansen, Ahmed Muaz,
  Willie~H Chang, Jenna Bergerson, Stanley J~F Laulederkind, Zafer Y\"{u}ksel,
  Sergi Beltran, Alexandra~F Freeman, Panagiotis~I Sergouniotis, Daniel Durkin,
  Andrea~L Storm, Marc Hanauer, Michael Brudno, Susan~M Bello, Murat Sincan,
  Kayli Rageth, Matthew~T Wheeler, Renske Oegema, Halima Lourghi, Maria G~Della
  Rocca, Rachel Thompson, Francisco Castellanos, James Priest, Charlotte
  Cunningham-Rundles, Ayushi Hegde, Ruth~C Lovering, Catherine Hajek, Annie
  Olry, Luigi Notarangelo, Morgan Similuk, Xingmin~A Zhang, David
  G{\'{o}}mez-Andr{\'{e}}s, Hanns Lochm\"{u}ller, H{\'{e}}l{\`{e}}ne Dollfus,
  Sergio Rosenzweig, Shruti Marwaha, Ana Rath, Kathleen Sullivan, Cynthia
  Smith, Joshua~D Milner, Doroth{\'{e}}e Leroux, Cornelius~F Boerkoel, Amy
  Klion, Melody~C Carter, Tudor Groza, Damian Smedley, Melissa~A Haendel, Chris
  Mungall, and Peter~N Robinson.
\newblock Expansion of the human phenotype ontology ({HPO}) knowledge base and
  resources.
\newblock {\em Nucleic Acids Research}, 47(D1):D1018--D1027, November 2018.

\bibitem{hpa}
Mathias Uhl{\'e}n, Linn Fagerberg, Bj{\"o}rn~M Hallstr{\"o}m, Cecilia Lindskog,
  Per Oksvold, Adil Mardinoglu, {\AA}sa Sivertsson, Caroline Kampf, Evelina
  Sj{\"o}stedt, Anna Asplund, et~al.
\newblock Tissue-based map of the human proteome.
\newblock {\em Science}, 347(6220):1260419, 2015.
\newblock [PubMed:\href{http://www.ncbi.nlm.nih.gov/pubmed/25613900}{25613900}]
  [doi:\href{http://dx.doi.org/10.1126/science.1260419}{10.1126/science.1260419}].

\bibitem{Raudvere2019}
Uku Raudvere, Liis Kolberg, Ivan Kuzmin, Tambet Arak, Priit Adler, Hedi
  Peterson, and Jaak Vilo.
\newblock g:profiler: a web server for functional enrichment analysis and
  conversions of gene lists (2019 update).
\newblock {\em Nucleic Acids Research}, 47(W1):W191--W198, May 2019.

\bibitem{Goldman2008}
Megan Goldman.
\newblock Lecture notes in stat c141/bioeng c141 - statistics for
  bioinformatics, 2008.

\bibitem{bonferroni1936}
Carlo Bonferroni.
\newblock Teoria statistica delle classi e calcolo delle probabilita.
\newblock {\em Pubblicazioni del R Istituto Superiore di Scienze Economiche e
  Commericiali di Firenze}, 8:3--62, 1936.

\bibitem{Noble2009}
William~S Noble.
\newblock How does multiple testing correction work?
\newblock {\em Nature Biotechnology}, 27(12):1135--1137, December 2009.

\bibitem{bhmethod}
Yoav Benjamini and Yosef Hochberg.
\newblock Controlling the false discovery rate - a practical and powerful
  approach to multiple testing.
\newblock {\em J. Royal Statist. Soc., Series B}, 57:289 -- 300, 11 1995.

\bibitem{Zumbo2014}
Bruno~D. Zumbo.
\newblock Univariate tests.
\newblock In {\em Encyclopedia of Quality of Life and Well-Being Research},
  pages 6819--6820. Springer Netherlands, 2014.

\bibitem{Tarca2007}
Adi~L Tarca, Vincent~J Carey, Xue wen Chen, Roberto Romero, and Sorin
  Dr{\u{a}}ghici.
\newblock Machine learning and its applications to biology.
\newblock {\em {PLoS} Computational Biology}, 3(6):e116, June 2007.

\bibitem{Quinlan1986}
J.~R. Quinlan.
\newblock Induction of decision trees.
\newblock {\em Machine Learning}, 1(1):81--106, March 1986.

\bibitem{Breiman2001}
Leo Breiman.
\newblock Random forests.
\newblock {\em Machine Learning}, 45(1):5--32, 2001.

\bibitem{Breiman2017}
Leo Breiman, Jerome~H. Friedman, Richard~A. Olshen, and Charles~J. Stone.
\newblock {\em Classification And Regression Trees}.
\newblock Routledge, October 2017.

\bibitem{Ottas2017}
Aigar Ottas, Dmytro Fishman, Tiia-Linda Okas, K\"{u}lli Kingo, and Ursel
  Soomets.
\newblock The metabolic analysis of psoriasis identifies the associated
  metabolites while providing computational models for the monitoring of the
  disease.
\newblock {\em Archives of Dermatological Research}, 309(7):519--528, July
  2017.

\bibitem{alexnet}
Alex Krizhevsky, Ilya Sutskever, and Geoffrey~E Hinton.
\newblock Imagenet classification with deep convolutional neural networks.
\newblock In F.~Pereira, C.~J.~C. Burges, L.~Bottou, and K.~Q. Weinberger,
  editors, {\em Advances in Neural Information Processing Systems 25}, pages
  1097--1105. Curran Associates, Inc., 2012.

\bibitem{Angermueller2016}
Christof Angermueller, Tanel P\"{a}rnamaa, Leopold Parts, and Oliver Stegle.
\newblock Deep learning for computational biology.
\newblock {\em Molecular Systems Biology}, 12(7):878, July 2016.

\bibitem{Jones2017}
William Jones, Kaur Alasoo, Dmytro Fishman, and Leopold Parts.
\newblock Computational biology: deep learning.
\newblock {\em Emerging Topics in Life Sciences}, 1(3):257--274, November 2017.

\bibitem{Kont2008}
Vivian Kont, Martti Laan, Kai Kisand, Andres Merits, Hamish~S. Scott, and
  P\"{a}rt Peterson.
\newblock Modulation of aire regulates the expression of tissue-restricted
  antigens.
\newblock {\em Molecular Immunology}, 45(1):25--33, January 2008.

\bibitem{Nagamine1997}
Kentaro Nagamine, P\"{a}rt Peterson, Hamish~S. Scott, Jun Kudoh, Shinsei
  Minoshima, Maarit Heino, Kai J.~E. Krohn, Maria~D. Lalioti, Primus~E. Mullis,
  Stylianos~E. Antonarakis, Kazuhiko Kawasaki, Shuichi Asakawa, Fumiaki Ito,
  and Nobuyoshi Shimizu.
\newblock Positional cloning of the {APECED} gene.
\newblock {\em Nature Genetics}, 17(4):393--398, December 1997.

\bibitem{Coutinho1984}
A.~Coutinho, G.~Pobor, S.~Pettersson, T.~Leandersson, S.~Forsgren, P.~Pereira,
  A.~Bandeira{\&}, and C.~Martinez-A.
\newblock T cell-dependent b cell activation.
\newblock {\em Immunological Reviews}, 78(1):211--224, April 1984.

\bibitem{Kluger2012}
Nicolas Kluger, Annamari Ranki, and Kai Krohn.
\newblock {APECED}: is this a model for failure of t cell and b cell tolerance?
\newblock {\em Frontiers in Immunology}, 3, 2012.

\bibitem{Landegren2019}
Nils Landegren, Lindsey~B Rosen, Eva Freyhult, Daniel Eriksson, Tove Fall,
  Gustav Smith, Elise M~N Ferre, Petter Brodin, Donald Sharon, Michael Snyder,
  Michail Lionakis, Mark Anderson, and Olle K\"{a}mpe.
\newblock Comment on '{AIRE}-deficient patients harbor unique high-affinity
  disease-ameliorating autoantibodies'.
\newblock {\em {eLife}}, 8, June 2019.

\bibitem{response2019}
Christina Hertel, Dmytro Fishman, Anna Lorenc, Annamari Ranki, Kai Krohn,
  P\"{a}rt Peterson, Kai Kisand, and Andrian Hayday.
\newblock Response to comment on ’aire-deficient patients harbor unique
  high-affinity disease-ameliorating autoantibodies’, 2019.

\bibitem{Fisher1922}
R.~A. Fisher.
\newblock On the interpretation of $\chi$ 2 from contingency tables, and the
  calculation of p.
\newblock {\em Journal of the Royal Statistical Society}, 85(1):87, January
  1922.

\bibitem{Rodero2017}
Mathieu~P. Rodero, J{\'{e}}r{\'{e}}mie Decalf, Vincent Bondet, David Hunt,
  Gillian~I. Rice, Scott Werneke, Sarah~L. McGlasson, Marie-Alexandra
  Alyanakian, Brigitte Bader-Meunier, Christine Barnerias, Nathalia Bellon,
  Alexandre Belot, Christine Bodemer, Tracy~A. Briggs, Isabelle Desguerre,
  Marie-Louise Fr{\'{e}}mond, Marie Hully, Arn~M.J.M. van~den Maagdenberg,
  Isabelle Melki, Isabelle Meyts, Lucile Musset, Nadine Pelzer, Pierre
  Quartier, Gisela~M. Terwindt, Joanna Wardlaw, Stewart Wiseman,
  Fr{\'{e}}d{\'{e}}ric Rieux-Laucat, Yoann Rose, B{\'{e}}n{\'{e}}dicte Neven,
  Christina Hertel, Adrian Hayday, Matthew~L. Albert, Flore Rozenberg,
  Yanick~J. Crow, and Darragh Duffy.
\newblock Detection of interferon alpha protein reveals differential levels and
  cellular sources in disease.
\newblock {\em The Journal of Experimental Medicine}, 214(5):1547--1555, April
  2017.

\bibitem{Fremond2017}
Marie-Louise Fr{\'{e}}mond, Carolina Uggenti, Lien~Van Eyck, Isabelle Melki,
  Vincent Bondet, Naoki Kitabayashi, Christina Hertel, Adrian Hayday,
  B{\'{e}}n{\'{e}}dicte Neven, Yoann Rose, Darragh Duffy, Yanick~J. Crow, and
  Mathieu~P. Rodero.
\newblock Brief report: Blockade of {TANK}-binding kinase
  1/{IKK}{$\varepsilon$} inhibits mutant stimulator of interferon genes
  ({STING})-mediated inflammatory responses in human peripheral blood
  mononuclear cells.
\newblock {\em Arthritis {\&} Rheumatology}, 69(7):1495--1501, June 2017.

\bibitem{Sng2019}
Joel Sng, Burcu Ayoglu, Jeff~W. Chen, Jean-Nicolas Schickel, Elise M.~N. Ferre,
  Salom{\'{e}} Glauzy, Neil Romberg, Manfred Hoenig, Charlotte
  Cunningham-Rundles, Paul~J. Utz, Michail~S. Lionakis, and Eric Meffre.
\newblock {AIRE} expression controls the peripheral selection of autoreactive b
  cells.
\newblock {\em Science Immunology}, 4(34):eaav6778, April 2019.

\bibitem{Bratland2011}
Eirik Bratland and Eystein~S. Husebye.
\newblock Cellular immunity and immunopathology in autoimmune addison's
  disease.
\newblock {\em Molecular and Cellular Endocrinology}, 336(1-2):180--190, April
  2011.

\bibitem{Herold2013}
Kevan~C. Herold, Dario A.~A. Vignali, Anne Cooke, and Jeffrey~A. Bluestone.
\newblock Type 1 diabetes: translating mechanistic observations into effective
  clinical outcomes.
\newblock {\em Nature Reviews Immunology}, 13(4):243--256, March 2013.

\bibitem{Okada2016}
Satoshi Okada, Anne Puel, Jean-Laurent Casanova, and Masao Kobayashi.
\newblock Chronic mucocutaneous candidiasis disease associated with inborn
  errors of {IL}-17 immunity.
\newblock {\em Clinical {\&} Translational Immunology}, 5(12):e114, December
  2016.

\bibitem{Agarwala2017}
NCBI~Resource Coordinators.
\newblock {Database resources of the National Center for Biotechnology
  Information}.
\newblock {\em Nucleic Acids Research}, 46(D1):D8--D13, 11 2017.

\bibitem{Yates2019}
Andrew~D Yates, Premanand Achuthan, Wasiu Akanni, James Allen, Jamie Allen,
  Jorge Alvarez-Jarreta, M~Ridwan Amode, Irina~M Armean, Andrey~G Azov, Ruth
  Bennett, Jyothish Bhai, Konstantinos Billis, Sanjay Boddu, Jos{\'{e}}~Carlos
  Marug{\'{a}}n, Carla Cummins, Claire Davidson, Kamalkumar Dodiya, Reham
  Fatima, Astrid Gall, Carlos~Garcia Giron, Laurent Gil, Tiago Grego, Leanne
  Haggerty, Erin Haskell, Thibaut Hourlier, Osagie~G Izuogu, Sophie~H Janacek,
  Thomas Juettemann, Mike Kay, Ilias Lavidas, Tuan Le, Diana Lemos,
  Jose~Gonzalez Martinez, Thomas Maurel, Mark McDowall, Aoife McMahon, Shamika
  Mohanan, Benjamin Moore, Michael Nuhn, Denye~N Oheh, Anne Parker, Andrew
  Parton, Mateus Patricio, Manoj~Pandian Sakthivel, Ahamed Imran~Abdul Salam,
  Bianca~M Schmitt, Helen Schuilenburg, Dan Sheppard, Mira Sycheva, Marek
  Szuba, Kieron Taylor, Anja Thormann, Glen Threadgold, Alessandro Vullo,
  Brandon Walts, Andrea Winterbottom, Amonida Zadissa, Marc Chakiachvili,
  Bethany Flint, Adam Frankish, Sarah~E Hunt, Garth IIsley, Myrto Kostadima,
  Nick Langridge, Jane~E Loveland, Fergal~J Martin, Joannella Morales,
  Jonathan~M Mudge, Matthieu Muffato, Emily Perry, Magali Ruffier, Stephen~J
  Trevanion, Fiona Cunningham, Kevin~L Howe, Daniel~R Zerbino, and Paul Flicek.
\newblock Ensembl 2020.
\newblock {\em Nucleic Acids Research}, November 2019.

\bibitem{Metsalu2015}
Tauno Metsalu and Jaak Vilo.
\newblock {ClustVis}: a web tool for visualizing clustering of multivariate
  data using principal component analysis and heatmap.
\newblock {\em Nucleic Acids Research}, 43(W1):W566--W570, May 2015.
\newblock [PubMed:\href{http://www.ncbi.nlm.nih.gov/pubmed/25969447}{25969447}]
  [PubMedCentral:
  \href{https://www.ncbi.nlm.nih.gov/pmc/articles/PMC4489295/}{PMC4489295}].

\bibitem{huprot}
J.~S. Jeong, L.~Jiang, E.~Albino, J.~Marrero, H.~S. Rho, J.~Hu, S.~Hu, C.~Vera,
  D.~Bayron-Poueymiroy, Z.~A. Rivera-Pacheco, L.~Ramos, C.~Torres-Castro,
  J.~Qian, J.~Bonaventura, J.~D. Boeke, W.~Y. Yap, I.~Pino, D.~J. Eichinger,
  H.~Zhu, and S.~Blackshaw.
\newblock {{R}apid identification of monospecific monoclonal antibodies using a
  human proteome microarray}.
\newblock {\em Mol. Cell Proteomics}, 11(6):O111.016253, Jun 2012.
\newblock [PubMed:\href{http://www.ncbi.nlm.nih.gov/pubmed/22307071}{22307071}]
  [doi:\href{https://doi.org/10.1074/mcp.O111.016253}{10.1074/mcp.O111.016253}]
  [PubMedCentral:
  \href{https://www.ncbi.nlm.nih.gov/pmc/articles/PMC3433917/}{PMC3433917}].

\bibitem{reshape2}
Hadley Wickham et~al.
\newblock Reshaping data with the reshape package.
\newblock {\em Journal of Statistical Software}, 21(12):1--20, 2007.
\newblock [doi:\href{http://dx.doi.org/10.18637/jss.v021.i12}].

\bibitem{d3}
M.~Bostock, V.~Ogievetsky, and J.~Heer.
\newblock {{D}$^{3}$: {D}ata-{D}riven {D}ocuments}.
\newblock {\em IEEE Trans Vis Comput Graph}, 17(12):2301--2309, Dec 2011.
\newblock
  [PubMed:\href{http://www.ncbi.nlm.nih.gov/pubmed/22034350}{22034350}].

\bibitem{Xu2016}
Zhaowei Xu, Likun Huang, Hainan Zhang, Yang Li, Shujuan Guo, Nan Wang, Shi hua
  Wang, Ziqing Chen, Jingfang Wang, and Sheng ce~Tao.
\newblock {PMD}: A resource for archiving and analyzing protein microarray
  data.
\newblock {\em Scientific Reports}, 6(1), January 2016.

\bibitem{Zhang2007}
Jun-Ming Zhang and Jianxiong An.
\newblock Cytokines, inflammation, and pain.
\newblock {\em International Anesthesiology Clinics}, 45(2):27--37, 2007.

\bibitem{HuynhThu2012}
V{\^{a}}n~Anh Huynh-Thu, Yvan Saeys, Louis Wehenkel, and Pierre Geurts.
\newblock Statistical interpretation of machine learning-based feature
  importance scores for biomarker discovery.
\newblock {\em Bioinformatics}, 28(13):1766--1774, April 2012.

\bibitem{Mann2003}
C~J Mann.
\newblock Observational research methods. research design {II}: cohort, cross
  sectional, and case-control studies.
\newblock {\em Emergency Medicine Journal}, 20(1):54--60, January 2003.

\bibitem{Melamed2018}
Alexander Melamed and Julian~N Robinson.
\newblock Case{\textendash}control studies can be useful but have many
  limitations.
\newblock {\em {BJOG}: An International Journal of Obstetrics {\&}
  Gynaecology}, 126(1):23--23, June 2018.

\bibitem{Kang2012}
Ju-Hee Kang, Hugo Vanderstichele, John~Q. Trojanowski, and Leslie~M. Shaw.
\newblock Simultaneous analysis of cerebrospinal fluid biomarkers using
  microsphere-based {xMAP} multiplex technology for early detection of
  alzheimer's disease.
\newblock {\em Methods}, 56(4):484--493, April 2012.

\bibitem{Kellar2002}
K~Kellar.
\newblock Multiplexed microsphere-based flow cytometric assays.
\newblock {\em Experimental Hematology}, 30(11):1227--1237, November 2002.

\bibitem{Strobl2009}
Carolin Strobl, James Malley, and Gerhard Tutz.
\newblock An introduction to recursive partitioning: Rationale, application,
  and characteristics of classification and regression trees, bagging, and
  random forests.
\newblock {\em Psychological Methods}, 14(4):323--348, December 2009.

\bibitem{Nelder1972}
J.~A. Nelder and R.~W.~M. Wedderburn.
\newblock Generalized linear models.
\newblock {\em Journal of the Royal Statistical Society. Series A (General)},
  135(3):370, 1972.

\bibitem{Gou2011}
Jianping Gou, Lan Du, Yuhong Zhang, and Taisong Xiong.
\newblock A new distance-weighted k -nearest neighbor classifier.
\newblock {\em J. Inf. Comput. Sci.}, 9, 11 2011.

\bibitem{Walsh2020}
Ian Walsh, Dmytro Fishman, Dario Garcia-Gasulla, Tiina Titma, The ELIXIR
  Machine~Learning focus group, Jen Harrow, Fotis~E. Psomopoulos, and Silvio
  C.~E. Tosatto.
\newblock Dome: Recommendations for machine learning validation in biology.
\newblock {\em Nature Methods}, 2020.

\end{thebibliography}

% == acknowledgement ==

\setcounter{secnumdepth}{-1}

\chapter{Acknowledgements}

Although it is hard to make a truly exhaustive list of all the people I met and who influenced me, below is my diligent attempt to give credit to as many of them as possible.\\

I met \textit{Jaak Vilo} on my first semester as a Master's student at the University of Tartu. He has been teaching one of the core courses in our curriculum - advanced algorithms. In the beginning, our expectations were not very high - in my previous university, no one would anticipate great teaching skills from the institute’s head. Yet, we were wrong. Jaak's lectures were and are interactive, not only he asked questions from the audience, but he was also genuinely interested in hearing the answers. I admired his ability to admit that he does not know everything and allowing students to correct him. His respectful and supportive attitude towards students inspired me. Later, I learned that Jaak is also the head of the bioinformatics research team - a group of people with a primarily technical background who use computer science to help improve our understanding of the human condition. Jaak also has been the one to invite me to apply for PhD in his group. I am incredibly grateful to Jaak for guiding me towards bioinformatics and teaching. I have never regretted my choice. \\

When \textit{Hedi Peterson} returned to Tartu from her post-doc in Switzerland, I was already in the middle of the first year of my PhD. Yet, very soon, she has become a go-to person and a support for all young PhD students in our group, including myself. She has successfully combined a scientific advisor, psychologist, and friend’s roles, helping us stay motivated, productive, and happy. Needless to say that we needed the last two roles as often as the first. \\
Work on this thesis was not always easy and did not come without moments of despair. On multiple such occasions, I approached Hedi asking if this would make sense to change the topic and maybe try something else. Yet every time, she would remain adamant and convinced me to keep trying. It is largely owing to her support at these darkest hours this thesis finally came to be. Moreover, all four publications included in this thesis have been carried out under her close supervision. \\
Hedi's contribution to the text of this thesis deserves a paragraph on its own. Although this work formally has only one author, if this would be a popular science book, Hedi Peterson would undoubtedly be its well-deserved co-author. Her clever and always-to-the-point edits have helped me to improve this text significantly. She has been through the entire writing process from the very early and barely readable drafts to the very end, fixing minor typos. Her almost superhuman capacity to find flaws both big and small and attention to details is as impressive as it is terrifying, making people who do not know her suspect that some extraterrestrial forces are at work. \\
I am fortunate that one day Hedi decided not to make an app but to make the \linebreak difference.
\\

When I started the work on Publication I, I had a very rudimentary understanding of basic biology and a completely non-existent comprehension of immunology. Although I certainly did not become an expert in immunology, yet everything I do know about the field, I owe to \textit{P{\"a}rt Peterson} and \textit{Kai Kisand}. Their seemingly endless patience when answering my borderline stupid questions on the one hand and a capacity to take very seriously my comments related to computer science and statistics on the other created a safe environment for interdisciplinary research so much needed for an early-stage PhD student. This tandem will remain in my memory as an example of perfect clinical collaborators, a source of wisdom and guidance. P{\"a}rt has also provided so much appreciated feedback on this thesis and defense presentation. He has edited a chapter related to immunity.  \\

After I first met Jaak and realised that by joining the bioinformatics group, one could learn to employ computer science methods for the good of all humanity, most of the hard work on turning me into someone at least partially resembling a bioinformatician took over \textit{Konstantin Tretjakov}. Kostja has supervised my Master's thesis, dedicating a lot of time to developing my writing and reasoning skills. Kostja has also significantly influenced me as a lecturer. I hope one day to solve at least one percent of all riddles that Kostja gave me.
\\

Learning from the best never stopped as later I met a bright-eyed group leader from Sanger Institute - \textit{Leopold Parts}. I am fortunate to have Leo as a senior member and a co-lead of the medical imaging research group in collaboration with PerkinElmer. Not only Leo served as a role model of a successful scientist and kept inspiring me throughout the years, but he has also actively helped and supported me as a mentor and unofficial fourth supervisor of this thesis. \\

I want to pay a special tribute to all three of my reviewers: \textit{Raivo Kolde}, \textit{Jessica Da Gama Duarte} and \textit{Fridtjof Lund-Johansen}. Their comments have significantly improved both the content and the presentation of this thesis. \\

Members of the bioinformatics and information technology (BIIT) research group will always have a special place in my mind and heart.  \textit{Priit (Lemps) Adler} - a mentor and a friend, I learn a lot from you. Big thank you to \textit{Liis Kolberg} and \textit{Uku Raudvere} for all the invaluable help with this thesis. I cannot thank enough to \textit{Kaido Lepik} for his cool-headed support in times of need as well as for the joint sports activities. \textit{Ivan Kuzmin} who has been my dear teacher of web development as well as an active maintainer of the PAWER tool we have developed for Publication III. He has also set in motion my interest in philosophy and showed a lot of new ways to look at the world. I want to thank \textit{Kaur Alasoo} for his comments on the PAWER paper - I am looking forward to our future joint paper. Thanks to \textit{Ahto Salumeets} for his valuable comments on the immunity chapter as well as great scientific discussions. I owe a lot in various ways to \textit{Elena S\"{u}gis}, \textit{Nurlan Kerimov}, \textit{Mari-Liis Allikivi}, \textit{Erik Jaaniso}, \textit{Joonas Puura}, \textit{Sulev Reisberg}, \textit{Sven Laur}, \textit{Anna Leontjeva}, \textit{Ilya Kuzovkin}, \textit{Meelis Kull}, \textit{Ulvi Talas}, \textit{Kateryna Peikova}, \textit{Balaji R}, \textit{Oliver Nisumaa}, \textit{Siyuan Gao}, \textit{Tambet Arak}, \textit{Tauno Metsalu}, \textit{Vijayachitra Modhukur} and \textit{Tõnis Tasa}. 
\\

I would like to express my deepest gratitude to PerkinElmer project family: \textit{Kaupo Palo}, \textit{Leopold Parts}, \textit{Olavi Ollikainen}, \textit{Hartwig Preckel}, \textit{Sten-Oliver Salumaa}, \textit{Mikhail Papkov}, \textit{Mohammed Ali}, \textit{T\~{o}nis Laasfeld}, \textit{Dmytro Urukov}, \textit{Kaspar Hollo}, \textit{Iaroslav Plutenko}, \textit{Tetiana Rabiichuk}, \textit{Tarun Khajuria}, \textit{Oleh Misko}, \textit{Oles Pryhoda}, \textit{Bohdan Petryshak}, \textit{Roman Ring}, \textit{Oliver Meikar} and \textit{Daniel Majoral}.  
\\

Warmest thanks to BioEndoCar project partners: \textit{Tea Lani\v{s}nik Ri\v{z}ner}, \textit{Andrea Romano}, \textit{Tamara Knific}, \textit{Eva Hafner}, \textit{Jerzy Adamski}, \textit{Janina Tokarz}, \textit{Christoph Schröder}, \textit{Camille Lowy} and \textit{Andrzej Semczuk}. This is one of the most medically relevant projects I have been part of.
\\

Many thanks to Autonomous Driving Lab researchers and students including but not limited to \textit{Tambet Matiisen}, \textit{Anne J\"{a}\"{a}ger}, \textit{Meelis Kull}, \textit{Ardi Tampuu}, \textit{Naveed Muhammad}, \textit{Alexander Nolte}, \textit{Karl Kruusamäe}, \textit{Raimundas Matulevicius}, \textit{Arun Singh},  \textit{Amnir Hadachi}, \textit{Edgar Sepp}, \textit{Navid Roshan}, \textit{Dmytro Zabolotnii}, \textit{Jan Aare van Gent}, \textit{Kertu Toompea}, \textit{Maxandre Ogeret}, \textit{Thomas Churchman} and \textit{Tanya Shtym}. I am proud to be part of this immensely cool and important project. \\

Special thanks to the admin team of the institute of computer science who have created an outstanding work environment: \textit{Jaak Vilo}, \textit{Mark Fishel}, \textit{Piret Orav}, \textit{Jaanika Seli}, \textit{Heili Kase}, \textit{Natali Belinska}, \textit{Ülle Holm}, \textit{Liivi Luik}, \textit{Anne Jääger}, \textit{Anastasiia Shevchenko}, \textit{Annet Muru}, \textit{Anni Suvi}, \textit{Henry Narits}, \textit{Martin Kaljula}, \textit{Reili Liiver}, \textit{Anneli Vainumäe}, \textit{Käroliin Jääger}, \textit{Saili Petti} and \textit{Daisy Alatare}. \\

Also, my profound appreciation to three brave human beings that together with me are trying to make the difference for all: \textit{Priit Salumaa}, \textit{Bohdan Petryshak} and \textit{Martin Reim}. \\

I am grateful to all students that I have ever had the pleasure to teach or supervise. Please, know that I have learned from you as much and even more than you have learned from me. \\

Thanks to my family and friends for their unconditional love and limitless support that kept me going.\\

Finally, last by definitely not least, thank you to my beloved wife - Lena. A political scientist by education and an artist at heart, she has willingly succumbed herself to the fields of machine learning, artificial intelligence, statistics, bioinformatics, and programming by agreeing to be the subject of my lecture and workshop rehearsals. She possesses a unique ability to add color, joy, and purpose to my life. It is my true pleasure to share a planet and an epoch with this caring and gentle being. \\

% == summary in estonian ==

\begin{otherlanguage}{estonian}
\chapter*{Sisukokkuv\~ote}
\addcontentsline{toc}{chapter}{Sisukokkuv\~ote (Summary in Estonian)}

\section*{Andmeanalüüsi töövoo loomine valkude automaatseks kirjeldamiseks immunoloogias}

Valgud on kõigi elusorganismide olulised koostisosad. Nendest keerukatest molekulidest sõltub suur hulk eluliselt olulisi funktsioone. Valkude kogus organismi rakkudes on rangelt reguleeritud, kuna liigne kogus või äge puudus võib põhjustada soovimatuid tagajärgi. Ebanormaalne valkude tase võib olla tõsise talitlushäire märk. Seetõttu võib võime täpselt hinnata valkude kontsentratsiooni kehas olla haiguse mehhanismide mõistmise võti.

Valgu mikrokiibid on populaarne viis valkude kontsentratsiooni mõõtmiseks vereproovist. Sel moel saab paralleelselt mõõta sadade või isegi tuhandete valkude kontsentratsioone. Ehkki valgukiipidel on palju ühist DNA-mikrokiipidega, ei sobi kõik DNA-mikrokiipide jaoks välja töötatud arvutusmeetodid erinevate bioloogiliste eelduste tõttu valgukiipidele. Seetõttu on valgukiipide kui andmete tootmise platvormi kõigi võimaluste tõhusaks kasutamiseks hädavajalikud spetsiaalselt neile kohandatud meetodid.

Klassikaline valguandmete analüüs on keeruline ja koosneb järjestikku rakendatud arvutusmeetodite seeriast. Analüüsi paikapidavuse tagamiseks on vajalikud meetodid tehnilise müra vähendamiseks, võõrväärtuste tuvastamiseks ja eemaldamiseks ning saadud signaali väärtuste normaliseerimiseks. Statistilisi teste ja masinõppe meetodeid kasutatakse nii individuaalsete valkude kui ka nende kombinatsioonide tuvastamiseks võttes arvesse valke, mille tasemed on katsetingimuste vahel piisavalt erinevad. Lõpuks aitavad funktsionaalse rikastamise analüüsi tööriistad viia sellised valgud kõige levinumate bioloogiliste funktsioonide konteksti. Käesolevas töös uurisime valgu mikrokiibi katsetest saadud andmete analüüsiks kasutatavaid arvutuslikke meetodeid ja optimeerisime nende andmete analüüsi töövoogu. Selle tulemusena töötasime välja veebitööriista, mis aitab kogu seda analüüsi poolautomaatselt läbi viia. Doktoritöös kirjeldatud meetodeid rakendasime praktikas mitmes valgu mikrokiipidega seotud teadustöös.

Käesolevas dissertatsioonis kirjeldame esmalt uuringut, kus valgu mikrokiipidega mõõdeti 1. Tüüpi autoimmuunse polüendokrinopaatia sündroomiga (APS1) patsientide vere autoantikehade sisaldust. APS1 patsientidel autoimmuunse reaktsiooni sihtmärgiks olevate valkude esialgse loendi määratlemiseks viisime läbi valgu mikrokiibi-spetsiifilise eeltöötluse analüüsi töövoo ja diferentsiaalanalüüsi.

Meie eesmärk oli APS1 seisundi ja autoimmuunsuse taga olevate mehhanismide sügavam mõistmine. Seetõttu uurisime esimeses artiklis tuvastatud valgu sihtmärke edasi. Töö käigus analüüsisime mitmeid avalikke andmebaase ja valkude funktsionaalse rikastamise andmekogusid, et teha kindlaks valgu sihtmärkide taga olevad ühised bioloogilised tegurid. Tulemuste kinnitamiseks viisime läbi funktsionaalse rikastamise analüüsi kasutades g:Profileri veebitööriista.

Saadud valgu mikrokiipide analüüsi kogemuse põhjal sidusime loodud andmeanalüüsi vahendid R-programmeerimiskeele põhiseks veebitööriistaks PAWER. PAWER, mis on loodud valgu mikrokiibi andmete analüüsi poolautomaatseks läbiviimiseks, on käesoleva doktoritöö põhitulemuseks. Selle intuitiivne kasutajaliides ja järkjärguline töövoog on loodud valgu mikrokiibi analüüsi hõlpsaks teostamiseks nii bioloogide kui bioinformaatikute poolt.

Käesoleva doktoritöö neljandas artiklis uurisime masinõppe mudelite ja varasemates artiklites käsitletud klassikaliste statistiliste meetodite koos rakendamise väärtust. Selles töös analüüsisime endometrioosi juhtkontrolluuringut. Eelnev statistiline analüüs näitas, et verest mõõdetud valgutasemete põhjal pole ükski üksik valk võimeline eristama endometrioosi põdevaid patsiente tervetest. Valkude kombinatsioonide ennustusvõime hindamiseks kasutasime erinevaid masinõppe meetodeid. Sarnaselt statistiliste testide tulemustega ei saavutanud masinõppe mudelid juhuslikkusest oluliselt erinevat tulemust. Seetõttu kinnitasid masinõppe tulemused hüpoteesi, et ei üksikute valkude mõõtmine ega ka valkude kombinatsioonid ei võimalda ennustada endometrioosi ja aidata haigust diagnoosida antud valimi baasil.

\end{otherlanguage}

% == publications ==
\vfill\newpage
\null 
\pagenumbering{gobble} % suppress page numbering
\newpage

\chapter{Publications}
\newpage

\vfill\newpage\null\newpage
\marginpar{
\begin{tikzpicture}
      \draw[fill,color=black] (0,0) rectangle (4cm,2cm);
      \draw[color=white] (1cm,1cm) node {\normalfont\huge\bfseries I};
\end{tikzpicture}
} 

\thispagestyle{empty}

\newpage

\phantomsection\addcontentsline{toc}{section}{AIRE-deficient patients harbor unique high-affinity disease-ameliorating autoantibodies}

\thispagestyle{empty}\null\vfill
\begin{flushleft}

S. Meyer, M. Woodward, C. Hertel, P. Vlaicu, Y. Haque, J. Kärner, A. Macagno, S. Onuoha, \textbf{D. Fishman}, H. Peterson, K. Metsküla, R. Uibo, K. Jäntti, K. Hoky-nar, A. Wolff, K. Krohn, A. Ranki, P. Peterson, K. Kisand, A. Hayday, A. Meloni, N. Kluger, E. Husebye, K. Trebusak Podkrajsek, T. Battelino, N. Bratanic, and A. Peet.\\
AIRE-deficient patients harbor unique high-affinity disease-ameliorating autoantibodies\\
\textbf{Cell}, 166(3):582–595, July 2016\\
\medskip
The article is reprinted with permission of the copyright owner.
\end{flushleft}

\includepdf[pages=-]{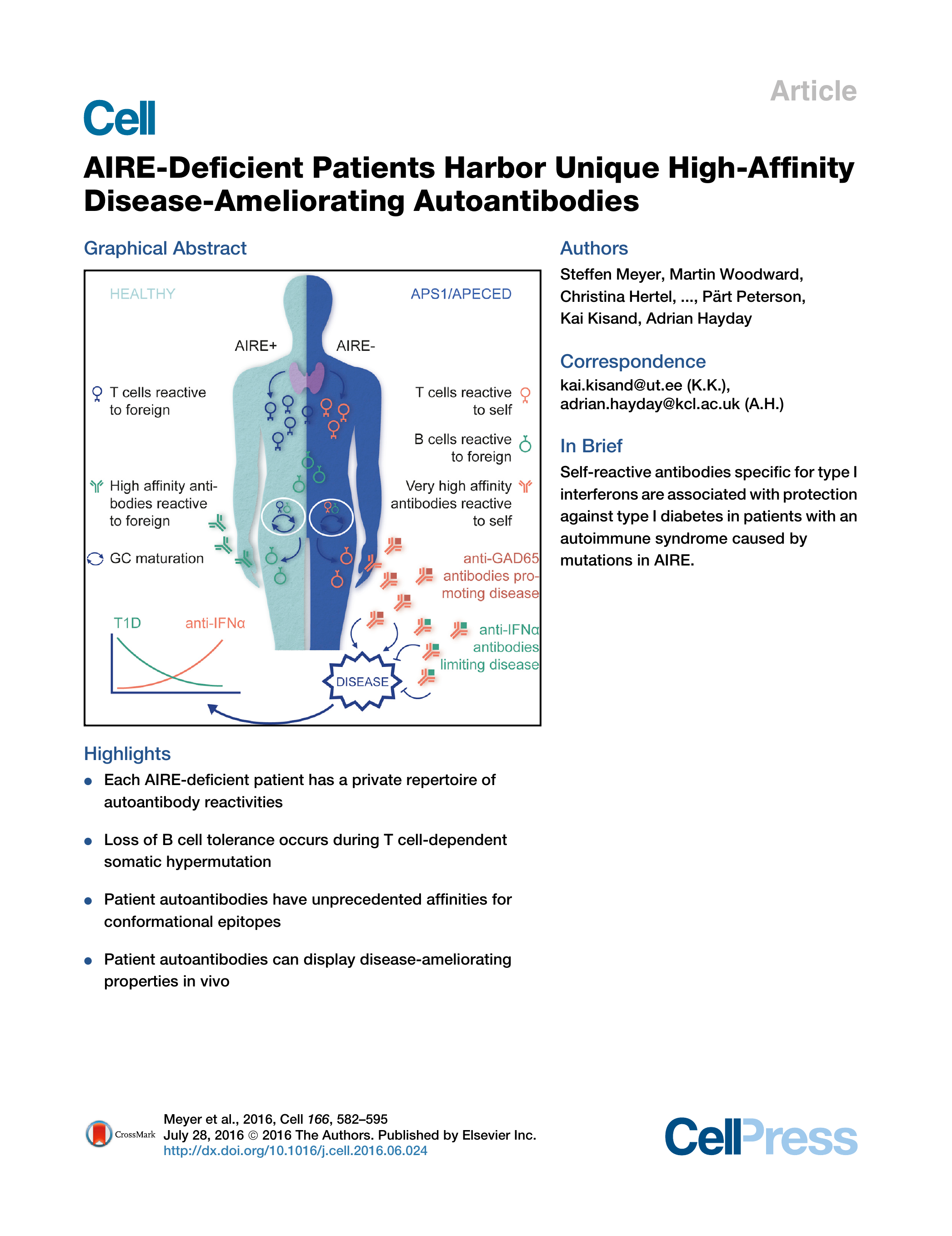}

\vfill\newpage\null\newpage
\marginpar{
\begin{tikzpicture}
      \draw[fill,color=black] (0,0) rectangle (4cm,2cm);
      \draw[color=white] (1cm,1cm) node {\normalfont\huge\bfseries II};
\end{tikzpicture}
} 

\newpage

\phantomsection\addcontentsline{toc}{section}{Autoantibody repertoirein APECED patients targets two distinct subgroups of proteins}
\thispagestyle{empty}\null\vfill
\begin{flushleft}

\textbf{D. Fishman}, K. Kisand, C. Hertel, M. Rothe, A. Remm, M. Pihlap, P. Adler, J. Vilo, A. Peet, A. Meloni, K. Trebusak Podkrajsek, T. Battelino, Ø. Bruserud, A. Wolff, E. Husebye, N. Kluger, K. Krohn, A. Ranki, H. Peterson, A. Hayday, and P. Peterson\\
Autoantibody repertoirein APECED patients targets two distinct subgroups of proteins\\
\textbf{Frontiers in Immunology}, 8, August 2017\\
\medskip
The article is reprinted with permission of the copyright owner.
\end{flushleft}

\includepdf[pages=-]{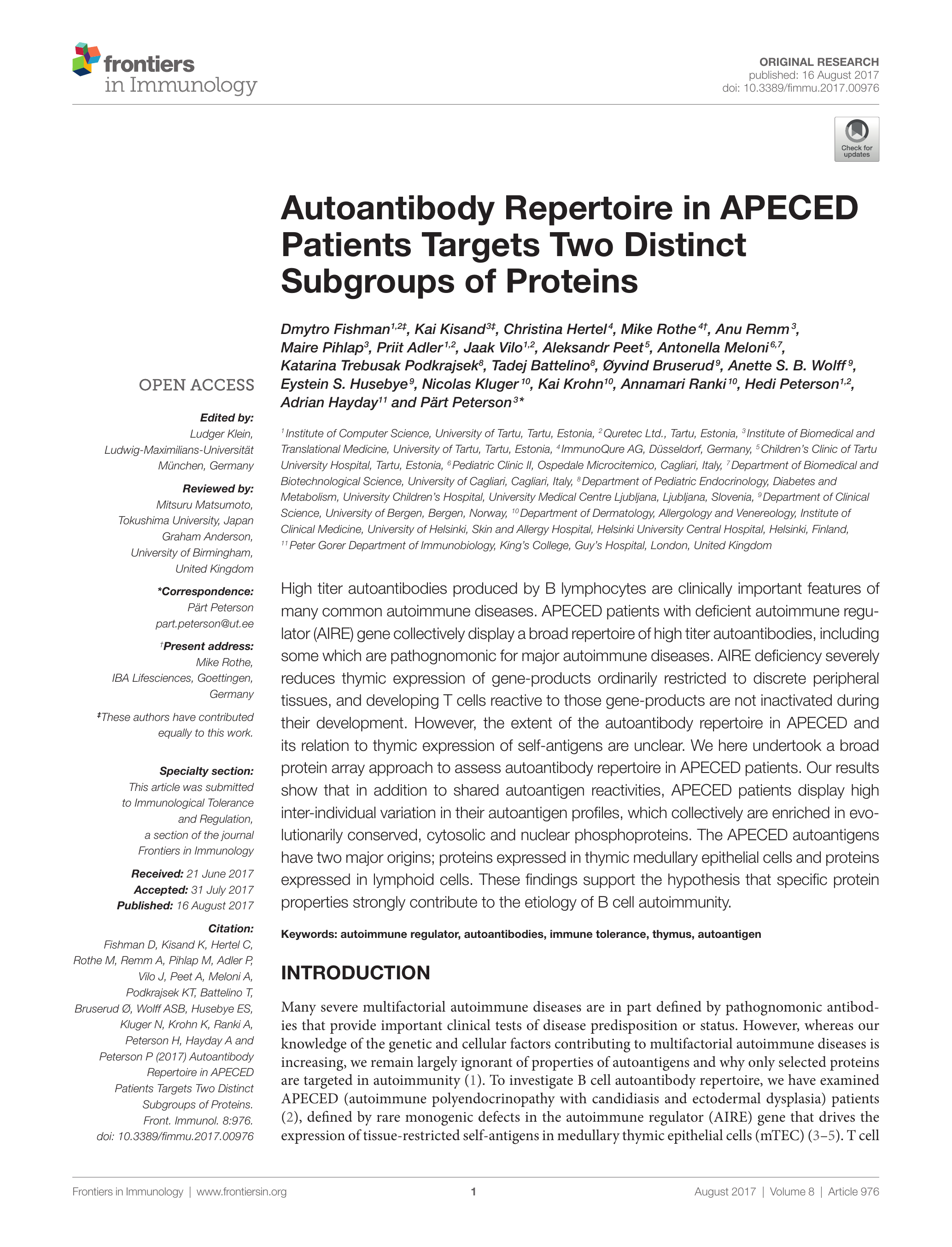}

\vfill\newpage\null\newpage
\marginpar{
\begin{tikzpicture}
      \draw[fill,color=black] (0,0) rectangle (4cm,2cm);
      \draw[color=white] (1cm,1cm) node {\normalfont\huge\bfseries III};
\end{tikzpicture}
} 

\newpage

\phantomsection\addcontentsline{toc}{section}{PAWER: Protein Array Web ExploreR}
\thispagestyle{empty}\null\vfill
\begin{flushleft}

\textbf{D. Fishman}, I. Kuzmin, P. Adler, J. Vilo, and H. Peterson\\
PAWER: Protein Array Web ExploreR\\
\textbf{BMC Bioinformatics}, 17, September 2020\\
\medskip
The article is reprinted with permission of the copyright owner.
\end{flushleft}

\includepdf[pages=-]{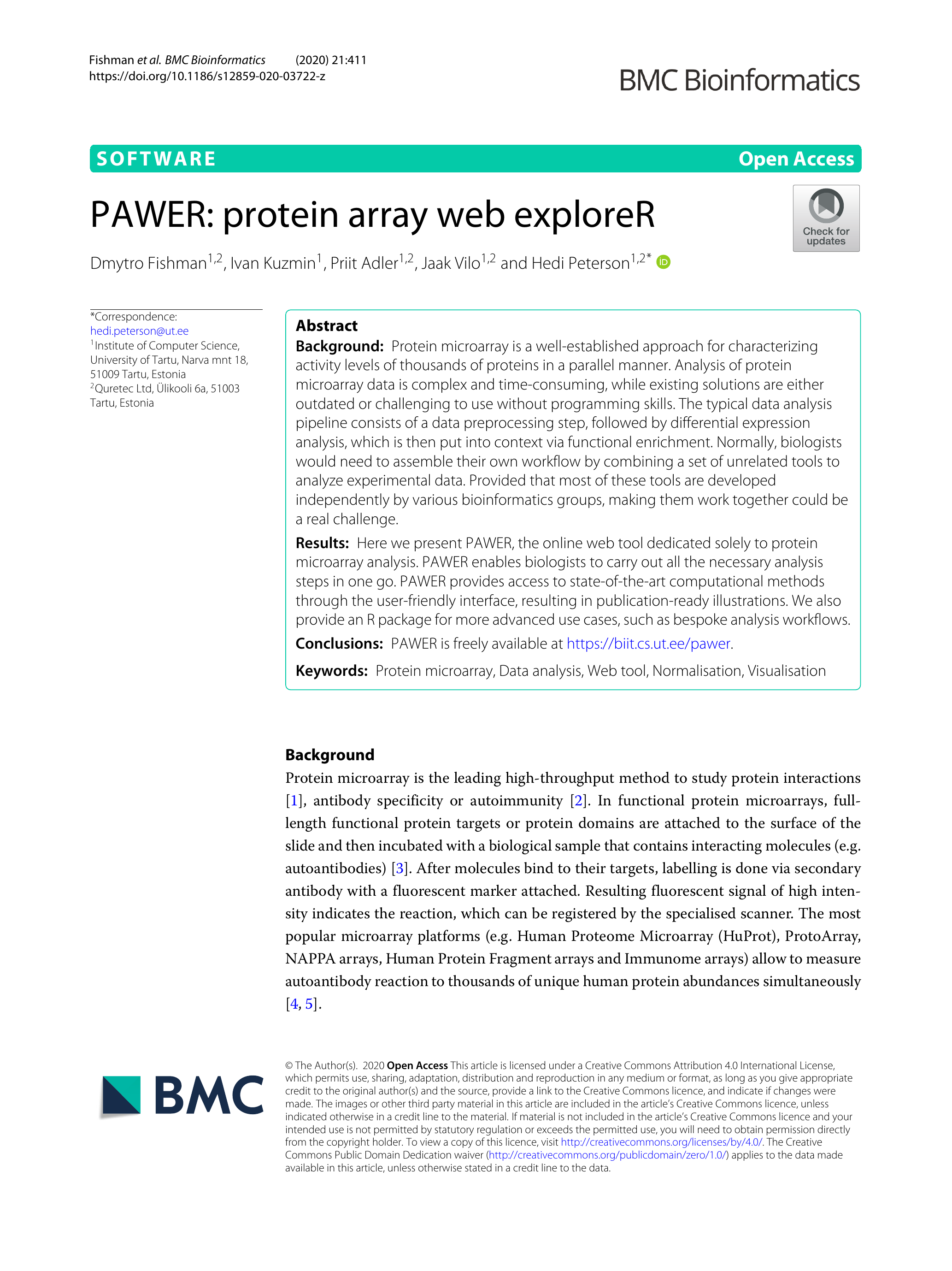}

\vfill\newpage\null\newpage
\marginpar{
\begin{tikzpicture}
      \draw[fill,color=black] (0,0) rectangle (4cm,2cm);
      \draw[color=white] (1cm,1cm) node {\normalfont\huge\bfseries IV};
\end{tikzpicture}
} 

\newpage

\phantomsection\addcontentsline{toc}{section}{Multiplex analysis of 40 cytokines do not allow separation between endometriosis patients and controls}
\thispagestyle{empty}\null\vfill
\begin{flushleft}

T. Knific, \textbf{D. Fishman}, A. Vogler, M. Gstöttner, R. Wenzl, H. Peterson and T. Lanisnik Rizner\\
Multiplex analysis of 40 cytokines do not allow separation between endometriosis patients and controls\\
\textbf{Scientific Reports}, 13 November 2019\\
\medskip
The article is reprinted with permission of the copyright owner.
\end{flushleft}

\includepdf[pages=-]{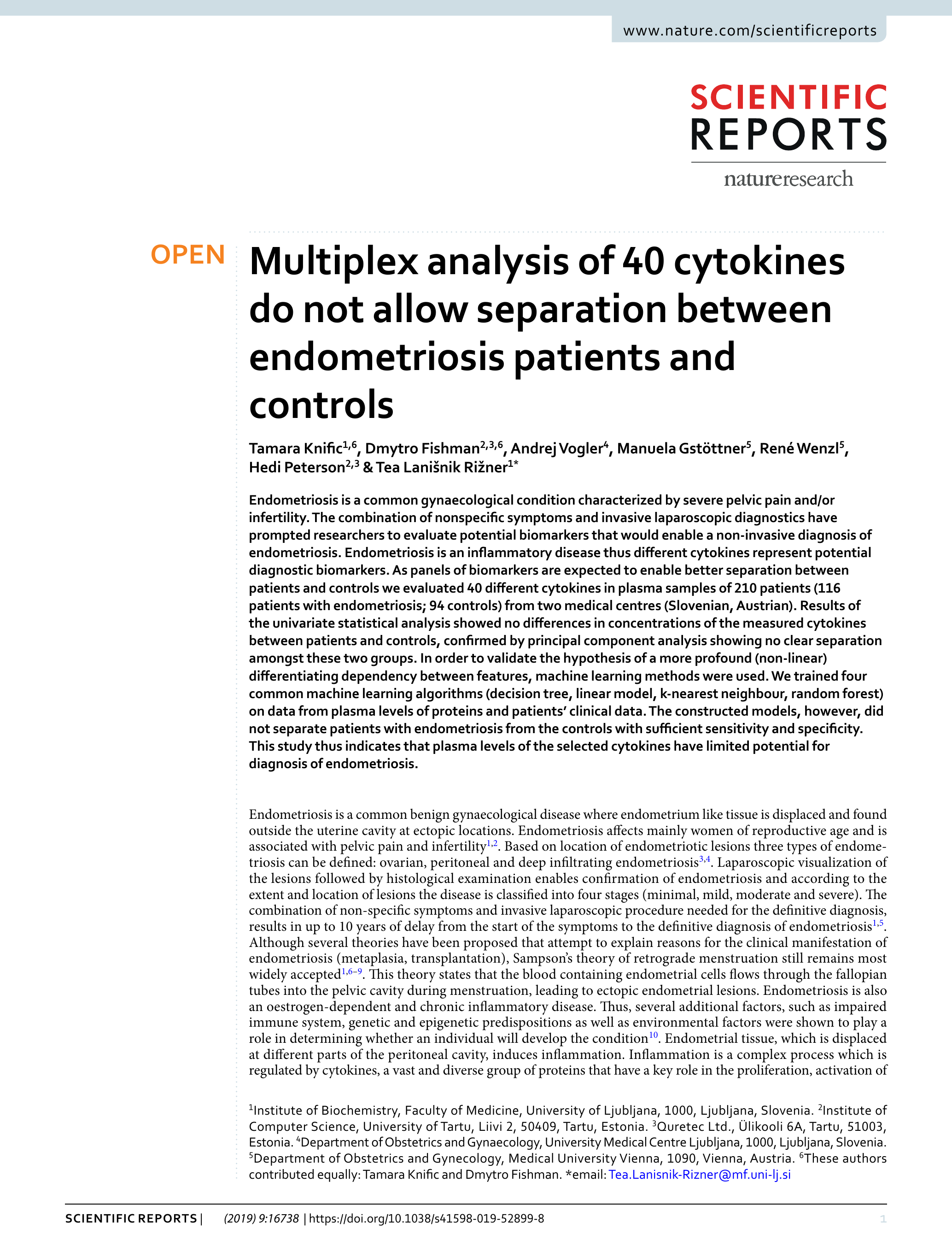}

\newpage

\pagenumbering{arabic} % unsuppress page numbering

\chapter{Curriculum Vitae}

\section*{Personal data}

\begin{tabular}{@{}l@{\hskip7mm}l}
Name:       &Dmytro Fishman\\
Date of birth:       &06.02.1991\\
Citizenship:       &Ukraine\\
Contact:       & dmytro.fishman@ut.ee \\
Current position:       &Junior Lecturer in Artificial Intelligence

\end{tabular}

\section*{Education}

\begin{tabular}{@{}l@{\hskip7mm}p{90mm}}
2013--2021         & Ph.D. Candidate, University of Tartu \\
2011--2013         & MSc. Software Engineering, University of Tartu and Tallinn Technical University\\
2007--2011         & BSc. Computer Science (\textit{cum laude}), National Technical University of Ukraine "Igor Sikorsky Kyiv Polytechnic Institute" 
\end{tabular}

\section*{Employment}

\begin{tabular}{@{}l@{\hskip7mm}p{90mm}}
2020--...          & Junior Lecturer of Artificial Intelligence, Institute of Computer Science, University of Tartu \\
2018--2020         & Assistant of Data Science, Institute of Computer Science, University of Tartu\\ 
2015--2018         & Junior Researcher Fellow in Bioinformatics, Institute of Computer Science, University of Tartu\\ 
2014--2015         & Research Project Specialist, Institute of Computer Science, University of Tartu \\
2013--2014         & Assistant of Informatics, Institute of Computer Science, University of Tartu
\end{tabular}

\section*{Honours \& awards}
\begin{itemize}
    \item Ustus Agur personal scholarship for the significant contribution to the development of Estonian information society, fall 2018;
    \item Raefond personal scholarship to outstanding international MSc students studying in the University of Tartu, fall 2012;

\end{itemize}

\section*{Teaching}
\begin{itemize}
    \item Lecturer for Machine Learning course at University of Tartu in fall 2020;
    \item Lecturer for Machine Learning course at Ukrainian Catholic University (Lviv, Ukraine) in spring 2020 and 2021;
    \item Lecturer for Advanced Machine Learning and Time Series Modelling course at Kyiv School of Economics (Kyiv, Ukraine) in spring 2020;
    \item Teaching assistant for Advanced Algorithms course at University of Tartu in 2013, 2014, 2016, 2017, 2018 and 2019;
    \item Teaching assistant for Data Mining course at University of Tartu in 2014, 2015 and 2017;
    \item Teaching assistant for Text Algorithms course at University of Tartu in fall 2013;
    \item Leading the seminars on Bioinformatics (in 2015 and 2016) as well as Special course in Machine Learning (in 2019).
\end{itemize}

\section*{Supervised theses}
\begin{itemize}
    \item Kaspar Hollo (MSc): Anomaly segmentation in microscopy images using deep learning.
	\item Navid Bamdad Roshan (MSc): Change detection in HD-maps using camera images for autonomous driving.
	\item Dmytro Zabolotnii (MSc): High-definition map generation for autonomous driving.
	\item Iaroslav Plutenko (MSc): Metadata for semantic segmentation by employing channel attention mechanism.
	\item Mihkel  Ilisson (MSc): Application of statistical analysis and machine learning methods for analysing blood metabolites.
    \item Richard Annilo (BSc): Using deep learning for medical sound analysis.
    \item Vladyslav Fediukov (MSc): Detection of Changes in Maps Using LiDAR Point Clouds.
    \item Oleh Misko (MSc): Ensembling and transfer learning for multi-domain microscopy image segmentation.
    \item Mikhail Papkov (MSc): Deep neural networks application for cell phenotyping in mixed cultures. Co-supervised with Leopold Parts.
    \item Oleksandr Pryhoda (BSc): Tissue segmentation in histological whole-slide images with deep learning.
    \item Yurii Toma (MSc): Predicting the Impact of Non-Coding Genetic Variants on Transcription Factor Binding with Machine Learning. Co-supervised with Kaur Alasoo.
    \item Sten-Oliver Salumaa (MSc): Convolutional Neural Networks for Cellular Segmentation. Co-supervised with Leopold Parts.
    \item Vitalii Peretiatko (MSc): Using Robust Rank Aggregation for prioritising autoimmune targets on protein microarrays. Co-supervised with Elena S\"{u}gis.
\end{itemize}

\section*{Scientific work}

Main fields of interest:
\begin{itemize}
  \item Bioinformatics
  \item Machine Learning and Deep Learning
  \item Computer Vision
\end{itemize}
\medskip

Served as a reviewer:
\begin{itemize}
  \item International Conference on Machine Learning (2020, 2021)
  \item International Conference on Learning Representations (2020)
  \item Nature Machine Intelligence, BMC Bioinformatics, Nature Scientific Reports, and Bioinformatics.
\end{itemize}

% == elulookirjeldus ==

\begin{otherlanguage}{estonian}
\chapter*{Elulookirjeldus}
\addcontentsline{toc}{chapter}{Elulookirjeldus (Curriculum Vitae in Estonian)}
\section*{Isikuandmed}

\begin{tabular}{@{}l@{\hskip7mm}l}
Nimi:              & Dmytro Fishman\\
Sünniaeg:          & 06.02.1991\\
Kodakondsus:       & Ukraina\\
E-mail:            & dmytro.fishman@ut.ee

\end{tabular}

\section*{Haridus}

\begin{tabular}{@{}l@{\hskip7mm}p{90mm}}
2013--2021         & { Tartu Ülikool, loodus- ja täppisteaduste valdkond, informaatika, doktoriõpe} \\
2011--2013         & {Tartu Ülikool, loodus- ja täppisteaduste valdkond, tarkvaratehnika, magistriõpe}\\
2007--2011         & {Ukraina Rahvuslik Tehnoloogiaülikool, Igor Sikorski nimeline Kiievi Polütehniline Instituut, bakalaureuseõpe (\textit{cum laude}})
\end{tabular}

\section*{Teenistuskäik}

\begin{tabular}{@{}l@{\hskip7mm}p{90mm}}
2020--...          & {Tartu Ülikool, loodus- ja täppisteaduste valdkond, arvutiteaduse instituut, andmeteaduse nooremlektor}\\
2018--2020         & {Tartu Ülikool, loodus- ja täppisteaduste valdkond, arvutiteaduse instituut, andmeteaduse assistent}\\ 
2015--2018         & {Tartu Ülikool, loodus- ja täppisteaduste valdkond, arvutiteaduse instituut, bioinformaatika nooremteadur}\\
2014--2015         & {Tartu Ülikool, loodus- ja täppisteaduste valdkond, arvutiteaduse instituut, teadusprojekti spetsialist} \\
2013--2014         & {Tartu Ülikool, loodus- ja täppisteaduste valdkond, arvutiteaduse instituut, informaatika assistent}
\end{tabular}

\section*{Teaduspreemiad ja tunnustused}
\begin{itemize}
    \item Ustus Aguri nimeline stipendium silmapaistva panuse eest Eesti infoühiskonna teerajajana, väljaandja Eesti Infotehnoloogia ja Telekommunikatsiooni Liit, sügis 2018; 
    \item Raefondi stipendium Tartu Ülikooli silmapaistvatele rahvusvahelistele tudengitele, sügis 2012;
\end{itemize}

\section*{Õppetöö}
\begin{itemize}
    \item Tartu Ülikool, loodus- ja täppisteaduste valdkond, arvutiteaduse instituut, lektor aines masinõpe, sügis 2020;
    \item Ukraina Katoliku Ülikool (Lviv, Ukraina), lektor aines masinõpe, kevad 2020 ja kevad 2021; 
    \item Kiievi Majanduskool (Kiiev, Ukraina), lektor aines masinõpe edasijõudnutele ja aegridade modelleerimine, kevad 2020;
    \item Tartu Ülikool, loodus- ja täppisteaduste valdkond, arvutiteaduse instituut, õppeassistent aines algoritmid edasijõudnutele, aastatel 2013, 2014, 2016, 2017, 2018, 2019;
    \item Tartu Ülikool, loodus- ja täppisteaduste valdkond, arvutiteaduse instituut, õppeassistent aines andmekaeve 2014, 2015, 2017;
    \item Tartu Ülikool, loodus- ja täppisteaduste valdkond, arvutiteaduse instituut, õppeassistent aines tekstialgoritmid, sügis 2013; 
    \item Seminaride juhendaja ainetes bioinformaatika aastatel 2015, 2016 ja erikursus masinõppes, 2019.
\end{itemize}

\section*{Juhendatud väitekirjad}
\begin{itemize}
    \item Kaspar Hollo (MSc): Anomaly segmentation in microscopy images using deep learning.
	\item Navid Bamdad Roshan (MSc): Change detection in HD-maps using camera images for autonomous driving.
	\item Dmytro Zabolotnii (MSc): High-definition map generation for autonomous driving.
	\item Iaroslav Plutenko (MSc): Metadata for semantic segmentation by employing channel attention mechanism.
	\item Mihkel  Ilisson (MSc): Application of statistical analysis and machine learning methods for analysing blood metabolites.
    \item Richard Annilo (BSc): Using deep learning for medical sound analysis.
    \item Vladyslav Fediukov (MSc): Detection of Changes in Maps Using LiDAR Point Clouds.
    \item Oleh Misko (MSc): Ensembling and transfer learning for multi-domain microscopy image segmentation.
    \item Mikhail Papkov (MSc): Deep neural networks application for cell phenotyping in mixed cultures. Kaasjuhendaja Leopold Parts.
    \item Oleksandr Pryhoda (BSc): Tissue segmentation in histological whole-slide images with deep learning.
    \item Yurii Toma (MSc): Predicting the Impact of Non-Coding Genetic Variants on Transcription Factor Binding with Machine Learning. Kaasjuhendaja Kaur Alasoo.
    \item Sten-Oliver Salumaa (MSc): Convolutional Neural Networks for Cellular Segmentation. Kaasjuhendaja Leopold Parts.
    \item Vitalii Peretiatko (MSc): Using Robust Rank Aggregation for prioritising autoimmune targets on protein microarrays. Kaasjuhendaja Elena S\"{u}gis.
\end{itemize}

\section*{Teadustöö}

Peamised uurimisvaldkonnad:
\begin{itemize}
  \item Bioinformaatika
  \item Masinõpe ja süvaõpe
  \item Arvutinägemine
\end{itemize}
\medskip

Tegevus retsensendina:
\begin{itemize}
  \item International Conference on Machine Learning (2020, 2021)
  \item International Conference on Learning Representations (2020)
  \item Nature Machine Intelligence, BMC Bioinformatics, Nature Scientific Reports, and Bioinformatics.
\end{itemize}
\end{otherlanguage}

\end{document}